\documentclass[review]{elsarticle}

\journal{Advances in Engineering Software}

\usepackage[margin=2cm]{geometry}
\usepackage{amsmath,amssymb}
\usepackage[table]{xcolor}
\usepackage{tikz}
\usepackage{pgfplots}
\usepackage{tabularx}
\usepackage{booktabs}
\usepackage{subcaption}
\usepackage{color, colortbl}
\usepackage{listings}
\usepackage{color}
\usepackage{multirow} 
\definecolor{mygreen}{RGB}{28,172,0}
\definecolor{mylilas}{RGB}{170,55,241}

\begin{document}

\begin{frontmatter}

\title{SEMDOT: Smooth-Edged Material Distribution for Optimizing Topology Algorithm}

\author[deakin]{Yun-Fei Fu}
\author[deakin]{Bernard Rolfe}
\author[monash]{Louis N. S. Chiu}
\author[deakin]{Yanan Wang}
\author[swinburne]{Xiaodong Huang}
\author[deakin]{Kazem Ghabraie\corref{cor}}

\address[deakin]{School of Engineering, Deakin University, Waurn Ponds, VIC 3217, Australia}
\address[monash]{Department of Materials Science and Engineering, Monash University, Clayton, VIC 3800, Australia}
\address[swinburne]{Faculty of Science, Engineering and Technology, Swinburne University of Technology, Hawthorn, VIC 3122, Australia}

\cortext[cor]{Corresponding author: email: k.ghabraie@deakin.edu.au, \quad Tel: +61 3 524 79574}

\begin{abstract}
Element-based topology optimization algorithms capable of generating smooth boundaries have drawn serious attention given the significance of accurate boundary information in engineering applications. The basic framework of a new element-based continuum algorithm is proposed in this paper. This algorithm is based on a smooth-edged material distribution strategy that uses  solid/void grid points assigned to each element.  Named Smooth-Edged Material Distribution for Optimizing Topology (SEMDOT), the algorithm uses elemental volume fractions which depend on the densities of grid points in the Finite Element Analysis (FEA) model rather than elemental densities. Several numerical examples are studied to demonstrate the application and effectiveness of SEMDOT. In these examples, SEMDOT proved to be capable of obtaining optimized topologies with smooth and clear boundaries showing better or comparable performance compared to other topology optimization methods. Through these examples, first, the advantages of using the Heaviside smooth function are discussed in comparison to the Heaviside step function. Then, the benefits of introducing multiple filtering steps in this algorithm are shown. Finally, comparisons are conducted to exhibit the differences between SEMDOT and some well-established element-based algorithms. The validation of the sensitivity analysis method adopted in SEMDOT is conducted using a typical compliant mechanism design case. In addition, this paper provides the Matlab code of SEMDOT for educational and academic purposes.
\end{abstract}


\end{frontmatter}

%\linenumbers

\section{Introduction}\label{Sec: Introduction}
Topology optimization basically aims to distribute a given amount of material within a predefined design domain such that optimal or near optimal structural performance can be obtained
\cite{zuo2015simple,chiu2018effect,pollini2020poor}. It often provides highly efficient designs that could not be obtained by simple intuition without assuming any prior structural configuration. Topology optimization as a design method has been greatly developed and extensively used since the pioneering paper on numerical topology optimization by \citet{bendsoe1988generating}. A number of topology optimization algorithms have been proposed based on different strategies: homogenization of microstructures \cite{bendsoe1988generating}, using elemental densities as design variables \cite{bendsoe1989optimal}, evolutionary approaches \cite{xie1993simple}, topological derivative \cite{sokolowski1999topological}, level-set (LS) \cite{allaire2002level,wang2003level}, phase field \cite{bourdin2003design}, moving morphable component (MMC) \cite{guo2014doing}, moving morphable void (MMV) \cite{zhang2017structural}, elemental volume fractions \cite{da2018evolutionary}, and using the floating projection \cite{huang2020smooth,huang2020smooth1}. In recent years, these topology optimization approaches have been applied in a wide range of distinct engineering problems, including frequency responses \cite{ma1993structural,zuo2012evolutionary}, stress problems \cite{le2010stress,de2015stress}, convection problems \cite{alexandersen2014topology,alexandersen2016large,asmussen2019poor}, structural failure problems \cite{nabaki2019evolutionary,wang2020robust}, large-scale problems \cite{aage2015topology,aage2013parallel,wang2020efficient}, nanophotonics \cite{jensen2011topology}, metamaterial design \citep{chen2019topological}, and manufacturing oriented methods \cite{liu2016survey,vatanabe2016topology,liu2019piecewise,langelaar2019topology,yu2020stress} have been presented in recent years.

Early proposed topology optimization algorithms are mainly element-based such as solid isotropic material with penalization (SIMP) algorithm \cite{sigmund1998numerical} and bi-directional evolutionary structural optimization (BESO) algorithm \cite{yang1999bidirectional}. SIMP uses the artificial power-law function between elemental densities and material properties to suppress intermediate elements to a solution with black and white elements, and BESO heuristically updates design variables using discrete values (0 and 1). As elements are not only involved in finite element analysis (FEA) but the formation of topological boundaries, zigzag (for example, BESO) or both zigzag and blurry boundaries (for example, SIMP) will be inevitably generated. Therefore, shape optimization or other post-processing methods have to be used to obtain accurate boundary information after topology optimization \cite{ghabraie2010shape1,ghabraie2010shape,liu2020post}. Given the significance of the accurate boundary representation, some proposed algorithms such as the level-set method, MMC-based method, elemental volume fractions based method, and using the floating projection have successfully solved the boundary issue. Even though there are a number of algorithms capable of forming smooth or high resolution boundaries, element-based algorithms that could combine the benefits of different methods are generally preferred because of their advantages of easy implementation and ability of creating holes freely across the design domain and obtaining final topologies that are not heavily dependent on the initial guess \citep{huang2020smooth}.

Elemental volume fractions based methods are originally from the evolutionary topology optimization (ETO) algorithms using the continuation method on the volume and BESO-based optimizer \cite{da2018evolutionary,chen2019topology,liu2019stress,li2019topological}. To provide a more easy-to-use, flexible, and efficient optimization platform, the authors proposed a new smooth continuum topology optimization algorithm through combining the benefits of the smooth representation in ETO and density-based optimization in SIMP \cite{fu2019AM, fu2019TOC, fu2019PAM, fu20203DTO}. The proposed algorithm is termed Smooth-Edged Material Distribution for Optimizing Topology (SEMDOT) based on its optimization mechanism. The basic idea of elemental volume fractions based algorithms (for example, ETO and SEMDOT) is to pursue the solid/void design of grid points that are assigned to each element through which smooth boundaries can be obtained. The concept of using design points within an element has been proposed and studied before ETO and SEMDOT. \citet{nguyen2010computational} presented a multiresolution topology optimization (MTOP) scheme that can generate high resolution designs with relatively low computational cost using three different meshes: the displacement mesh, density mesh, and design variable mesh. In MTOP, Gauss points were used for the integration of the stiffness matrix. Afterwards, \citet{park2015multi} first used MTOP to obtain high resolution designs for 3D multi-material problems. However, \citet{gupta2018qr} pointed out that the complexity of MTOP would restrict its attractiveness to students and scholars. Instead of using design points within each element, \citet{kang2011structural} proposed a pointwise density-based interpolation method where the density field is constructed from design points within a certain circular influence domain of the point. Unlike ETO and SEMDOT, design points in the method proposed by \citet{kang2011structural} are not involved in forming smooth boundaries, so the non-smooth boundary issue persists in it.

Compared with some newly developed or improved algorithms capable of generating smooth boundaries, SEMDOT can be easily integrated with some existing methods that were established based on SIMP to achieve specific performance goals. An example is the combination of SEMDOT and Langelaar's additive manufacturing (AM) filter \cite{langelaar2017additive} which can successfully generate smooth self-supporting topologies \cite{fu2019AM,fu2019PAM,fu2020optimizing}. Other than Langelaar's AM filter, some other strategies regarding AM restrictions proposed by \citet{van2018continuous} and \citet{zhang2019topology} can also be considered in SEMDOT for the support-free design. However, extra efforts have to be made for MMCs-based methods to obtain self-supporting designs \cite{guo2017self}. In addition, the effectiveness of SEMDOT in solving 3D optimization problems is recently demonstrated by \citet{fu20203DTO}, through solving a number of benchmark problems and a comparison with a well-established large-scale topology optimization framework, TopOpt (proposed by \citet{aage2015topology}). 

Even though the theoretical framework of SEMDOT was built and demonstrated by authors, other benefits and distinctions of SEMDOT compared with some current element-based algorithms have not been thoroughly discussed. Furthermore, details of SEMDOT algorithm and some of its subtle differences with methods like ETO which translate into more robust performance have not been discussed before. 

In this work, the reason behind using the Heaviside smooth function in SEMDOT instead of the Heaviside step function that is extensively used in ETO algorithms is explained. Effects of different combinations of filter radii on performance, convergence, and topologies, which have not been discussed in authors' previous works, are investigated, and the rationality of the sensitivity analysis strategy used in SEMDOT is proved through a compliant mechanism design case. Numerical comparisons with other element-based topology optimization methods are thoroughly conducted. Finally, the Matlab code of SEMDOT is released to the topology optimization community facilitating the replication of the results presented in this paper, and also for future use.

An overview of this paper is as follows. Section \ref{Sec:Formulation} explains the mathematical framework of SEMDOT. Section \ref{Sec: NE} conducts several numerical examples to exhibit the benefits and distinctions of SEMDOT compared with a number of element-based topology optimization algorithms. Concluding remarks are drawn in Section \ref{Sec: Conclusions}.  Finally, the Matlab code of SEMDOT is presented in the Appendix.

\section{Formulation} \label{Sec:Formulation}

\subsection{Mathematical descriptions of smooth-edged material distribution strategy}
\begin{figure}[htbp!]
	\centering
	\includegraphics[width=400 pt]{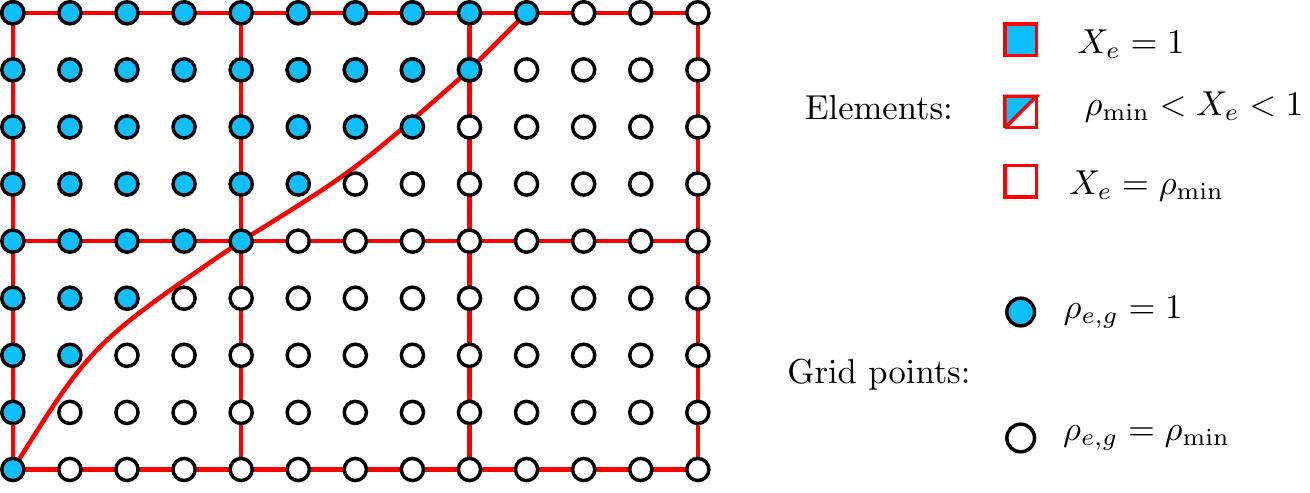}
	\caption{Illustration of smooth-edged material distribution}
	\label{Fig: Illustration}
\end{figure}

The smooth-edged material distribution strategy in SEMDOT is to form a clear topological boundary based on the the solid/void design of grid points that are assigned to each element, as illustrated in Figure \ref{Fig: Illustration}. In Figure \ref{Fig: Illustration}, the density of the $g$th grid point in the $e$th element $\rho_{e,g}$ is assigned 1 to represent a solid grid point or a small artificial value $\rho_{\min}$ (typically 0.001) to represent a void grid point. In SEMDOT, densities at grid points ($\rho_{e,g}$) are actually the design variables. Following the power-law model in SIMP, the material interpolation scheme at grid points is expressed by	
\begin{equation}
	E_e(\rho_{e,g})=\rho_{e,g}^pE^1
	\label{Eq:ERHO}
\end{equation}
where $E_e(\rho_{e,g})$ is the function of the Young's modulus with respect to grid point densities, $E^1$ is the Young's modulus of the base material, and $p$ is a penalty coefficient.

These grid point densities are not directly involved in Finite Element Analysis (FEA). Instead, in SEMDOT, element-based variables are used in the FEA model. Elemental volume fractions that depend on the densities at grid points are defined as 
\begin{equation}
X_{e}=\dfrac{1}{N} \sum\limits_{g=1}^N \rho_{e,g}
\label{Eq: Vef}
\end{equation}
where $X_e$ is the volume fraction of the $e$th element and $N$ is the total number of grid points in each element. Even though the number of grid points is much higher than elements, because grid points are not involved in FEA, SEMDOT can maintain a proper balance between the smoothness of topological boundaries and computational cost.

As illustrated in Figure \ref{Fig: Illustration} and Equation \ref{Eq: Vef}, in solid or void elements, grid point densities are homogeneously distributed. Based on Equation \ref{Eq:ERHO}, for solid and void elements we can write
\begin{equation}
E_e(X_e)=X_e^pE^1 \ \ \ X_e \in \{\rho_{\min}, 1\}
\label{Eq:EXe}
\end{equation}

In SEMDOT, the Young's modulus and stiffness matrices of solid and void elements are calculated using the standard SIMP expressions based on Equation \ref{Eq:EXe}. However, unlike SIMP, the intermediate (boundary) elements in SEMDOT are non-homogenised combination of solid and void materials (bi-material elements), as illustrated in Figure \ref{Fig: Illustration}. Because of this representation, elemental material properties are not well-defined functions of $X_e$ \citep{li2019topological}. Therefore, in SEMDOT such properties are approximated using a linear interpolation between the two phases of solid and void in the form, $\square = (1-X_e)\square_\text{void} + X_e\square_\text{solid}$ where $X_e$ and $(1-X_e)$ can be regarded as weighting factors of solid and void materials, respectively. This linear interpolation scheme is used in this version of SEMDOT because of its simplicity. Based on the above discussion, the elemental stiffness matrix can be estimated as
\begin{equation}
\begin{split}
\mathbf{K}_e(X_e)=
& (1-X_e)\mathbf{K}_e^0+X_e\mathbf{K}_e^1\\
& =(1-X_e)\rho_{\min}^p\mathbf{K}_e^1+X_e\mathbf{K}_e^1, \ \ \ X_e \in [\rho_{\min}, 1]
\label{Eq: KXe}
\end{split}
\end{equation} 
where $\mathbf{K}_e(X_e)$ is the function of the stiffness matrix with respect to the elemental volume fraction $X_e$,  $\mathbf{K}_e^0$ is the stiffness matrix of the void element , and $\mathbf{K}_e^1$ is the stiffness matrix of the solid element.

The sensitivity measures the effectiveness of altering elemental volume fractions on reducing or increasing the objective function through which the search direction of optimization can be determined \cite{ghabraie2015improved,ghabraie2015eso}. Instead of calculating the sensitivities with respect to $\rho_{e,g}$, we can find the sensitivities with respect to $X_e$ as auxiliary variables and then find $\rho_{e,g}$ values based on $X_e$ in a later stage.  

Assuming compliance as the objective function (see Section \ref{sc:problemstatement}), for void and solid elements, based on Equation \ref{Eq:EXe}, the sensitivities with respect to $X_e$ can be calculated as $\partial C(X_e)/\partial X_e\vert_{X_e=\rho_{\min}}=-p\rho_{\min}^{p-1}\mathbf{u}_e^\mathbf{T} \mathbf{K}_e^1 \mathbf{u}_e$ and $\partial C(X_e)/\partial X_e \vert_{X_e=1}=-p\mathbf{u}_e^\mathbf{T} \mathbf{K}_e^1 \mathbf{u}_e$, respectively. On the other hand, calculating the accurate sensitivities of boundary (intermediate) elements in SEMDOT with respect to $X_e$ is impossible as these boundary elements are not homogeneous. Therefore, similar to what noted for material properties above, sensitivities of boundary elements are approximated as a linear combination of the sensitivities of void elements with a weighting factor of $(1-X_e)$ and sensitivities of solid elements with a weighting factor of $X_e$, i.e.
\begin{equation}
\begin{split}
\dfrac{\partial C(X_e)}{\partial X_e}  \approx & (1-X_e)\dfrac{\partial C(X_e)}{\partial X_e} \bigg\vert_{X_e=\rho_{\min}} + X_e\dfrac{\partial C(X_e)}{\partial X_e} \bigg\vert_{X_e=1} \\
& = -p[(1-X_e)\rho_{\min}^{p-1}+X_e]\mathbf{u}_e^\mathbf{T} \mathbf{K}_e^1 \mathbf{u}_e, \ \ \ X_e \in [\rho_{\min}, 1]
\end{split}
\label{Eq: boundaryES}
\end{equation}
where $C$ is the objective function and $\mathbf{u}_e$ is the displacement vector of the $e$th element.  It is noted that Equation \ref{Eq: boundaryES} represents the sensitivity analysis of the compliance minimization problem. The validity of this approximation can be verified through numerical examples in Section \ref{Sec: NE}.

\subsection{Topology optimization problems}	\label{sc:problemstatement}
The minimum compliance (the maximum stiffness) optimization problem, one of the most popular test cases for topology optimization, is used to comprehensively demonstrate the algorithm mechanism of SEMDOT and conduct numerical comparisons in this paper. The corresponding optimization problem can be stated as
\begin{equation}
\begin{split}
\min:\, & C(X_e)=\mathbf{f^T}\mathbf{u}\\ 
\mathrm{subject\,\, to:}\,\, & \mathbf{K}(X_e)\mathbf{u}=\mathbf{f} \\
& \dfrac{\sum\limits_{e=1}^M X_e V_e}{\sum\limits_{e=1}^M V_e} -V^* \leq 0 \\
&0 < \rho_{\min} \leq X_e \leq 1; \ e=1, 2, \cdots, M
\end{split}
\label{Eq: Objectivef}
\end{equation}
where $\mathbf{f}$ and $\mathbf{u}$ are global force and displacement vectors, respectively; $\mathbf{K}$ is the global stiffness matrix;  $V_e$ is the volume of the $e$th element; $V^*$ is the prescribed value of the allowable volume; and $M$ is the total number of elements in the design domain.

The compliant mechanism design case is also considered to further demonstrate the effectiveness of the sensitivity analysis method (Equation \ref{Eq: boundaryES}) adopted in SEMDOT. In compliant mechanism design, single-piece flexible structures transfer an input force or displacement to another point through elastic deformation  \citep{bensvarphie2003topology,huang2014topology,zhu2020design}. The corresponding optimization problem can be stated as the following mathematical program:
	
\begin{equation}
\begin{split}
\min:\, & C(X_e)=-u_{out}={-}\mathbf{L^Tu}{=\mathbf{\tilde{u}}^\mathbf{T} \mathbf{K} \mathbf{u}} \\ 
\mathrm{subject\,\, to:}\,\, & \mathbf{K}(X_e)\mathbf{u}=\mathbf{f}_{in} \\
& \dfrac{\sum\limits_{e=1}^M X_e V_e}{\sum\limits_{e=1}^M V_e} -V^* \leq 0 \\
&0 < \rho_{\min} \leq X_e \leq 1; \ e=1, 2, \cdots, M
\end{split}
\label{Eq: ObjCMs}
\end{equation}
where $\mathbf{L}$ is a unit length vector with zeros at all degrees of freedom except at the output point where it is one, $u_{out}$ is the output port displacement, { $\mathbf{\tilde{u}}$ is the dummy load displacement vector calculated by solving $\mathbf{K}\mathbf{\tilde{u}}={-}\mathbf{L}$,} and $\mathbf{f}_{in}$ is the input force vector. Similar to Equation \ref{Eq: boundaryES}, sensitivities for compliant mechanism design can be estimated as
\begin{equation}
\begin{split}
\dfrac{\partial C(X_e)}{\partial X_e}  \approx & (1-X_e)\dfrac{\partial C(X_e)}{\partial X_e} \bigg\vert_{X_e=\rho_{\min}} + X_e\dfrac{\partial C(X_e)}{\partial X_e} \bigg\vert_{X_e=1} \\
& = p[(1-X_e)\rho_{\min}^{p-1}+X_e]{ \mathbf{\tilde{u}}_{e}^\mathbf{T}} \mathbf{K}_e^1 \mathbf{u}_e, \ \ \ X_e \in [\rho_{\min}, 1]
\end{split}
\label{Eq: boundaryCMD}
\end{equation}
{where $\mathbf{\tilde{u}}_{e}$ is the dummy load displacement vector of the $e$th element.}

\subsection{Filtering}
Using filters in topology optimization is an effective way to ensure regularity or existence of topological designs \cite{lazarov2011filters}. The basic idea of filters is to substitute a (possibly) non-regular function with its regularization \cite{lazarov2011filters}. The filter that has the same form of the density filter is used for regularization of topology optimization problems in SEMDOT. The filtering technique for elemental volume fractions is expressed as \cite{sigmund2007morphology}:

\begin{equation}
\tilde{X}_e=\dfrac{\sum\limits_{l=1}^{N_e} \omega_{el} X_l}{\sum\limits_{l=1}^{N_e}  \omega_{el}}	
\label{Eq: Filter}
\end{equation}
where $\tilde{X}_e$ is the filtered elemental volume fraction, $N_e$ is the neighborhood set of elements within the filter domain for the $e$th element that is a circle centered at the centroid of this element with a predefined filter radius $r_{\min}$, and $\omega_{el}$ is a linear weight factor defined as	
\begin{equation}
\omega_{el}=\max (0, r_{\min}-\Delta(e,l))
\label{Eq: FilterW}
\end{equation}
where $\Delta(e,l)$ is the center-to-center distance of the $l$th element within the filter domain to the $e$th element. It is noted that the filter for elemental volume fractions (Equations \ref{Eq: Filter} and \ref{Eq: FilterW}) can be substituted by other filters that were developed based on standard SIMP such as the sensitivity filter presented by \citet{sigmund1997design11} and the partial differential equation (PDE) filters proposed by \citet{lazarov2011filters}. It should be noted that SEMDOT aims to obtain a topological design with intermediate (gray) elements only along  boundaries instead of pursuing a pure black and white (0/1) design, which means intermediate elements are useful for the determination of smooth boundaries. Consequently, some filters suppressing intermediate elements to black and white (0/1) elements (for example, Heaviside projection filter \cite{guest2004achieving}, morphology based filters \cite{sigmund2007morphology}, and volume preserving Heaviside projection scheme \cite{xu2010volume}) do not suit SEMDOT.

Even though the filtering technique is applied on $X_e$, sensitivities with respect to $X_e$ are still required by the optimizer to update $X_e$. Sensitivities of the objective functional with respect to $X_e$ can then be calculated based on the chain rule as
\begin{equation}
	\dfrac{\partial C(\tilde{X}_e)}{\partial X_l}=\sum\limits_{e=1}^{N_l} \dfrac{\partial C(\tilde{X}_e)}{\partial \tilde{X}_e} \dfrac{\partial \tilde{X}_e}{\partial X_l} = \sum\limits_{e=1}^{N_l} \dfrac{\omega_{el}}{\sum\limits_{\varsigma=1}^{N_e} \omega_{e\varsigma}} \dfrac{\partial C(\tilde{X}_e)}{\partial \tilde{X}_e} 
\end{equation}

To obtain the densities of grid points, nodal densities should be obtained first in SEMDOT. Nodal densities can be obtained using a heuristic filter similar to the one presented in BESO \cite{huang2007convergent}:
\begin{equation}
\rho_n =\dfrac{\sum\limits_{e=1}^M \omega_{ne} \tilde{X}_e}{\sum\limits_{e=1}^M  \omega_{ne}}	
\label{Eq: FEN}
\end{equation}
where $\rho_n$ is the density of the $n$th node and $\omega_{ne}$ is the weight factor defined as
\begin{equation}
\omega_{ne}=\max (0, \Upsilon_{\min}-\Delta(n,e))
\end{equation}
where $\Delta(n,e)$ is the distance between the $n$th node and the center of the $e$th element and $\Upsilon_{\min}$ is the heuristic filter radius. It is important that the heuristic filter radius $\Upsilon_{\min}$ in SEMDOT is set to a value not less than 1, otherwise topological designs cannot be obtained for most test cases.

\subsection{Generation of smooth topological boundaries}
The density at the grid point $\rho(\zeta, \eta)$ in the $e$th element can be obtained through linear interpolation of nodal densities $\rho_n$. Considering a four-node element as an example, the density of the grid point $\rho(\zeta, \eta)$ is expressed by
\begin{equation}
\rho(\zeta, \eta)=\sum\limits_{\gamma=1}^4 N^\gamma (\zeta, \eta)\rho_n^\gamma  \ \ \ \mathrm{and} \ \ \ \rho(\zeta, \eta) \in \rho(x,y)
\label{Eq: Griddensity}
\end{equation}  
where $(\zeta, \eta)$ is the local coordinate of the grid point, $\rho_n^\gamma$ is the density for the $\gamma$th node of the element, and  $N^\gamma(\zeta, \eta)$ is an appropriate shape function. 

Theoretically, the solid/void design of grid points can be implemented by either Heaviside step function or Heaviside smooth function. The Heaviside step function is expressed by \citep{guest2004achieving}
\begin{equation}
\rho_{e,g}=
\begin{cases}
1   & \mathrm{if}  \ \rho_{e,g} > \Psi \\ 
\rho_{\mathrm{min}}  & \mathrm{if} \ \rho_{e,g} \leq \Psi 
\end{cases}
\label{Eq: Heavisidestep}
\end{equation}
where $\Psi$ is a threshold value. 

The tanh-based expression of the Heaviside smooth function proposed by \citet{wang2011projection} is used in SEMDOT, which is 
\begin{equation}
\rho_{e,g}=\dfrac{\tanh(\beta \cdot \Psi)+\tanh[\beta \cdot (\rho(x,y)-\Psi)]}{\tanh(\beta \cdot \Psi)+\tanh[\beta \cdot (1.0-\Psi)]}
\label{Eq: Heavisidesmooth}
\end{equation}
where $\beta$ is a scaling parameter that controls the steepness and is updated by 

\begin{equation}
\beta_k=\beta_{k-1} + \Lambda
\label{Eq: Hupdate}
\end{equation}
where the subscripts denote the iteration number and $\Lambda$ is the evolution rate for $\beta$. 

Once grid point densities are calculated, filtered elemental volume fractions are updated for the next round of FEA through summing up the grid points for each element:
\begin{equation}
\tilde{X}_e^{\text new}=\dfrac{1}{N} \sum\limits_{g=1}^N \rho_{e,g}^{\text new}
\label{Eq: Xenew}
\end{equation}
where $\rho_{e,g}^{\text new}$ is the density of the grid point obtained by the Heaviside step or smooth function.

The relationship between $\tilde{X}_e$ and $\tilde{X}_e^{\text new}$ can be simply established as:
\begin{equation}
\delta_e=\tilde{X}_e^{\text new}-\tilde{X}_e
\label{Eq:Delta}
\end{equation}
where $\delta_e$ is the deviation of $\tilde{X}_e$ and $\tilde{X}_e^{\text new}$. Equation \ref{Eq:Delta} was mentioned in \citep{fu2019AM}, whereas no details were provided there. Hence comprehensive discussions regarding this relationship will be presented in Section \ref{Sec: DEVF}.

The shape of the topological design is represented by a level-set function $\Phi(x,y)$:
\begin{equation}
\Phi(x,y)=
\begin{cases}
\rho(x,y)-\Psi>0  \ \ \ \  \text{for solid region} \\
\rho(x,y)-\Psi=0 \ \ \ \   \text{for boundary} \\
\rho(x,y)-\Psi<0 \ \ \ \  \text{for void region}
\end{cases}
\label{Eq: levelset}
\end{equation}
where $(x,y)$ is the global coordinate of grid points, $\Phi(x,y)$ is the level-set function for grid points, and $\rho(x,y)$ is the density of the grid point at $(x,y)$. In SEMDOT, the threshold value $\Psi$ is determined iteratively using the bi-section method such that the target volume constrain can be satisfied iteratively. Unlike the direct sensitivity-based level-set function presented by \citet{da2018evolutionary}  and \citet{liu2019stress}, Equation \ref{Eq: levelset} uses the densities of grid points that are determined originally based on sensitivity analysis (Equation \ref{Eq: boundaryES}). Therefore, Equation \ref{Eq: levelset} can be regarded as an indirect sensitivity-based or grid point density based level-set function.

\subsection{Convergence criteria} \label{Sec: CONC}
For BESO-based methods, the optimization procedure terminates when the average change of the objective function values in recent iterations is less than a prescribed tolerance value \cite{xia2016bi}. As previously mentioned by \citet{sigmund2013topology}, the convergence criterion of BESO could prematurely terminate the optimization procedure, because design variables may be in an oscillating state switching between 0 and 1 even though the objective function value based convergence criterion is satisfied. This oscillating state would result in a solution far from the optimum. On the other hand, the optimization procedure of standard SIMP is terminated when the maximum variation of design variables within two successive iterations is less than a prescribed tolerance \cite{sigmund200199}. However, when solving an optimization problem with a large number of elements, convergence difficulties can be observed in SIMP, as is discussed in Section \ref{Sec: NC}.

If the Heaviside step function is considered in SEMDOT, the optimization procedure terminates when the overall topological alteration is less than its predefined tolerance value, which can be stated as:
\begin{equation}
\dfrac{\sum\limits_{e=1}^M |X_e^k-X_e^{k-1}|}{\sum\limits_{e=1}^M X_e^k} \leq  \tau
\label{Eq: Terror}
\end{equation}
where $\tau$ is the tolerance value for the overall topological alteration. This convergence criterion (Equation \ref{Eq: Terror}) is based on the overall measure of the variation of design variables compared to the local measure used in SIMP, and its better performance in determining convergency had been demonstrated by \citet{fu2019AM,fu2019PAM,fu20203DTO}. It is noted that Equation \ref{Eq: Terror} can also be used in SIMP or BESO and is likely to improve their determination of termination point.

When the Heaviside smooth function is considered, intermediate elements that are not along the boundary will inevitably appear during the optimization process. Intermediate elements that are not along the boundary are the elements with the maximum grid point density less than 1 and the minimum density larger than $\rho_{\min}$. To measure the accuracy of the level-set function (Equation \ref{Eq: levelset}) representing the smooth topological boundary, an indicator termed topological boundary error is defined as the ratio of the number of the intermediate elements that are not along the boundary to the total number of elements. The topological boundary error convergence criterion is therefore stated as:
\begin{equation}
\dfrac{N_v}{M} \leq \epsilon;
\label{Eq: Berror}
\end{equation}
where $N_v$ is the number of intermediate elements that are not along boundaries, $M$ is the total number of elements, and $\epsilon$ is the tolerance value for the topological boundary error. When the topological boundary error becomes negligible, intermediate elements are all distributed along the boundary, meaning that the level-set function  accurately represents the current design.

In comparison to the Heaviside smooth function (Equation \ref{Eq: Heavisidesmooth}), the Heaviside step function (Equation \ref{Eq: Heavisidestep}) will not cause the topological boundary error, but it can cause numerical instabilities in SEMDOT, as is discussed in Section \ref{Sec: EHF}. 

\subsection{Optimization procedure}
\begin{figure}[htbp!]
	\centering
	\includegraphics[width=300 pt]{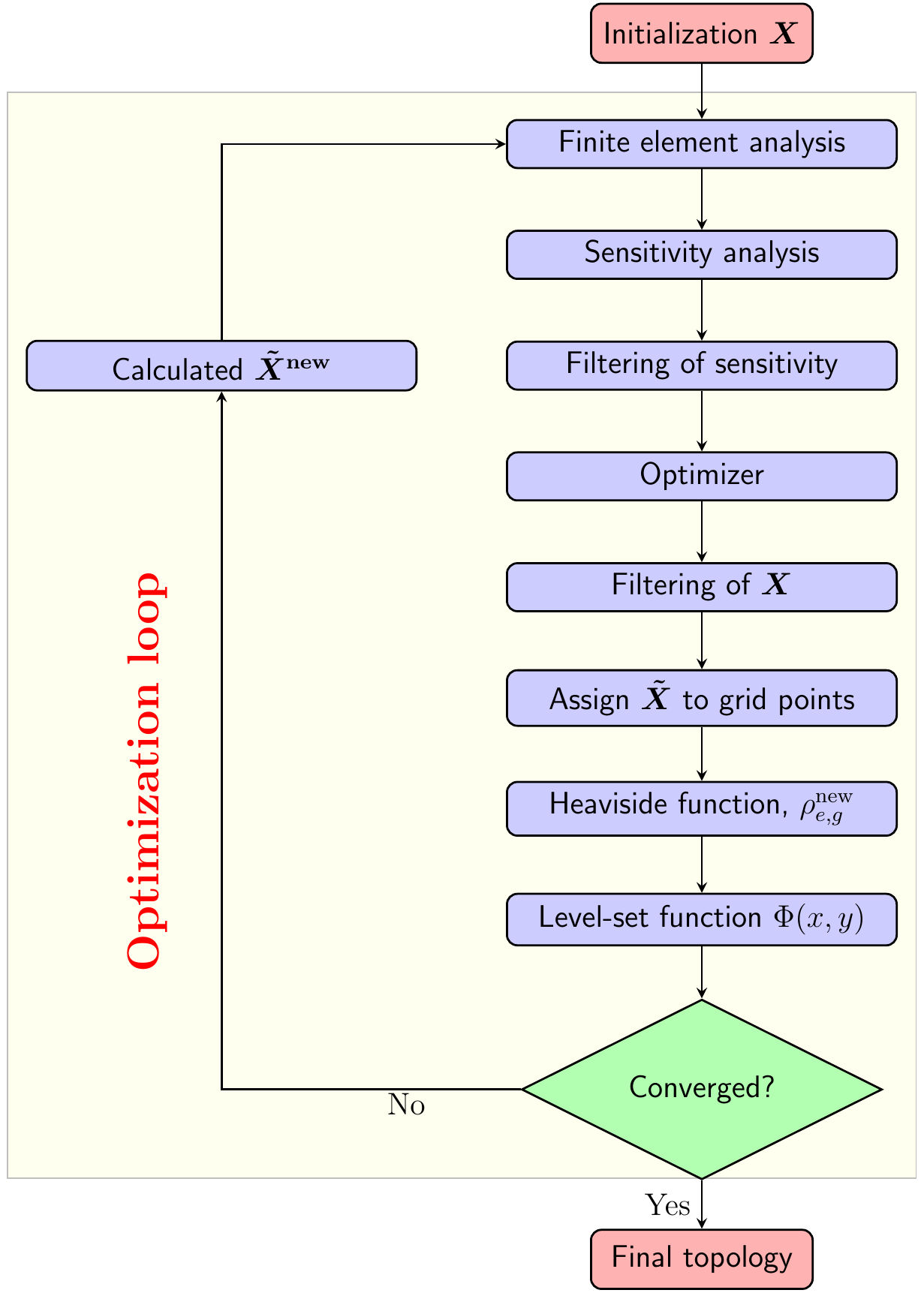}
	\caption{Flowchart of the SEMDOT method}
	\label{Fig: FC}
\end{figure}

The optimization procedure of SEMDOT mainly consists of two parts: the implementation of structural changes based on elemental volume fractions and the generation of smooth topological boundaries based on the solid/void design of grid points. The improved and simplified flowchart of SEMDOT, which is based on the flowchart presented by \citet{fu2019PAM}, is illustrated in Figure \ref{Fig: FC}.

%%%%%%%%%%%%%%%%%%%%%%%%%%%%%%%%%%%%%%%%%%
\section{Numerical Experiments} \label{Sec: NE}
Benchmark 2D optimization problems are solved to demonstrate the validity of SEMDOT and exhibit the differences between SEMDOT and some existing algorithms: SIMP, BESO, and ETO. The prescribed value of the allowable volume $V^{*}$ is set to 0.3. For all numerical examples, an isotropic linear elastic material model is assumed with Young's modulus of $E=1$ MPa and Poisson's ratio of $\mu=0.3$, and equally sized four-node plane-stress elements are used. Following the parametric studies in \citep{fu20203DTO}, $\beta_0=0.5$ and $\Lambda=0.5$ are employed in the Heaviside smooth function (Equation \ref{Eq: Heavisidesmooth}), and the penalty coefficient of $p=1.5$ is used in SEMDOT. The tolerance values of $\epsilon=0.001$ and $\tau=0.001$ are used in the convergence criteria. Following the parametric studies in \citep{fu2019TOC}, a grid with 10$\times$10 points in each element is used.  Unless otherwise stated, the heuristic filter radius $\Upsilon_{\min}$ is set to 1 time element width ($\Upsilon_{\min}$=1). In addition, the method of moving asymptotes (MMA) proposed by \citet{svanberg1993method} is used to update design variables, and default parameters in MMA are adopted.

\subsection{Comparisons between Heaviside step and smooth functions} \label{Sec: EHF}
\begin{figure}[htbp!]
	\centering
	\includegraphics[scale=0.8]{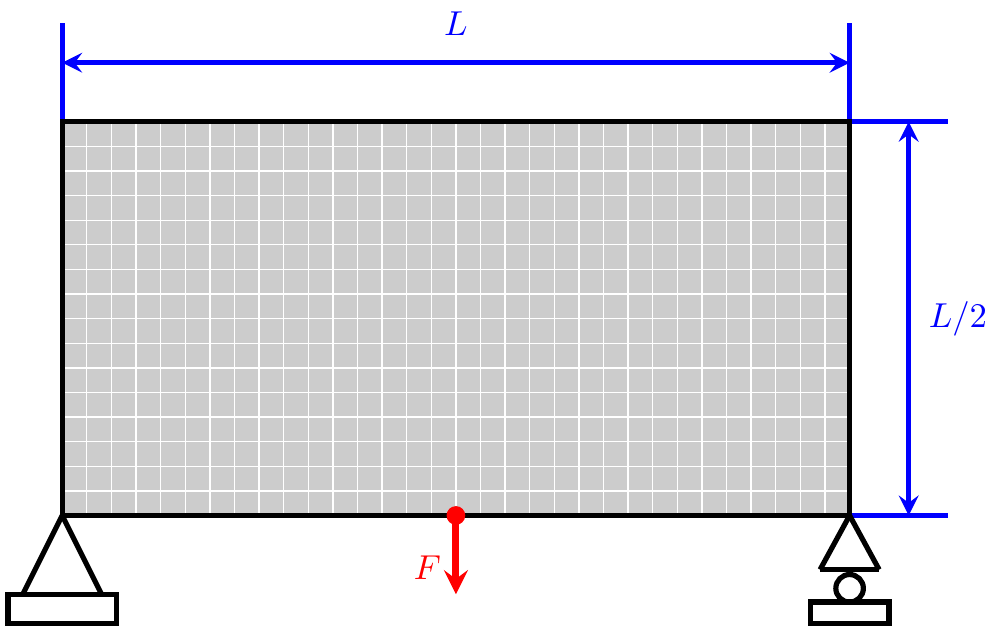}
	\caption{Design domain of a simply supported deep beam}
	\label{Fig: Sbeam}
\end{figure}

A simply supported deep beam subjected to a unit vertical load ($F=-1\mathrm{N}$) at its bottom center is considered to investigate the influences of the Heaviside step and smooth functions on the topological design, performance, and convergency of SEMDOT. The design domain and boundary condition are shown in Figure \ref{Fig: Sbeam}. The bottom left corner is hinged, and the vertical displacement at the bottom right corner is prevented. A $180\times90$ mesh is used, and the filter radius $r_{\min}$ is set to 2 time elements width ($r_{\min}$=2). 

\begin{figure}[htbp!]
	\centering
	\begin{subfigure}{\textwidth}
	\centering
	\includegraphics[scale=0.7]{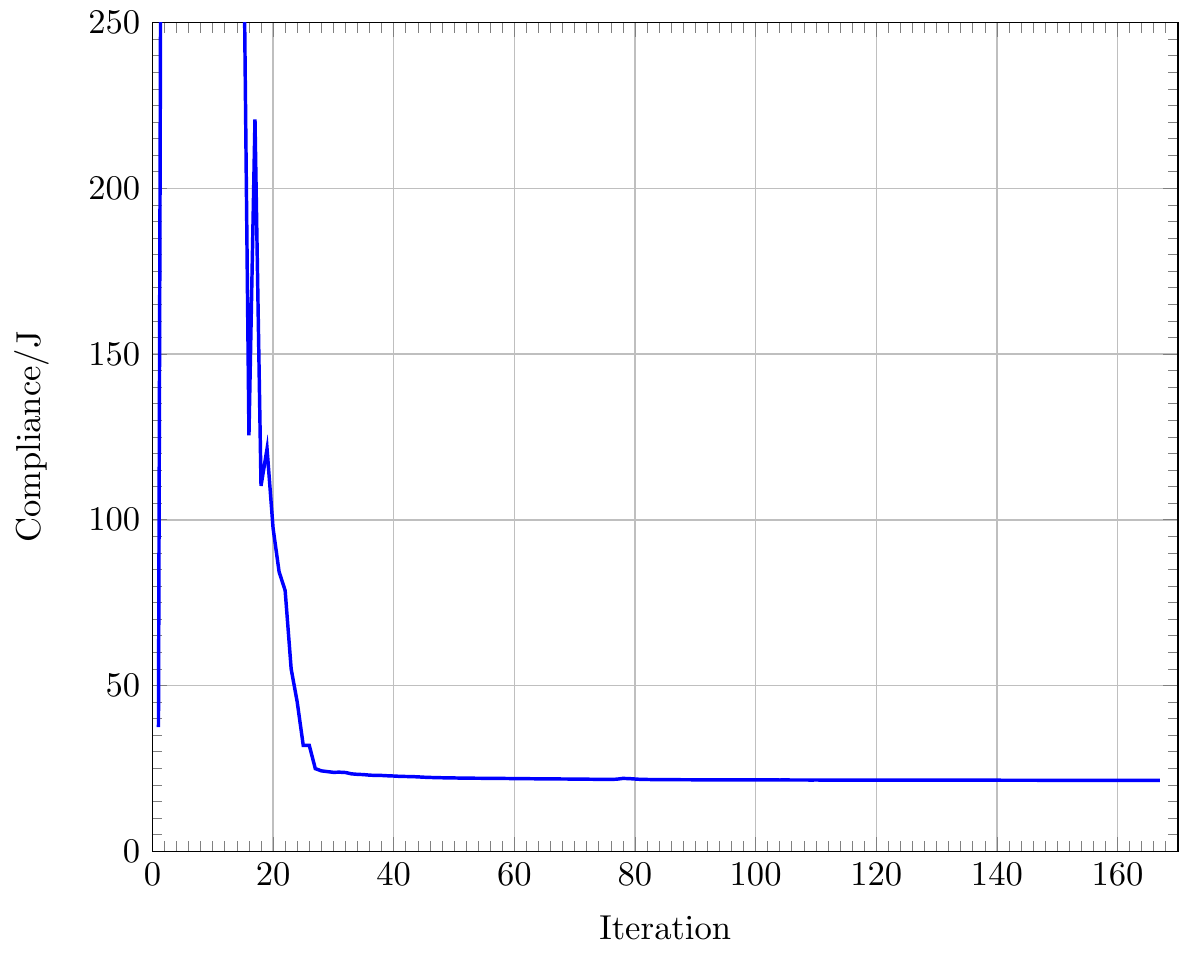}
	\caption{Compliance}
	\label{sFig: HStC}
	\end{subfigure}
\\
	\begin{subfigure}{0.32\textwidth}
	\centering
	\includegraphics[width=145 pt]{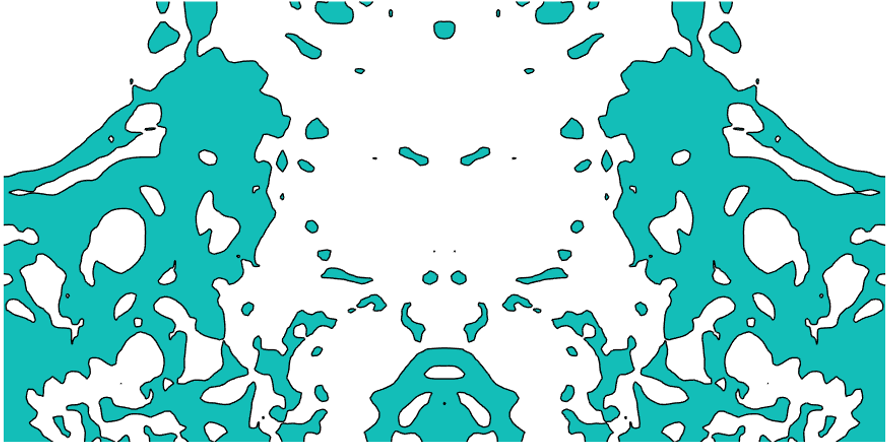}
	\caption{Iteration 10}
	\label{sFig: HST10}
	\end{subfigure}
	\begin{subfigure}{0.32\textwidth}
	\centering
	\includegraphics[width=145 pt]{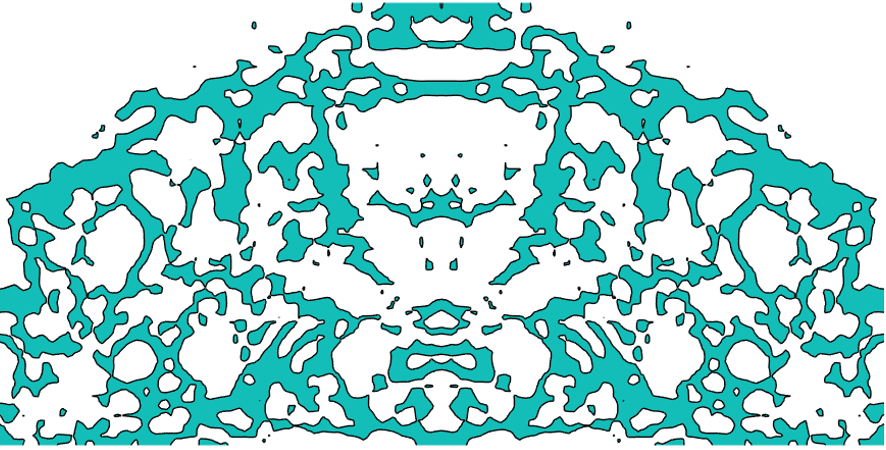}
	\caption{Iteration 20}
	\label{sFig: HST20}
	\end{subfigure}
	\begin{subfigure}{0.32\textwidth}
	\centering
	\includegraphics[width=145 pt]{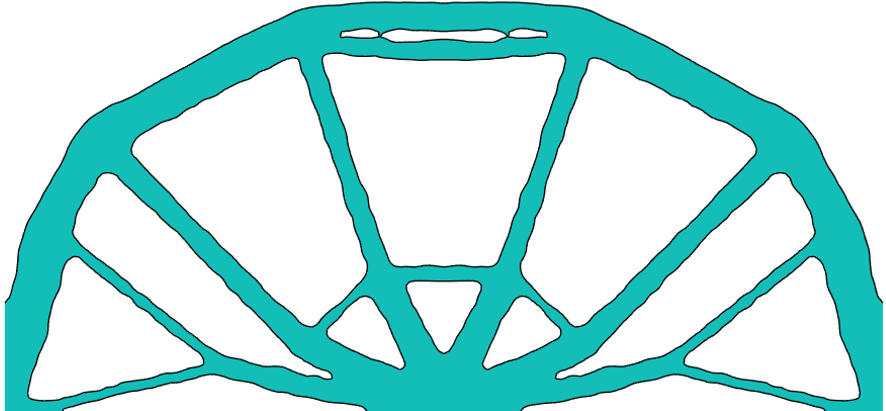}
	\caption{Iteration 40}
	\label{sFig: HST40}
	\end{subfigure}
\\
	\begin{subfigure}{0.32\textwidth}
	\centering
	\includegraphics[width=145 pt]{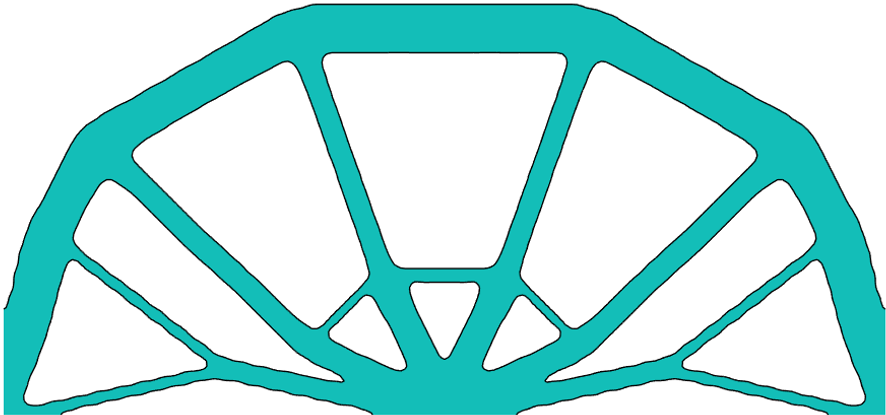}
	\caption{Iteration 60}
	\label{sFig: HST60}
	\end{subfigure}
	\begin{subfigure}{0.32\textwidth}
	\centering
	\includegraphics[width=145 pt]{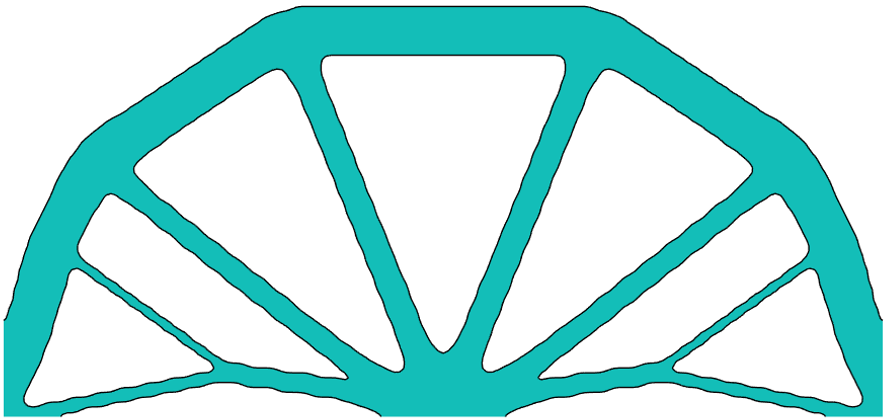}
	\caption{Iteration 100}
	\label{sFig: HST100}
	\end{subfigure}
	\begin{subfigure}{0.32\textwidth}
	\centering
	\includegraphics[width=145 pt]{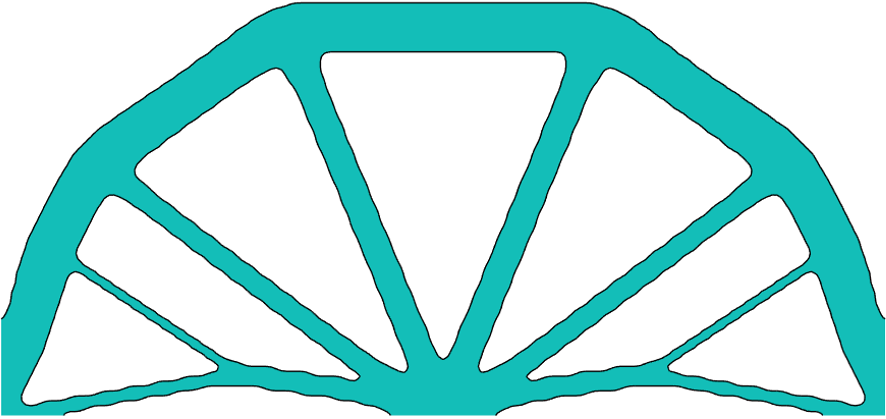}
	\caption{Final topology}
	\label{sFig: HST167}
	\end{subfigure}
	\caption{Compliance and optimized topology obtained with Heaviside step function for simply supported deep beam case}
\end{figure}

\begin{figure}[htbp!]
	\centering
	\begin{subfigure}{0.49\textwidth}
		\centering
		\includegraphics[scale=0.7]{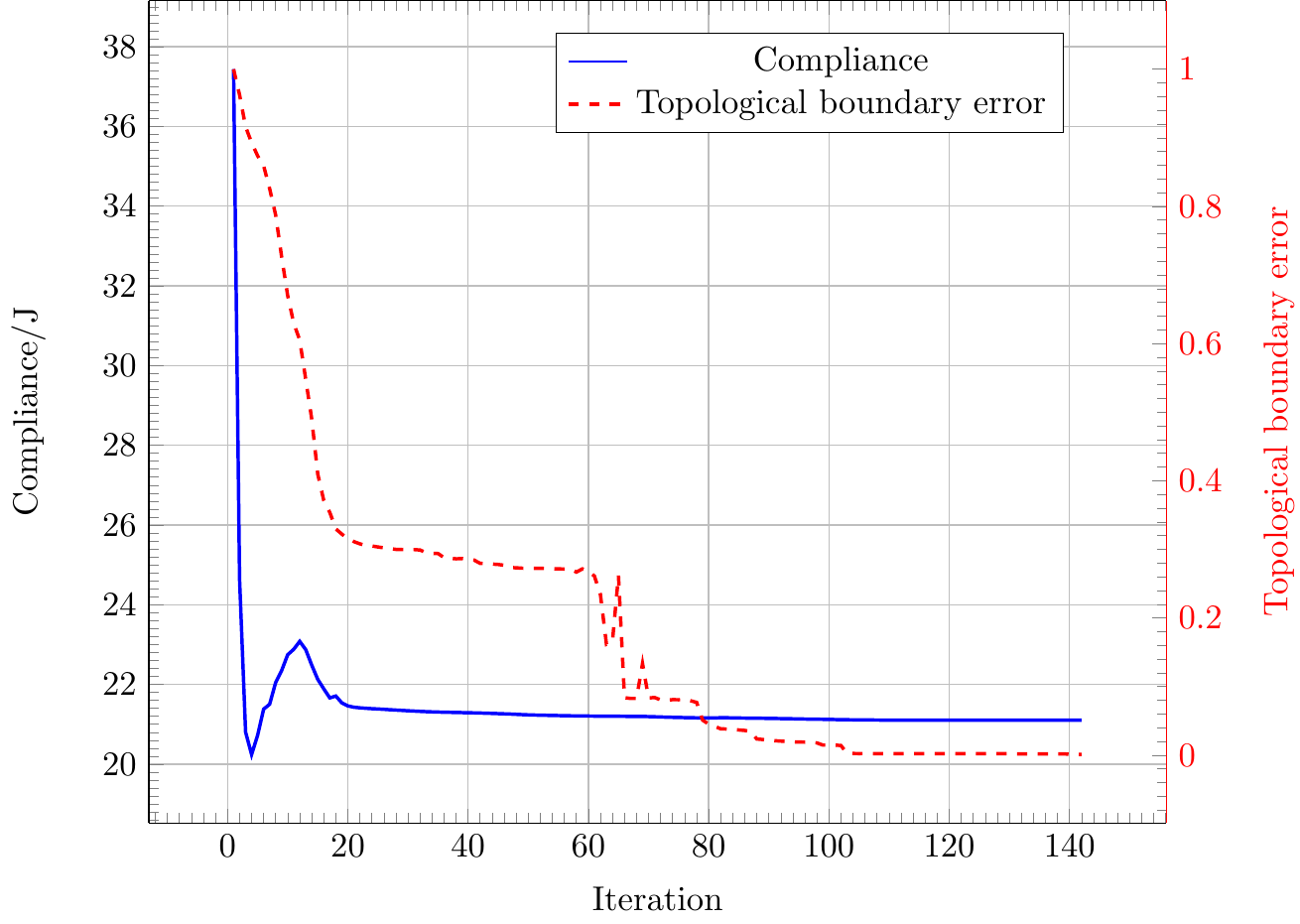}
		\caption{Convergence process}
		\label{sFig: HSC}
	\end{subfigure}
	\begin{subfigure}{0.49\textwidth}
		\centering
		\includegraphics[scale=0.4]{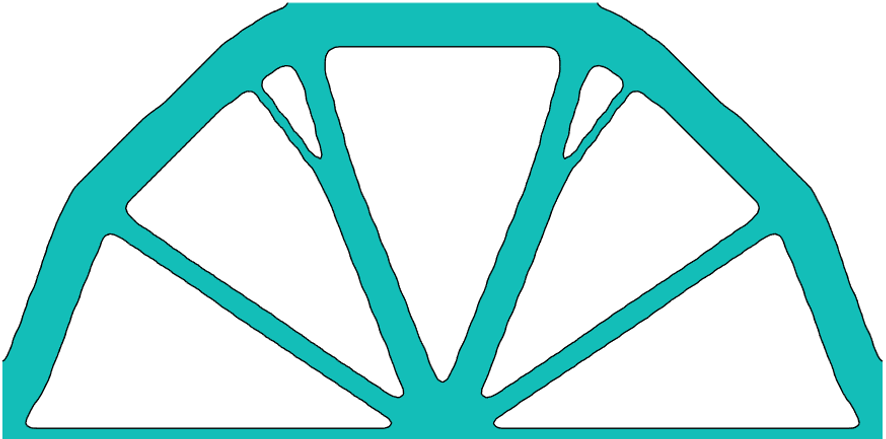}
		\caption{Optimized topology}
		\label{sFig: HSF}
	\end{subfigure}
	\caption{Convergence process and optimized topology obtained with Heaviside smooth function for simply supported deep beam case}
\end{figure}

The convergence history in Figure \ref{sFig: HStC} shows that great fluctuations appear in the initial 20 iterations when the Heaviside step function is used. These fluctuations are associated with scattered material resulting in complicated topologies (Figures \ref{sFig: HST10} and \ref{sFig: HST20}), meaning that the Heaviside step function has the difficulty in extracting topological boundaries at the early stages of optimization. Afterwards, the optimization process steadily converges at 21.3875 J after 167 iterations and later topologies are reasonable (Figures \ref{sFig: HST40} to \ref{sFig: HST167}). Compared to the Heaviside step function, the Heaviside smooth function does not cause any numerical instabilities, and therefore the whole optimization process steadily converges at 21.1049 J after 142 iterations (Figure \ref{sFig: HSC}). As shown in Figure \ref{sFig: HSC}, the topological boundary error gradually decreases to almost 0\%  when the converged topology is obtained. Two different final topologies obtained by Heaviside step and smooth functions are shown in Figures \ref{sFig: HST167} and \ref{sFig: HSF}, respectively. In this case, the Heaviside smooth function performs better than the Heaviside step function both in compliance and convergence. 

\begin{figure}[htbp!]
	\centering
	\includegraphics[scale=0.8]{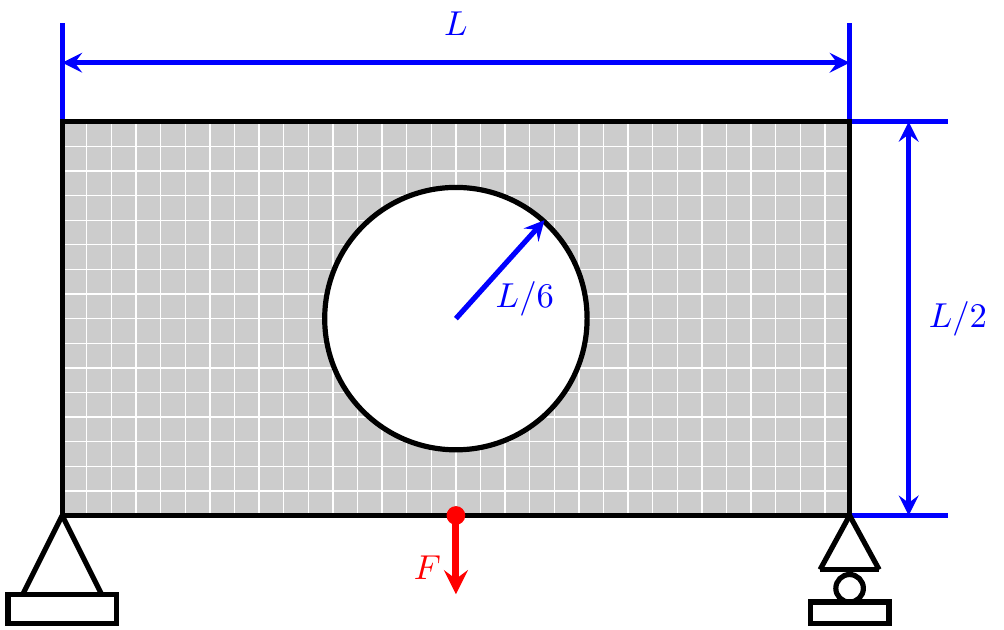}
	\caption{Design domain of a simply supported deep beam with a fixed hole}
	\label{Fig: HSbeam}
\end{figure}

In some engineering problems, certain areas of the design domain are required to be void (non-design areas) during the whole optimization process. Another version of this problem with a non-designable circular hole with a radius of $L/6$ and a center located at ($L/2$, $L/4$) as illustrated in Figure \ref{Fig: HSbeam} is used to further test the performance of Heaviside step and smooth functions in SEMDOT. All parameter settings remain unchanged.

\begin{figure}[htbp!]
	\centering
	\begin{subfigure}{\textwidth}
		\centering
		\includegraphics[scale=0.85]{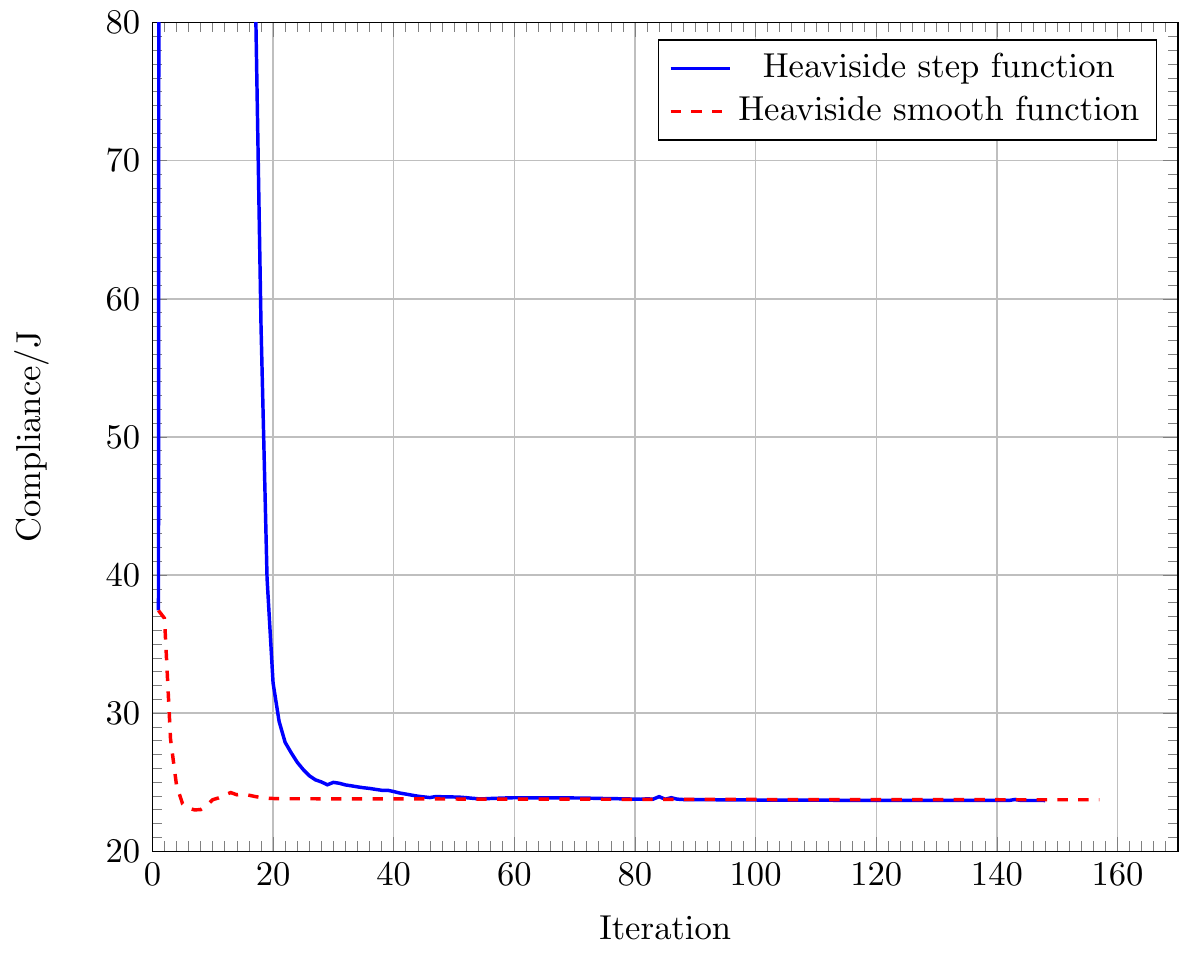}
		\caption{Convergence histories}
		\label{sFig: DCSST}
	\end{subfigure}
\\
	\begin{subfigure}{0.49\textwidth}
		\centering
		\includegraphics[width=210 pt]{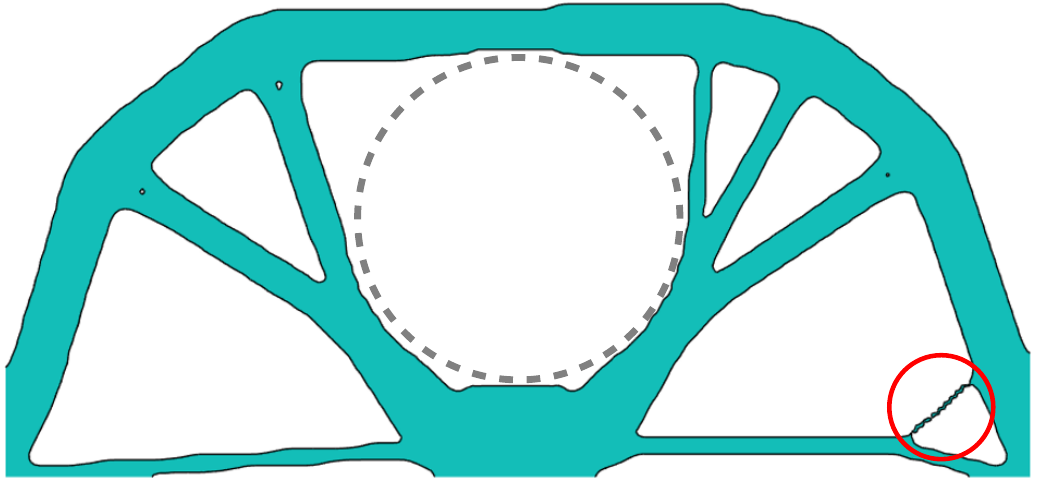}
		\caption{Optimized topology obtained with Heaviside step function}
		\label{sFig: DHST}
	\end{subfigure}
	\begin{subfigure}{0.49\textwidth}
		\centering
		\includegraphics[width=200 pt]{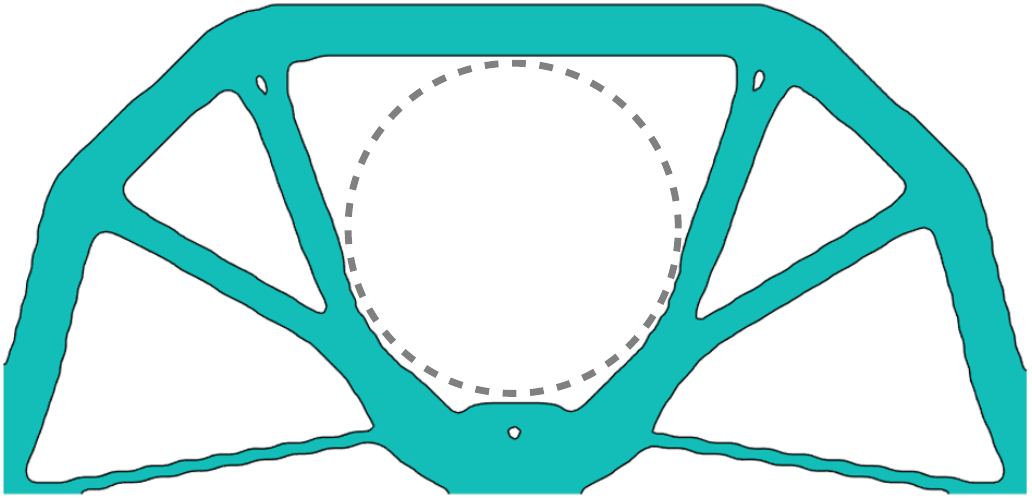}
		\caption{Optimized topology obtained with Heaviside smooth function}
		\label{sFig: DHS}
	\end{subfigure}
	\caption{Comparisons of performance, convergency, and topological designs between Heaviside step and smooth functions for simply supported deep beam case with a fixed hole}
\end{figure}

Figure \ref{sFig: DCSST} shows that SEMDOT using the Heaviside step function converges after 148 iterations, which is a little less than when the Heaviside smooth function is used (157). However, like the previous case, great fluctuations appear at the early stages of optimization when the Heaviside step function is used. Converged compliance is 23.6741 J for the Heaviside step function and 23.7306 J for the Heaviside smooth function, so the difference is negligible (around 0.2\%). In this case,  the Heaviside step function performs a little better than the Heaviside smooth function in compliance and convergence. However, the asymmetric topology obtained by the Heaviside step function has several tiny holes and one thin bar shown by a red circle in Figure \ref{sFig: DHST}. This design cannot be easily manufactured even with AM \cite{fu2019AM}. By contrast, the symmetric topology obtained by the Heaviside smooth function has better manufacturability (Figure \ref{sFig: DHS}). 

\begin{figure}[htbp!]
	\centering
		\begin{subfigure}{\textwidth}
		\centering
		\includegraphics[scale=0.85]{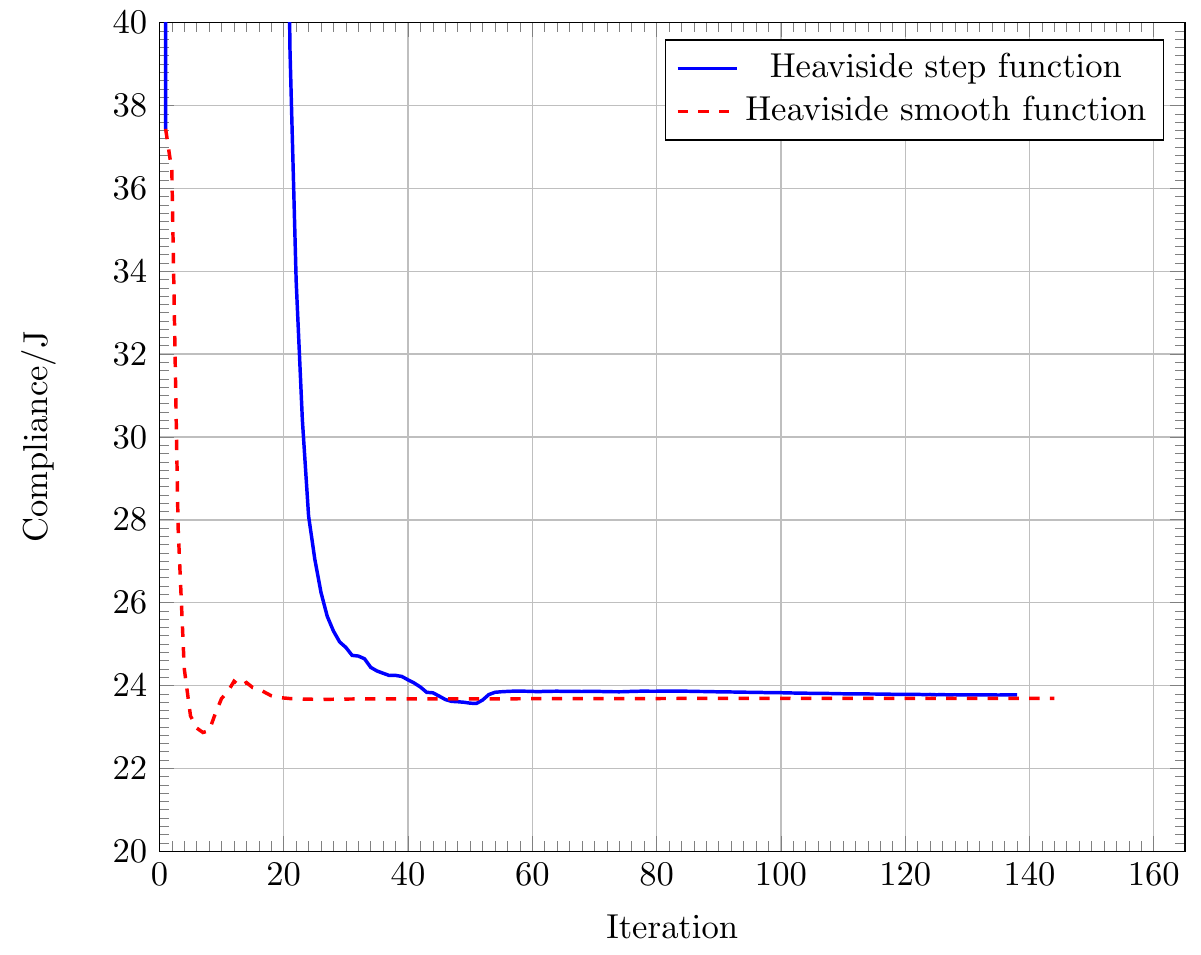}
		\caption{Convergence histories}
		\label{sFig: HDCSST}
	\end{subfigure}
	\\	
	\begin{subfigure}{0.49\textwidth}
		\centering
		\includegraphics[width=215 pt]{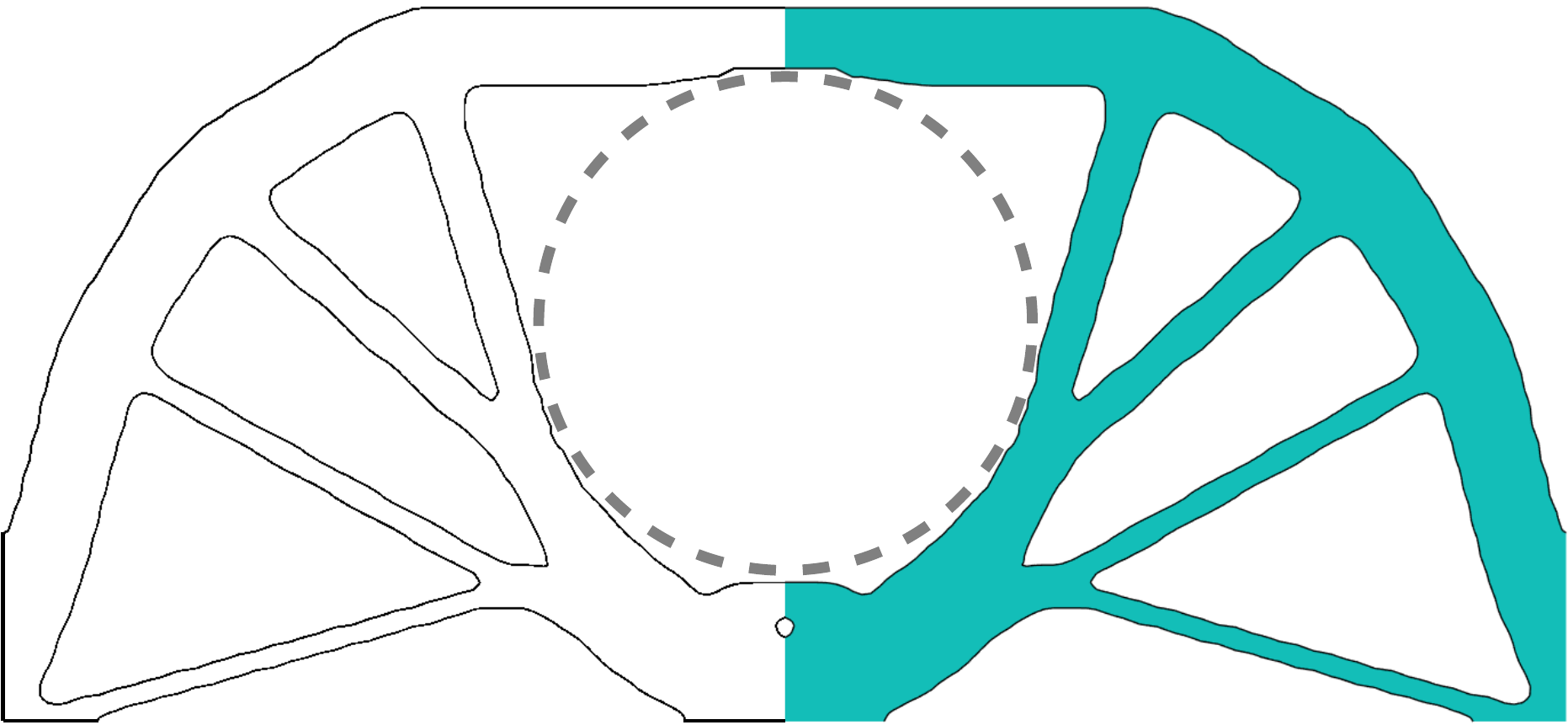}
		\caption{Optimized topology obtained with Heaviside step function}
		\label{sFig: HDHST}
	\end{subfigure}
	\begin{subfigure}{0.49\textwidth}
		\centering
		\includegraphics[width=210 pt]{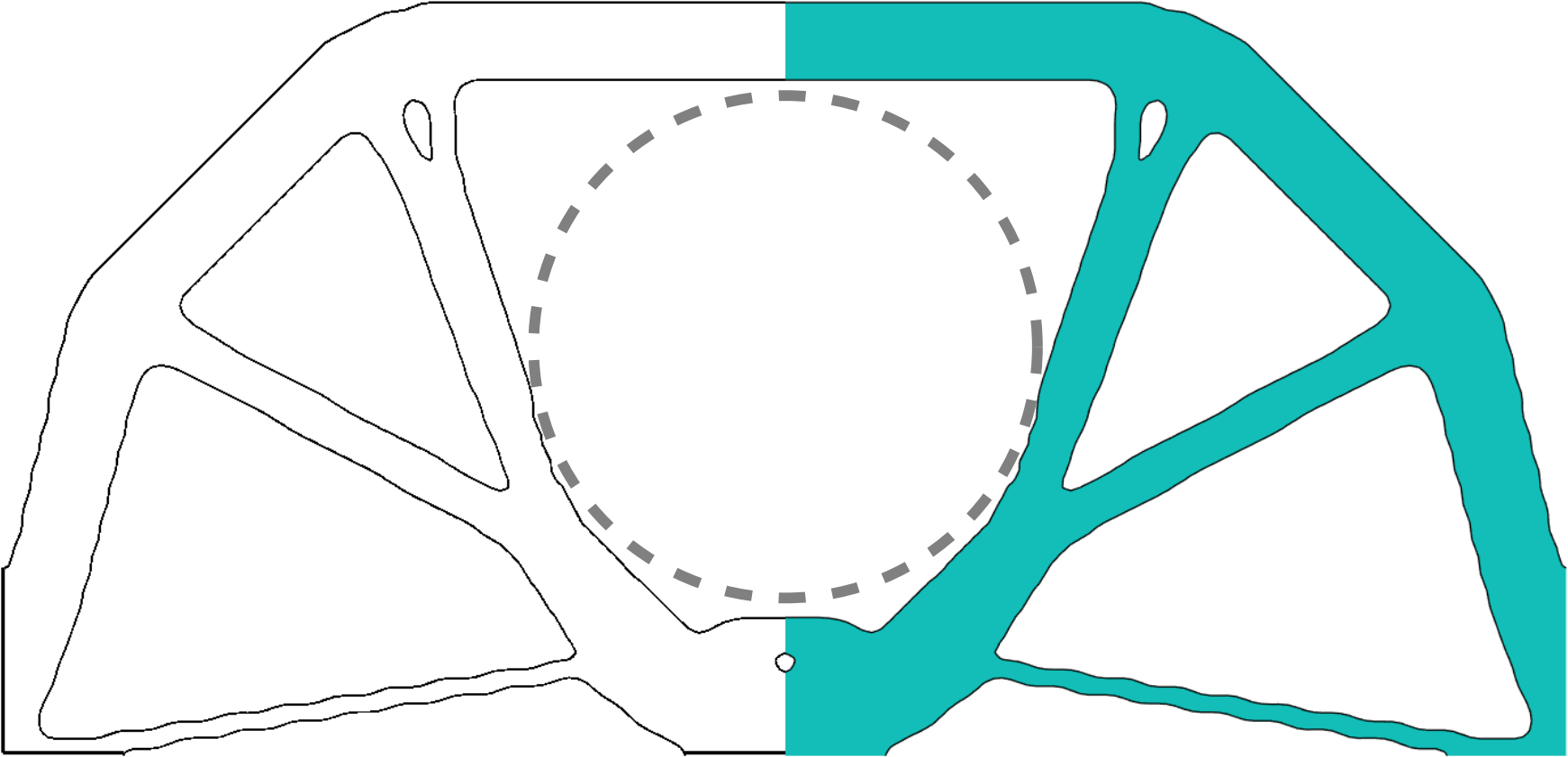}
		\caption{Optimized topology obtained with Heaviside smooth function}
		\label{sFig: HDHS1}
	\end{subfigure}
	\caption{Comparisons of performance, convergency, and topological designs between Heaviside step and smooth functions considering symmetry condition}
	\label{Fig: CPCTSY}
\end{figure}

Considering the symmetry condition in the case shown in Figure \ref{Fig: HSbeam}, comparisons of performance, convergency, and topological designs between Heaviside step and smooth functions  are shown in Figure \ref{Fig: CPCTSY}. Identical to the previous discussion, the Heaviside smooth function needs a longer convergence process (157)  than the Heaviside step function (138). Converged performance of the Heaviside smooth function (23.686 J) is slightly better than the Heaviside step function (23.772 J). In both cases, the performance changes before and after imposing symmetry are insignificant. In terms of topological designs, a clear difference can be observed between Figures \ref{sFig: DHST} and \ref{sFig: HDHST}, but Figures \ref{sFig: DHS} and \ref{sFig: HDHS1} are almost identical. Therefore, it can be concluded that the Heaviside step function is more susceptible to the non-designable passive area compared to the Heaviside smooth function in SEMDOT. The asymmetric topology obtained by the Heaviside step function (Figure \ref{sFig: DHST}) is because of the disturbance caused by the non-designable passive area during the determination of the topological boundary.

Based on the above discussions, it is concluded that the Heaviside smooth function is more suitable than the Heaviside step function for SEMDOT despite requiring an additional termination criterion (Equation \ref{Eq: Berror}). Therefore, the Heaviside smooth function will be used for the rest of numerical experiments in this paper rather than the Heaviside step function.

\subsection{Deviation of filtered elemental volume fractions} \label{Sec: DEVF}
\begin{figure}[htbp!]
	\centering
	\includegraphics[scale=1]{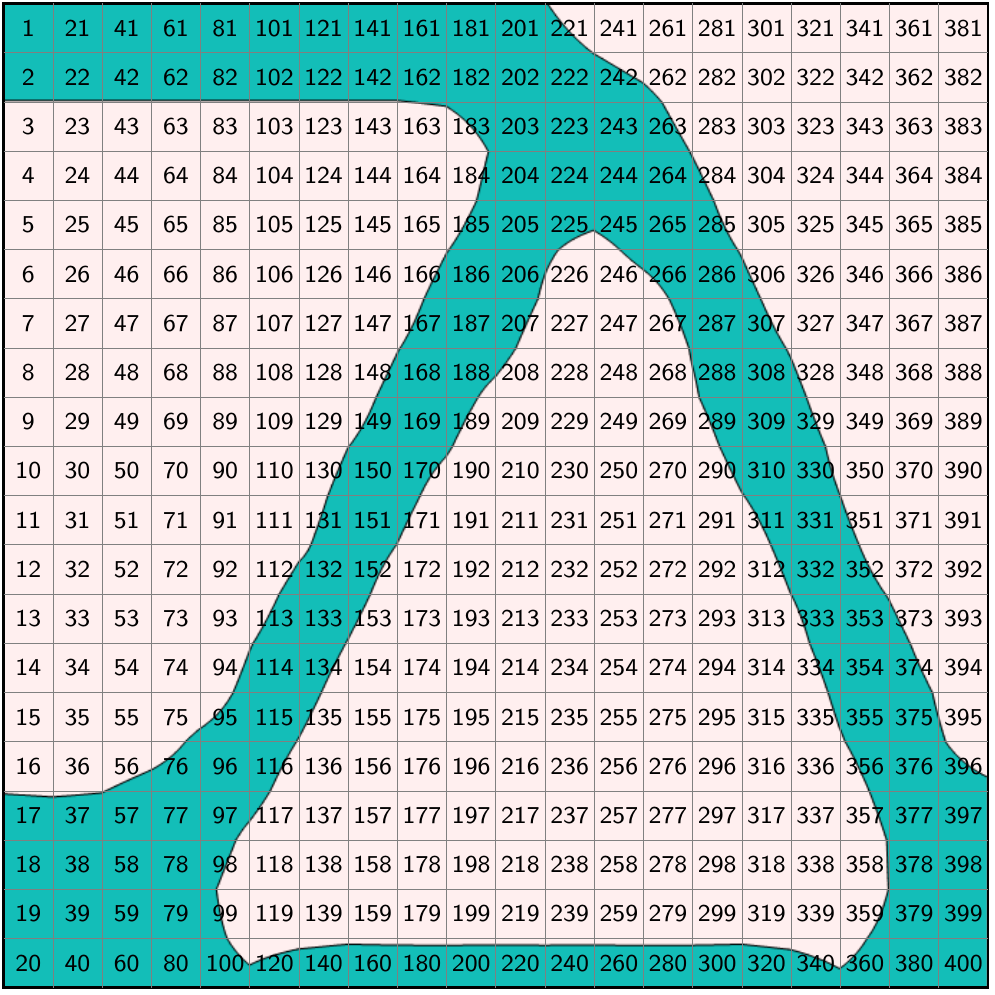}
	\caption{Optimized topology with element numbers}
	\label{Fig: DXe}
\end{figure}

\begin{figure}[htbp!]
	\centering
	\includegraphics[width=350 pt]{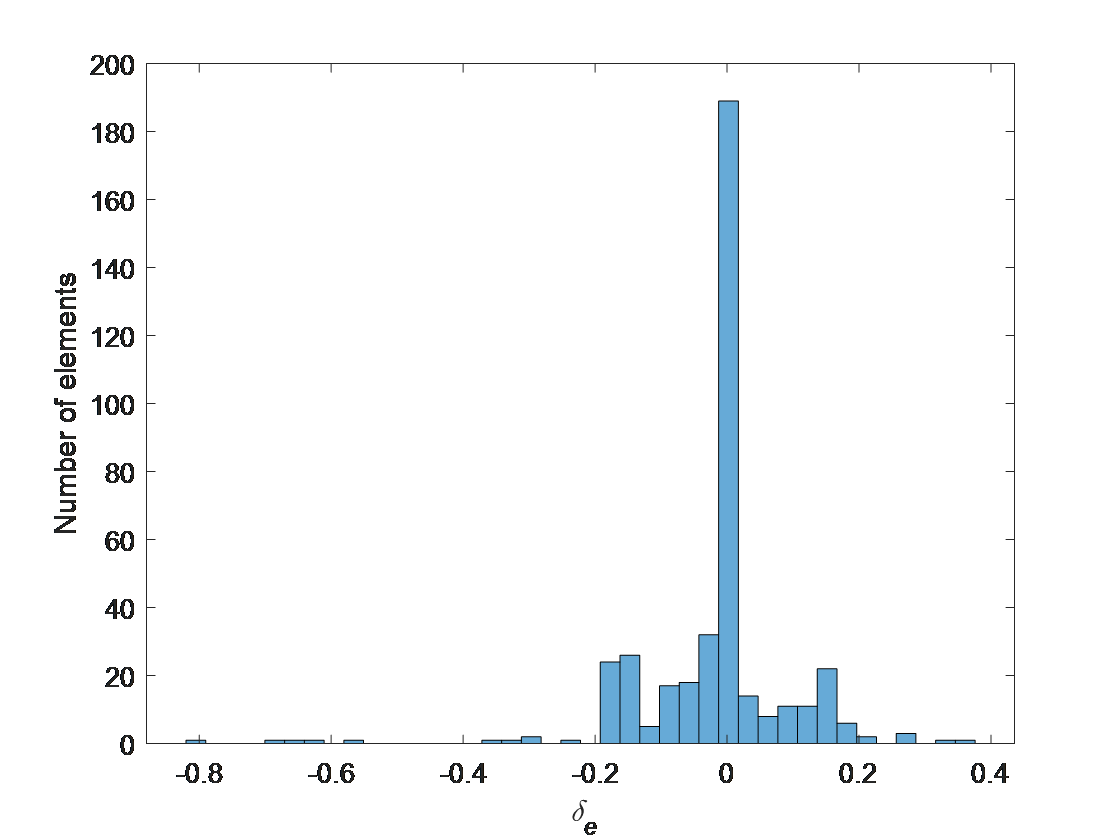}
	\caption{Histogram of $\delta_e$ with passive area}
	\label{sFig: HDDe}
\end{figure}

Following Section \ref{Sec: EHF}, a coarse mesh of 20$\times$20 is used for the simply supported deep beam case considering both the passive area and symmetry condition to demonstrate the relationship between $\tilde{X}_e$ and $\tilde{X}_e^{\text new}$ (Equation \ref{Eq:Delta}). In this case, the filter radius $r_{\min}$ is set to 1.5 time elements width ($r_{\min}$=1.5). The optimized topology with element numbers is shown in Figure \ref{Fig: DXe}. The histogram of the difference between $\tilde{X}_e$ and $\tilde{X}_e^{\text new}$ ($\delta_e$) for the final topology is shown in Figure \ref{sFig: HDDe}, which reveals that the deviation $\delta_e$ mainly concentrates on the range from -0.2 to 0.2, and there is no difference between $\tilde{X}_e$ and $\tilde{X}_e^{\text new}$ for the majority of elements ($\delta_e$=0).

\begin{figure}[htbp!]
	\centering
	\begin{subfigure}{0.49\textwidth}
		\centering
		\includegraphics[width=235 pt]{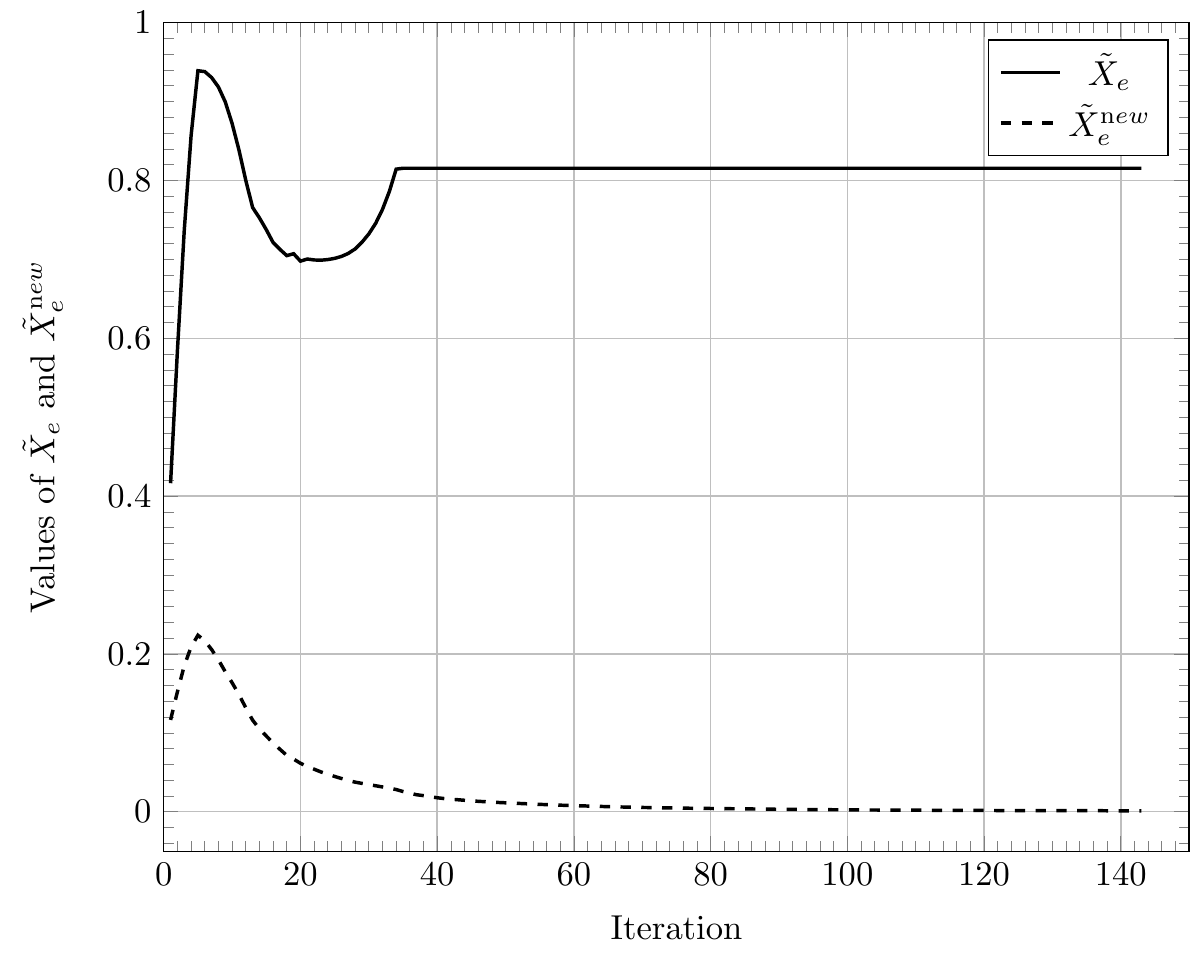}
		\caption{Element 16}
		\label{sFig: E16}
	\end{subfigure}
	\begin{subfigure}{0.49\textwidth}
		\centering
		\includegraphics[width=235 pt]{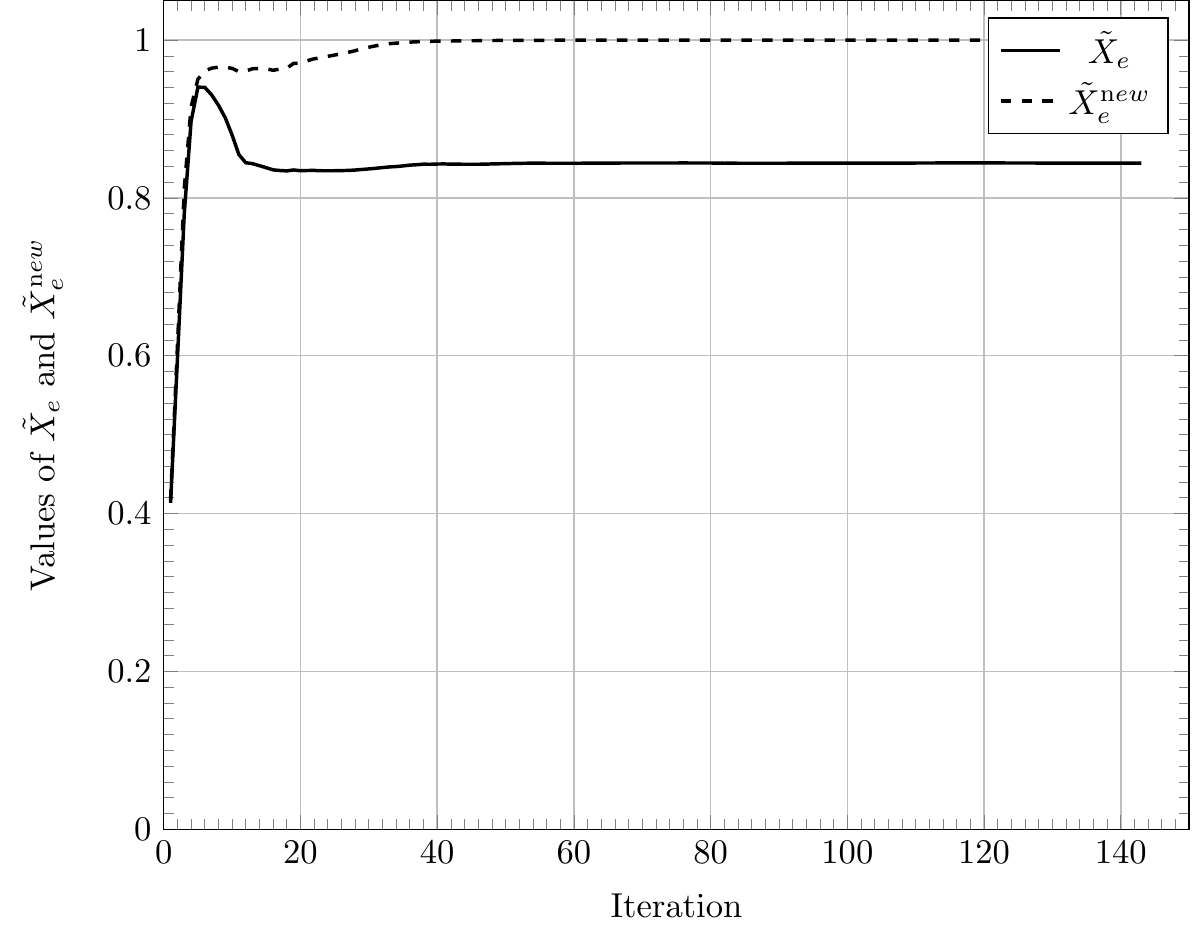}
		\caption{Element 79}
		\label{sFig: E79}
	\end{subfigure}
\\
	\begin{subfigure}{0.49\textwidth}
		\centering
		\includegraphics[width=235 pt]{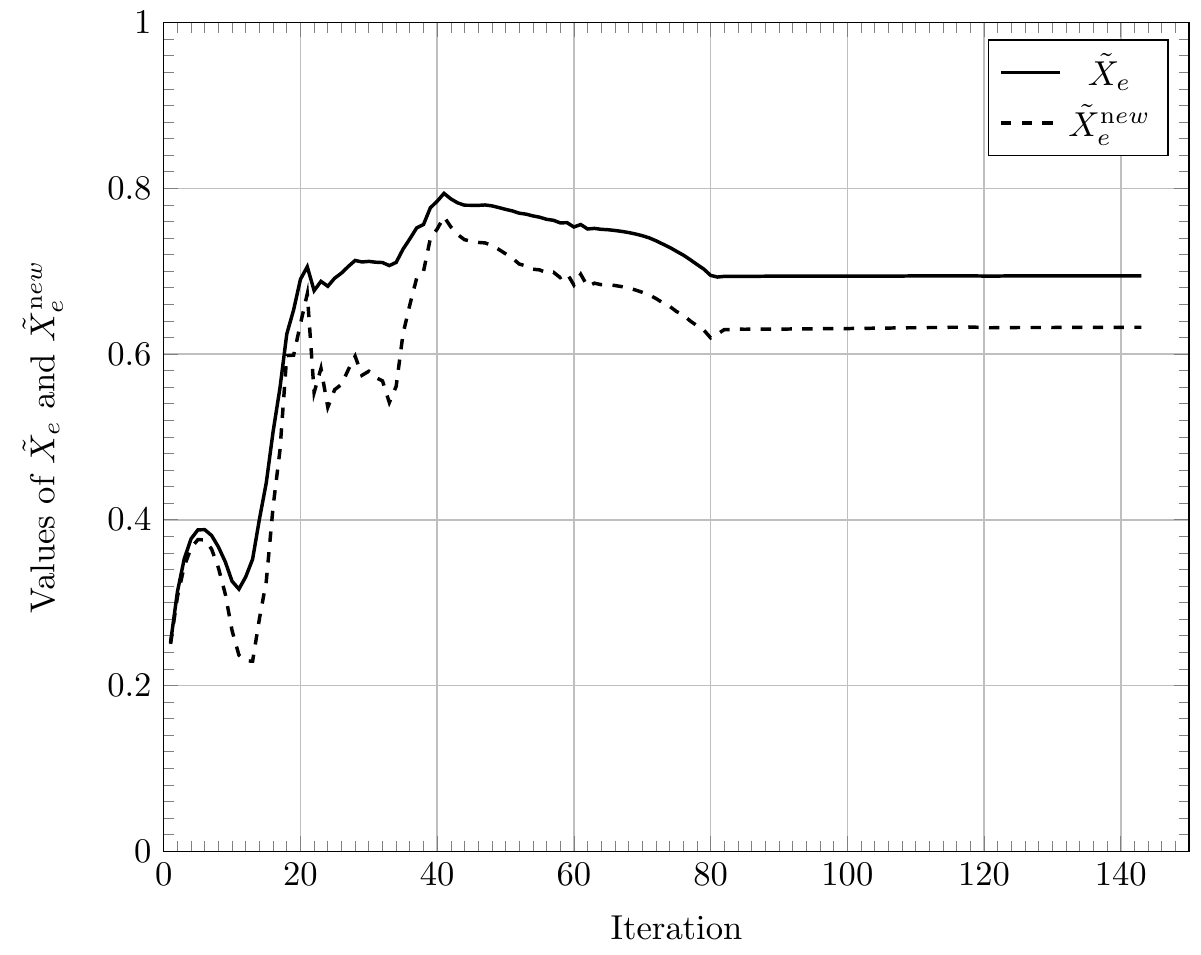}
		\caption{Element 185}
		\label{sFig: E185}
	\end{subfigure}
	\begin{subfigure}{0.49\textwidth}
		\centering
		\includegraphics[width=235 pt]{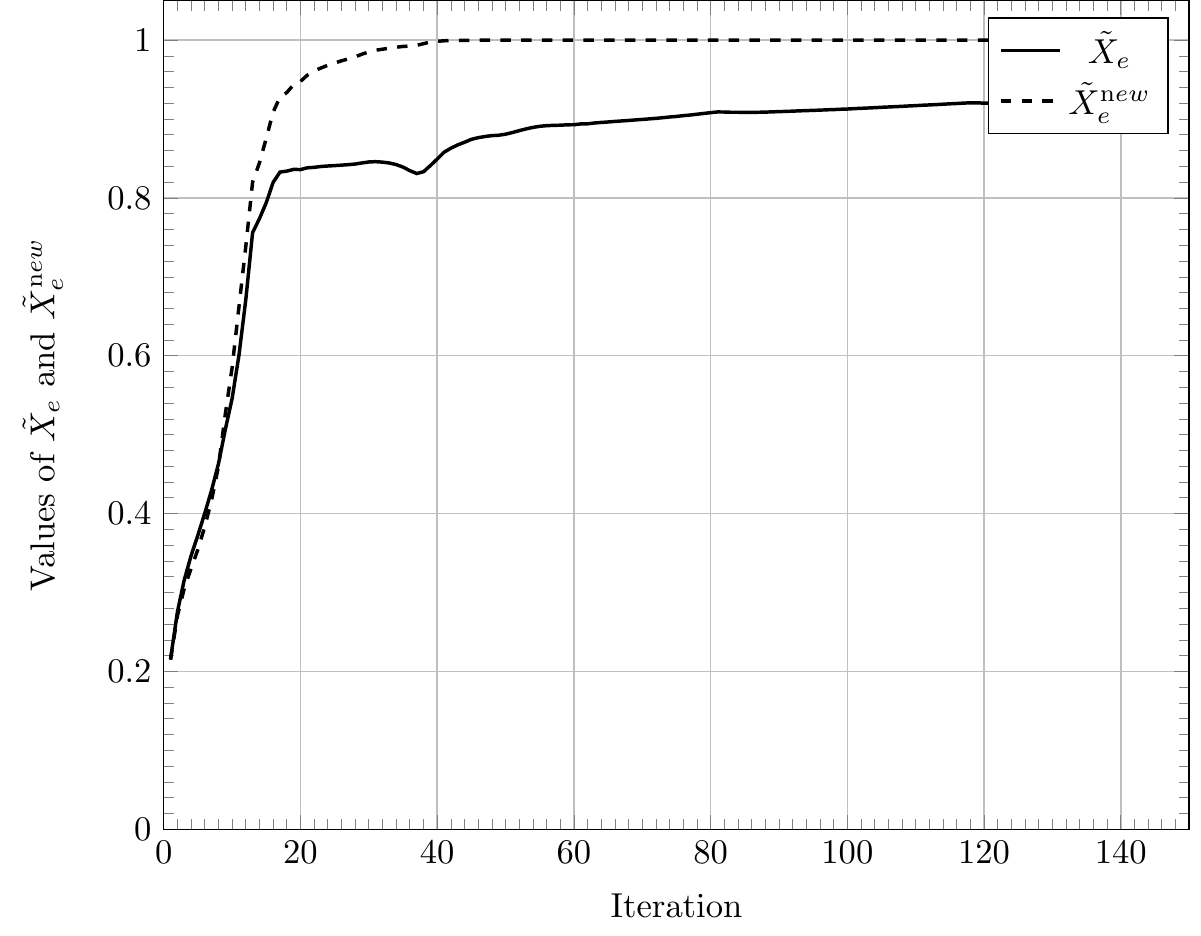}
		\caption{Element 202}
		\label{sFig: E202}
	\end{subfigure}
\\
	\begin{subfigure}{0.49\textwidth}
		\centering
		\includegraphics[width=235 pt]{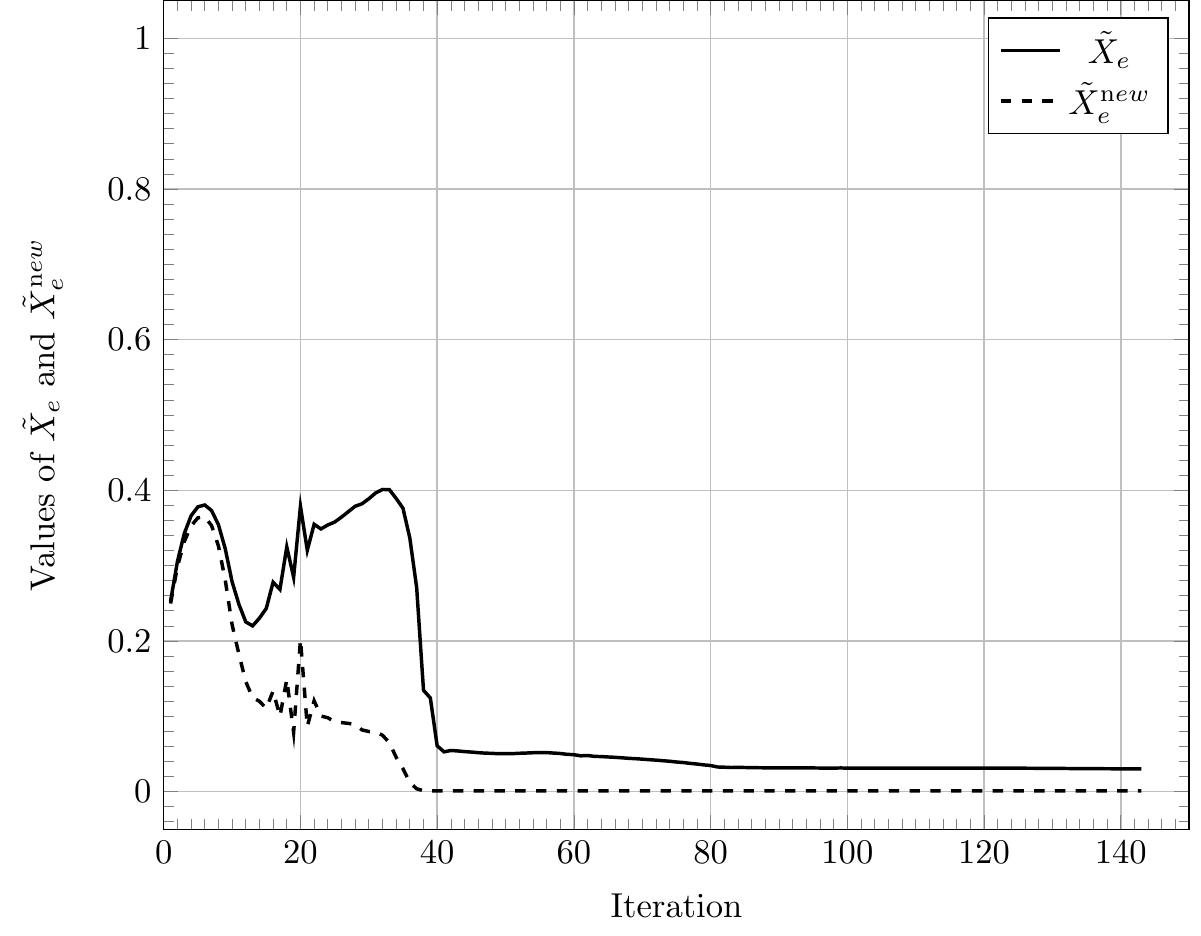}
		\caption{Element 247}
		\label{sFig: E247}
	\end{subfigure}
	\begin{subfigure}{0.49\textwidth}
		\centering
		\includegraphics[width=235 pt]{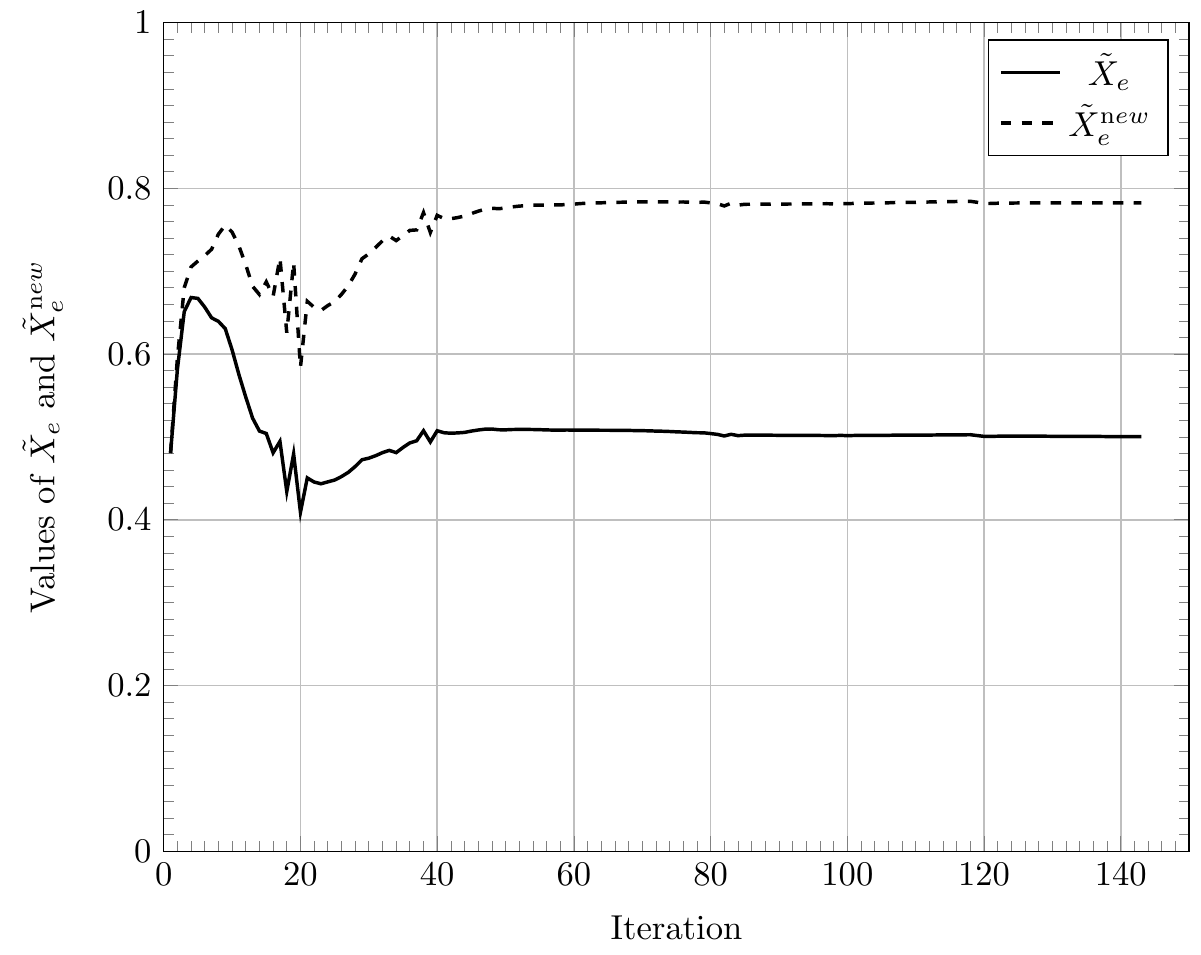}
		\caption{Element 360}
		\label{sFig: E360}
	\end{subfigure}
	\caption{Values of $\tilde{X}_e$ and $\tilde{X}_e^{\text new}$ for different elements}
	\label{Fig:VXENE}
\end{figure}

To explain the relationship between $\tilde{X}_e$ and $\tilde{X}_e^{\text new}$ in each iteration, values of $\tilde{X}_e$ and $\tilde{X}_e^{\text new}$ for elements 16, 79, 195, 202, 247, and 360 are shown in Figure \ref{Fig:VXENE}. The biggest difference ($\delta_e$=-0.8142 in the final iteration) appears in element 16, as illustrated in Figure \ref{sFig: E16}. Even though element 16 has a relatively high value of $\tilde{X}_e$ during the optimization process, its $\tilde{X}_e^{\text new}$ goes down to 0.001. This is because element 16 is in the passive area, meaning that it needs to be artificially suppressed to the void element. For the rest of elements that are not in the passive area, $\tilde{X}_e^{\text new}$ is approximately proportional to $\tilde{X}_e$, as illustrated in Figures \ref{sFig: E79}, \ref{sFig: E185}, \ref{sFig: E202}, \ref{sFig: E247}, and \ref{sFig: E360}. The function of Equation \ref{Eq: Filter} is to filter elemental volume fractions $X_e$ without considering the target volume constraint, so there is $\sum\limits_{e=1}^M \tilde{X}_e/V$= 0.3125$>V^*/V$ in the final iteration. Therefore, the use of $\tilde{X}_e$ for the next round of FEA will result in the overestimation of structural performance at least in this case. By contrast, $\tilde{X}_e^{\text new}$ is calculated from the newly obtained grid point densities that are determined based on the target volume constraint (refer to Equations \ref{Eq: Heavisidesmooth} and \ref{Eq: Xenew}), so $\sum\limits_{e=1}^M \tilde{X}_e^{\text new}/V$=0.3$=V^*/V$ can be guaranteed iteratively.

\begin{figure}[htbp!]
	\centering
	\includegraphics[width=350 pt]{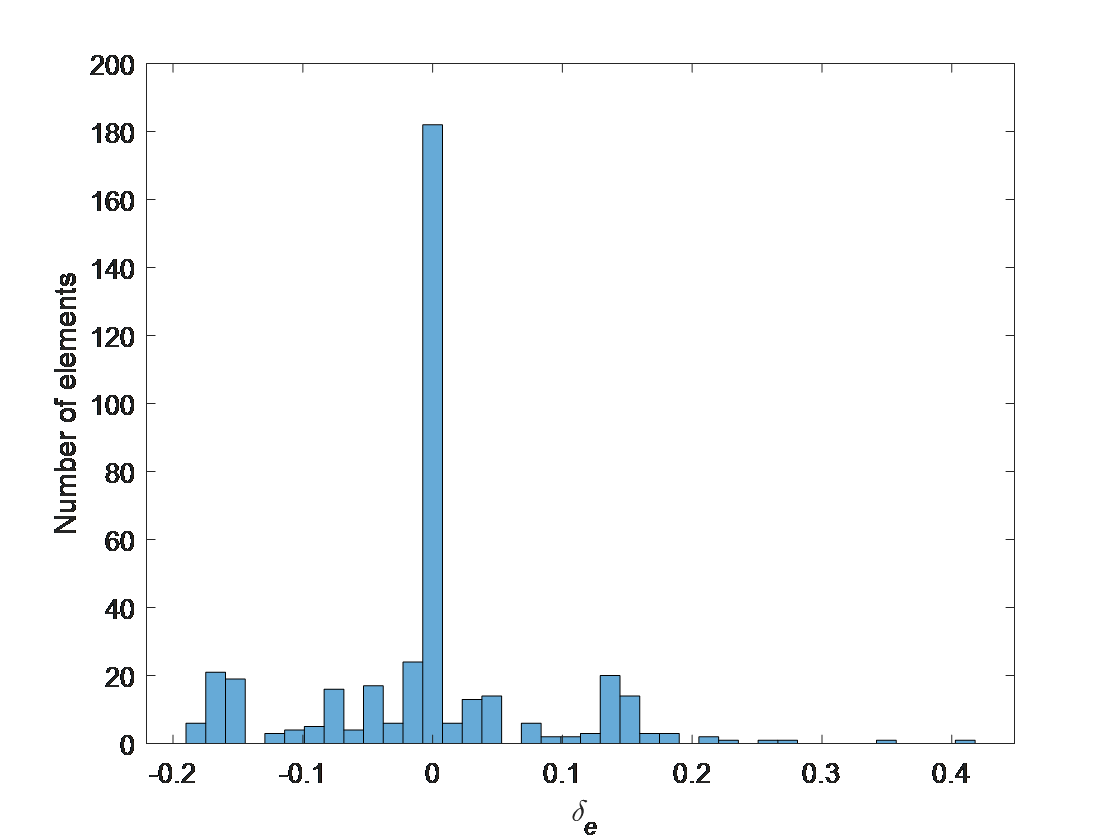}
	\caption{Histogram of $\delta_e$ without passive area}
	\label{sFig: HDDOe}
\end{figure}

When the passive area is not considered in the test case, there is no  big deviation such as the one observed in element 16 (Figure \ref{sFig: E16}), and the deviation $\delta_e$ mainly falls in the range of -0.2 to 0.2 (Figure \ref{sFig: HDDOe}). 

\subsection{Effects of multiple filtering steps}
\begin{figure}[htbp!]
	\centering
	\includegraphics[scale=0.75]{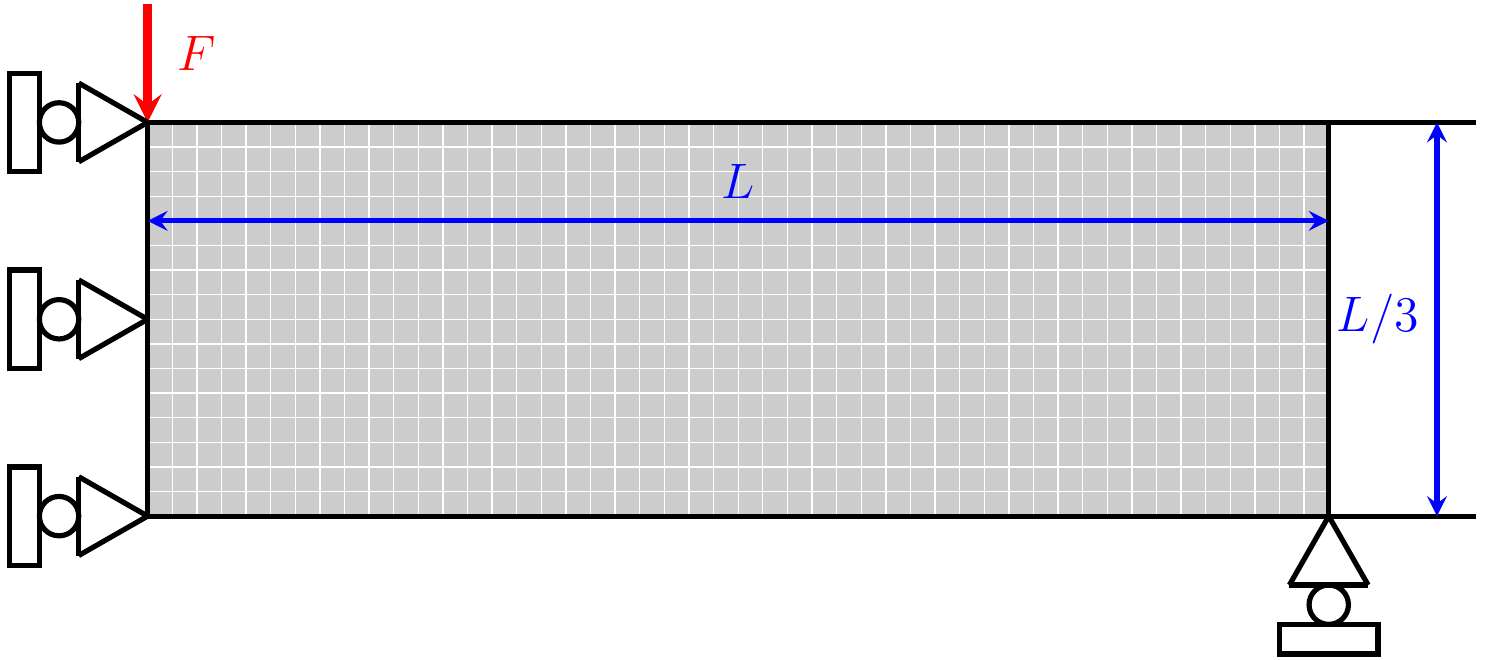}
	\caption{Design domain of half an MBB beam}
	\label{Fig: MBB}
\end{figure}

 Two combinations of filter radii (i.e., $r_{\min}=3$, $\Upsilon_{\min}=3$ and $r_{\min}=3$, $\Upsilon_{\min}=1$) were used in authors' previous works \cite{fu2019AM,fu2019TOC,fu2019PAM,fu20203DTO}. Other than those two combinations, different combinations of filter radii can also be considered in optimizing topologies with SEMDOT, which provides more design freedom for designers. The so-called Messerschmidt–Bölkow–Blohm (MBB) beam is used here to demonstrate the effects of the two filter radii, $r_{\min}$ and $\Upsilon_{\min}$, on performance, convergency, and topological designs. The design domain and boundary conditions are shown in Figure \ref{Fig: MBB}. Only half of the MBB beam is considered as the design domain due to symmetry. As illustrated in Figure \ref{Fig: MBB}, the symmetric boundary condition is applied to the left side; the vertical displacement at the bottom right corner is restricted; and a unit vertical load (\textit{F}=-1 N) is applied at the top left corner. The design domain is discretized by a 150$\times$50 finite element mesh.

\begin{figure}[htbp!]
	\centering
	\begin{subfigure}{0.49\textwidth}
		\centering
		\includegraphics[scale=0.9]{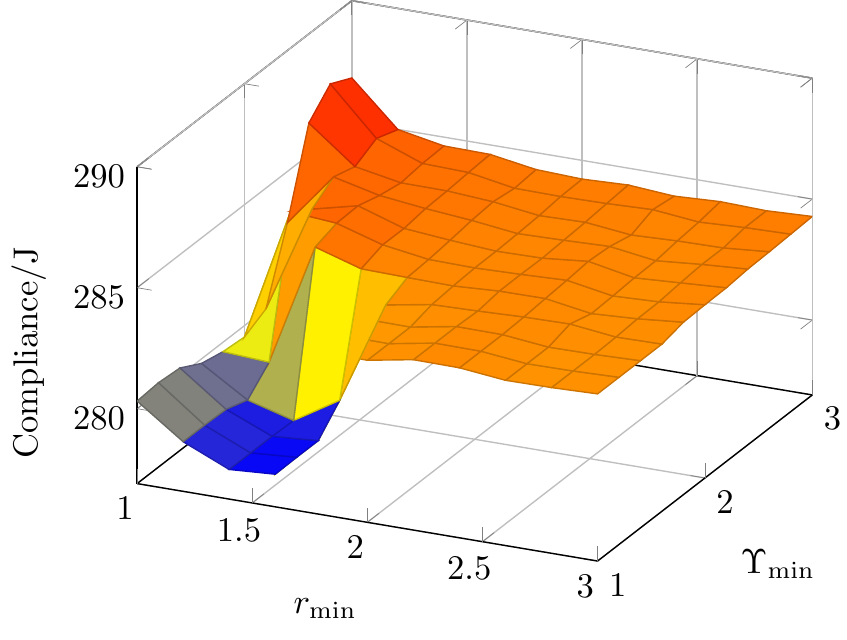}
		\caption{Compliance}
		\label{sFig: Com}
	\end{subfigure}
	\begin{subfigure}{0.49\textwidth}
		\centering
		\includegraphics[scale=0.9]{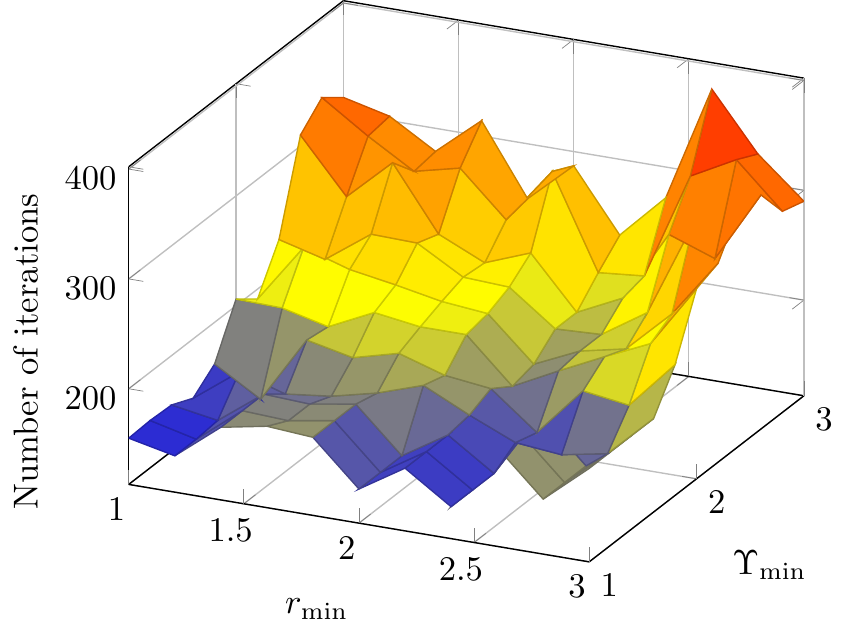}
		\caption{Number of iterations}
		\label{sFig: NOI}
	\end{subfigure}
	\caption{Compliance and number of iterations for different values of $r_{\min}$ and $\Upsilon_{\min}$ solving MBB beam case}
\end{figure}

Figure \ref{sFig: Com} shows that the highest compliance (287.2474 J) is obtained when the combination of  $r_{\min}=1$ and $\Upsilon_{\min}=2.8$ is used. This drops to 283.7538 J when the combination of $r_{\min}=2.8$ and $\Upsilon_{\min}=1$ is used, meaning that $\Upsilon_{\min}$ will cause worse results than  $r_{\min}$. Generally, increasing either $r_{\min}$ or $\Upsilon_{\min}$ can contribute to the rise of compliance, and then a relatively stable value can be reached when either $r_{\min}$ or $\Upsilon_{\min}$ is large enough. Figure \ref{sFig: NOI} shows that using high values of $\Upsilon_{\min}$ will prolong the convergence process more than using high values of $r_{\min}$, and the highest number of iterations (378) is obtained when $r_{\min}=2.6$ and $\Upsilon_{\min}=3$.

\begin{figure}[htbp!]
	\centering
	\begin{subfigure}{0.32\textwidth}
		\centering
		\includegraphics [width=145 pt] {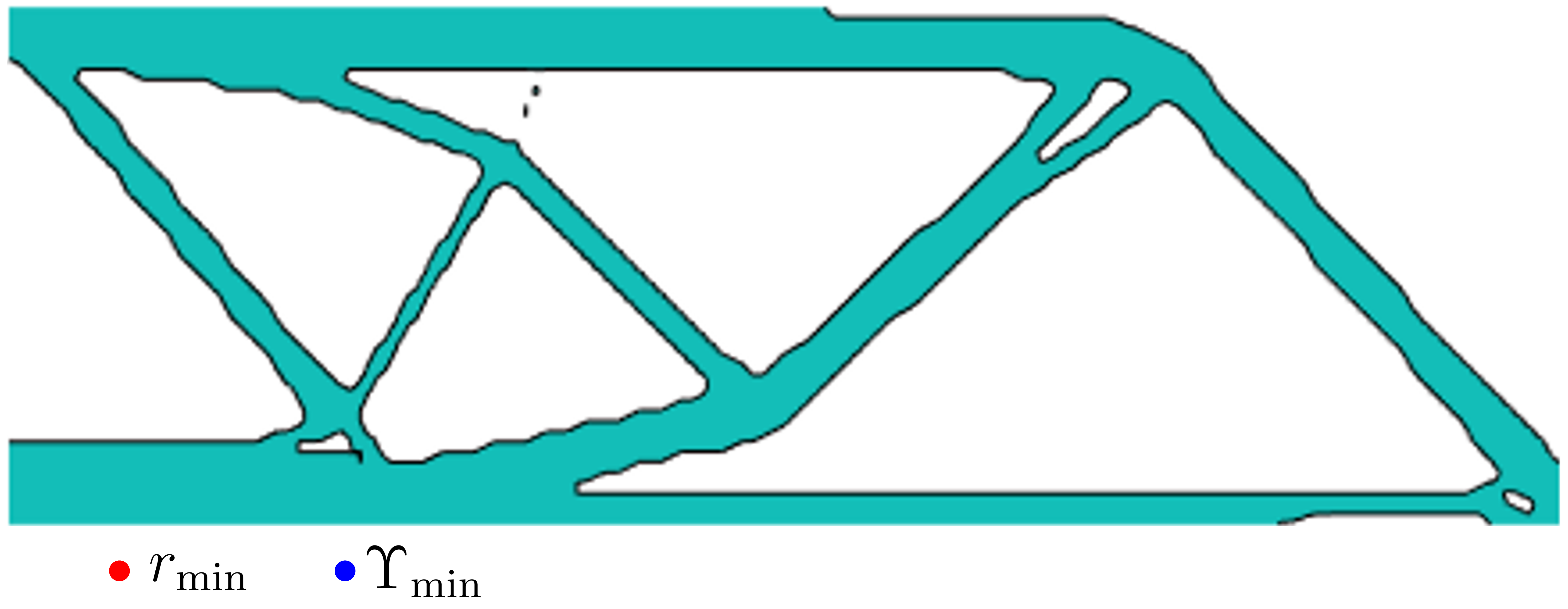}
		\caption{$r_{\min}=1$, $\Upsilon_{\min}=1$}
		\label{sFig: 2DC11}
	\end{subfigure}
	\begin{subfigure}{0.32\textwidth}
		\centering
		\includegraphics [width=145 pt] {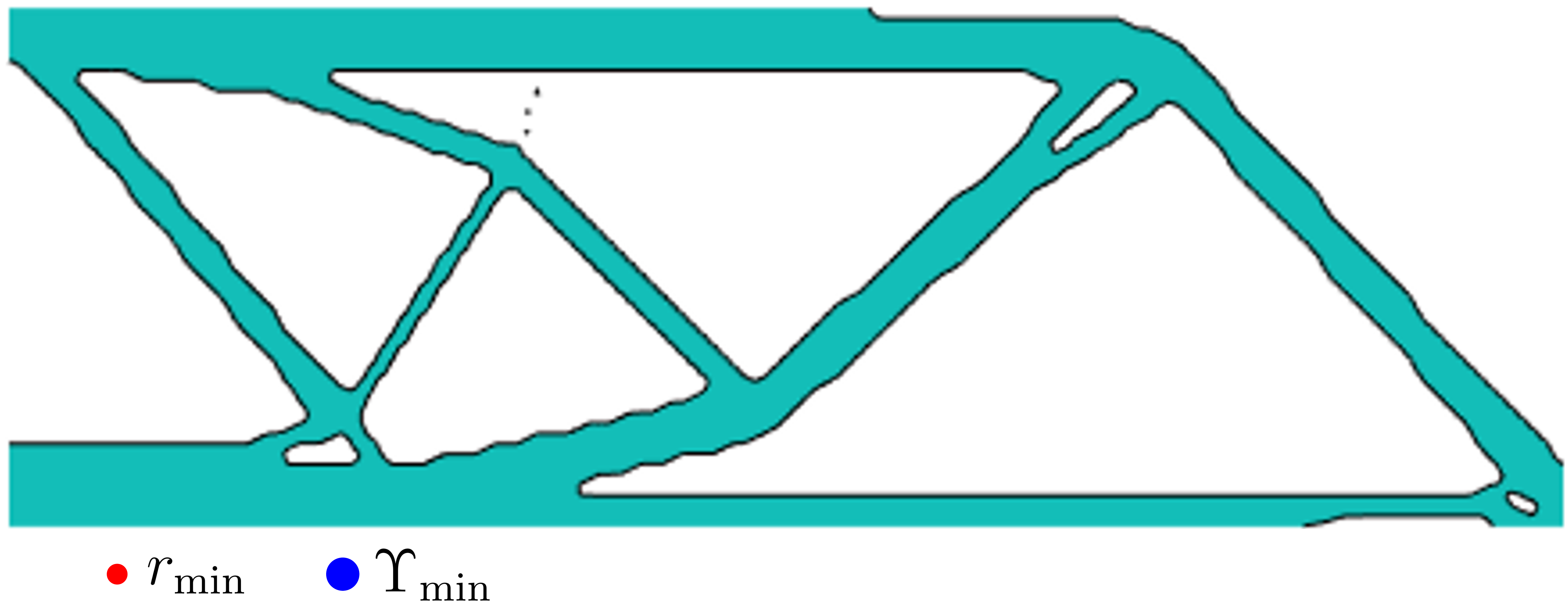}
		\caption{$r_{\min}=1$, $\Upsilon_{\min}=1.6$}
		\label{sFig: 2DC116}
	\end{subfigure}
	\begin{subfigure}{0.32\textwidth}
		\centering
		\includegraphics [width=145 pt] {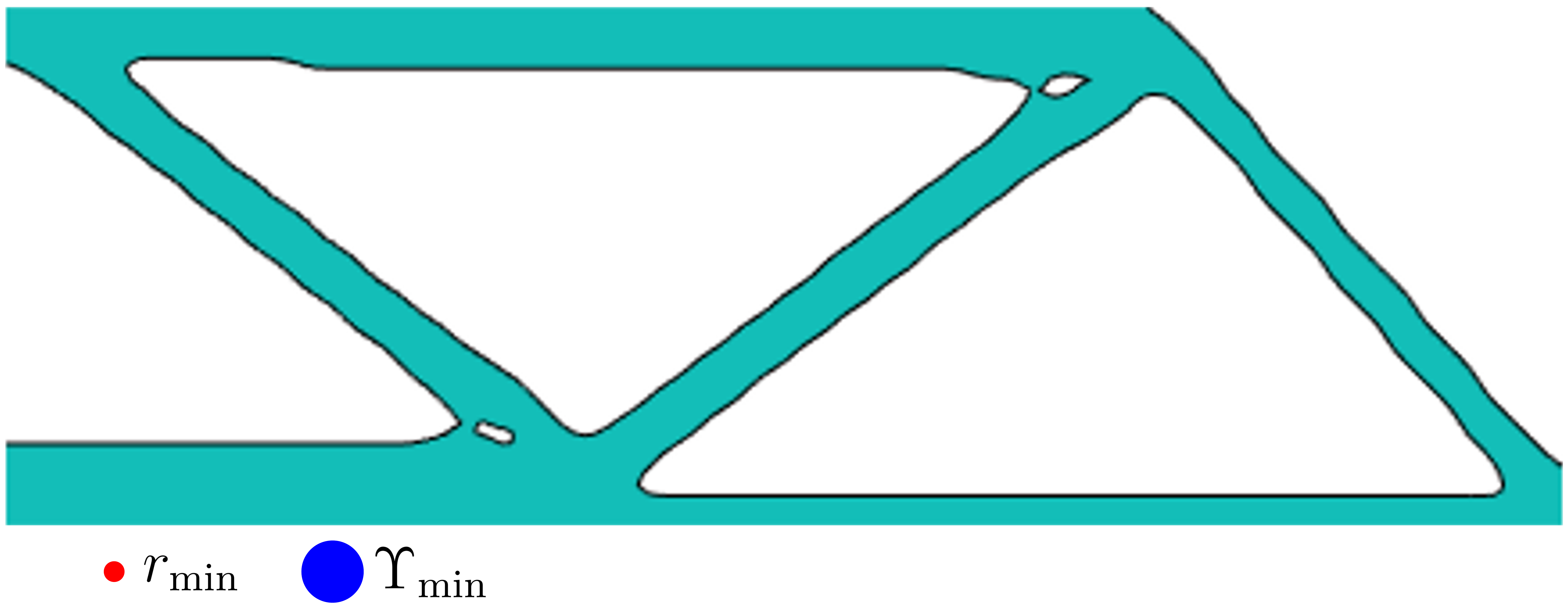}
		\caption{$r_{\min}=1$, $\Upsilon_{\min}=3$}
		\label{sFig: 2DC13}
	\end{subfigure}
	\\
	\begin{subfigure}{0.32\textwidth}
		\centering
		\includegraphics [width=145 pt] {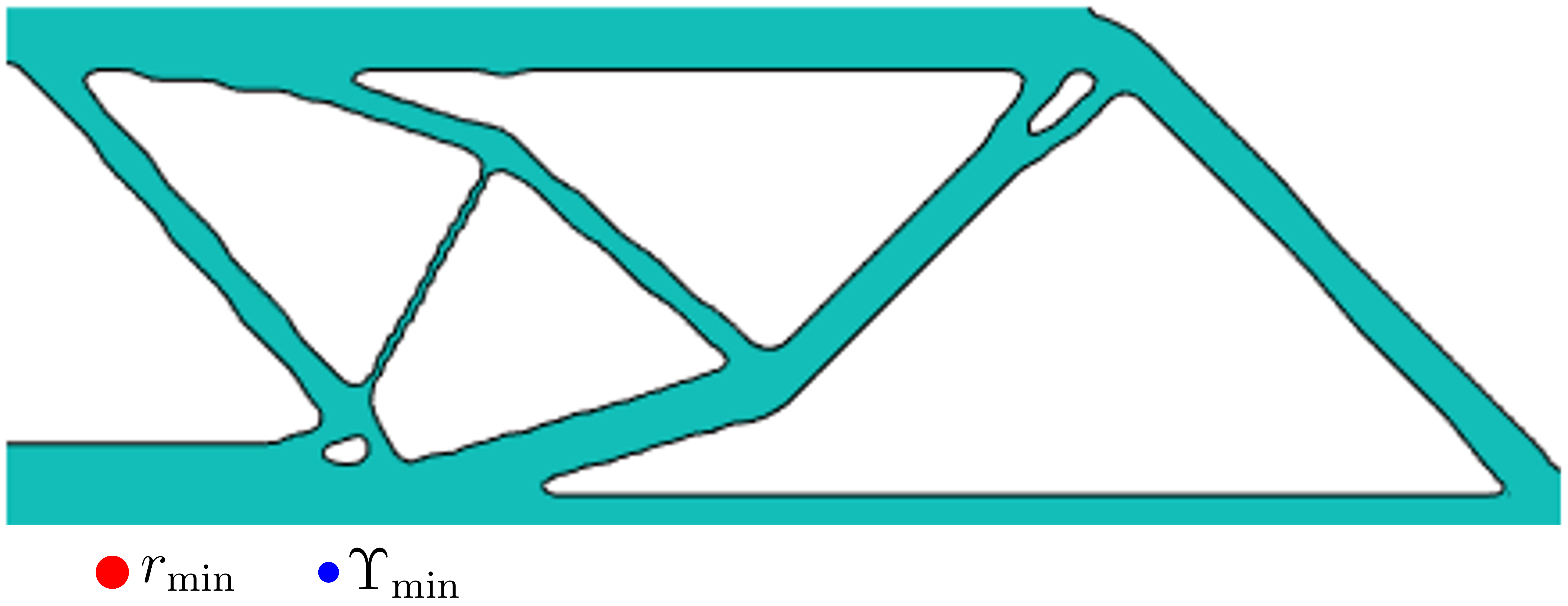}
		\caption{$r_{\min}=1.6$, $\Upsilon_{\min}=1$}
		\label{2DC151}
	\end{subfigure}
	\begin{subfigure}{0.32\textwidth}
		\centering
		\includegraphics [width=145 pt] {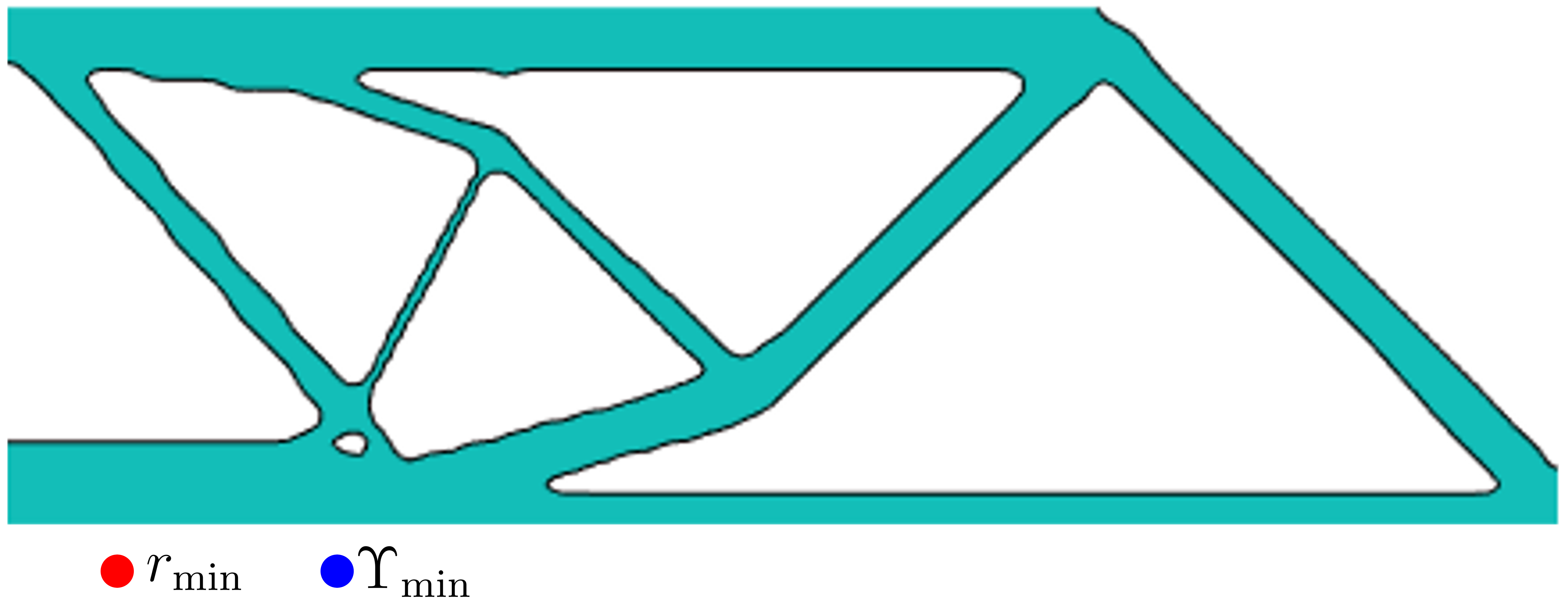}
		\caption{$r_{\min}=1.6$, $\Upsilon_{\min}=1.6$}
		\label{2DC1515}
	\end{subfigure}
	\begin{subfigure}{0.32\textwidth}
		\centering
		\includegraphics [width=145 pt] {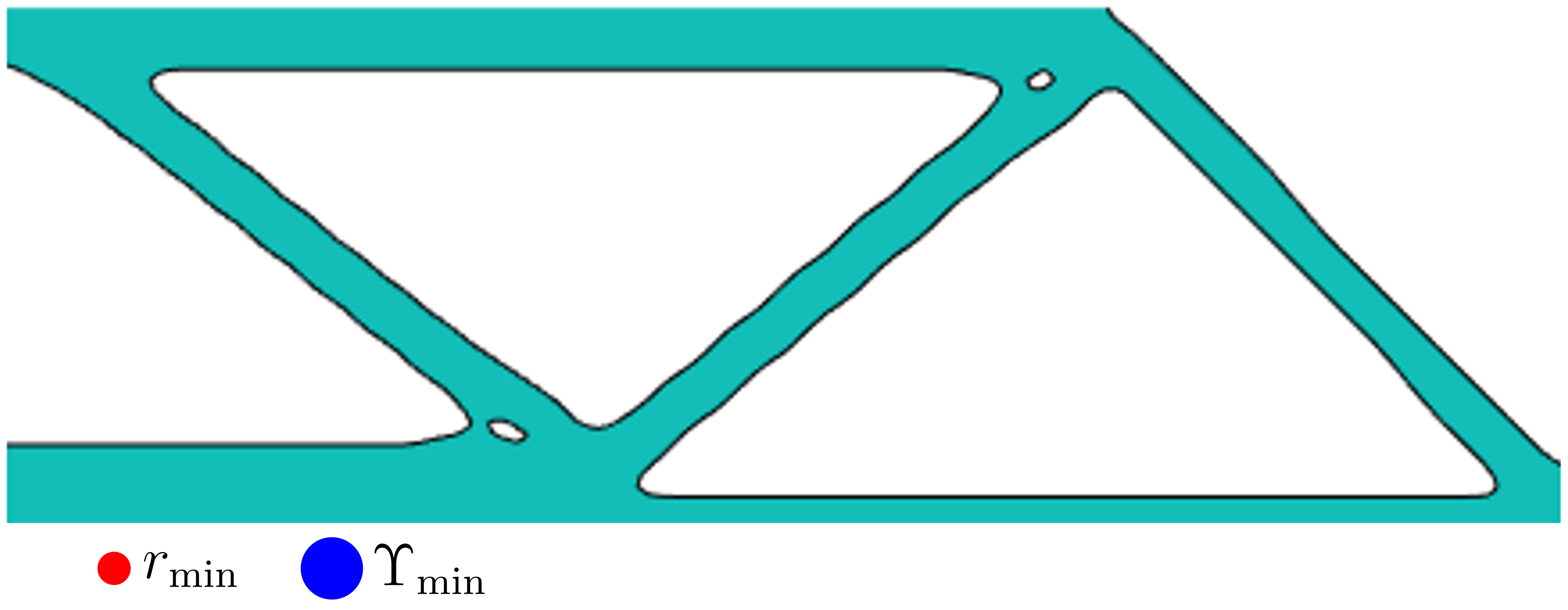}
		\caption{$r_{\min}=1.6$, $\Upsilon_{\min}=3$}
		\label{2DC153}
	\end{subfigure}
	\\
	\begin{subfigure}{0.32\textwidth}
		\centering
		\includegraphics [width=145 pt] {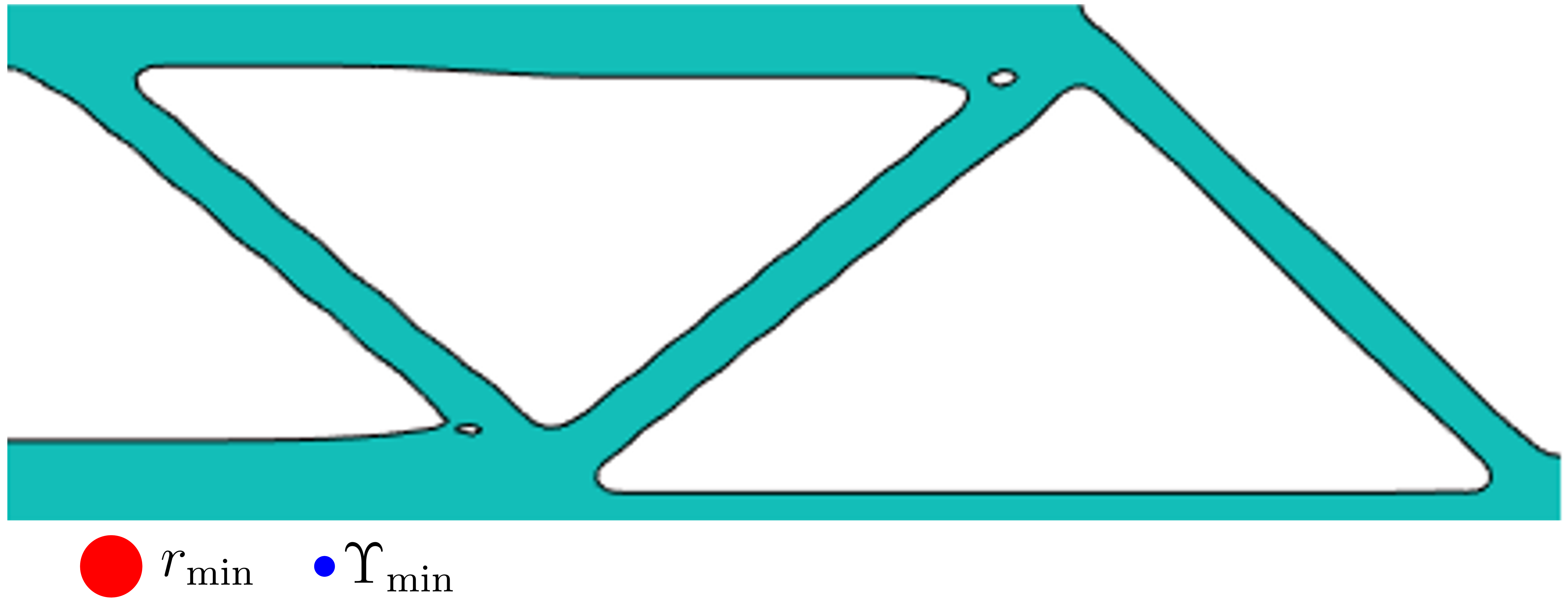}
		\caption{$r_{\min}=3$, $\Upsilon_{\min}=1$}
		\label{2DC31}
	\end{subfigure}
	\begin{subfigure}{0.32\textwidth}
		\centering
		\includegraphics [width=145 pt] {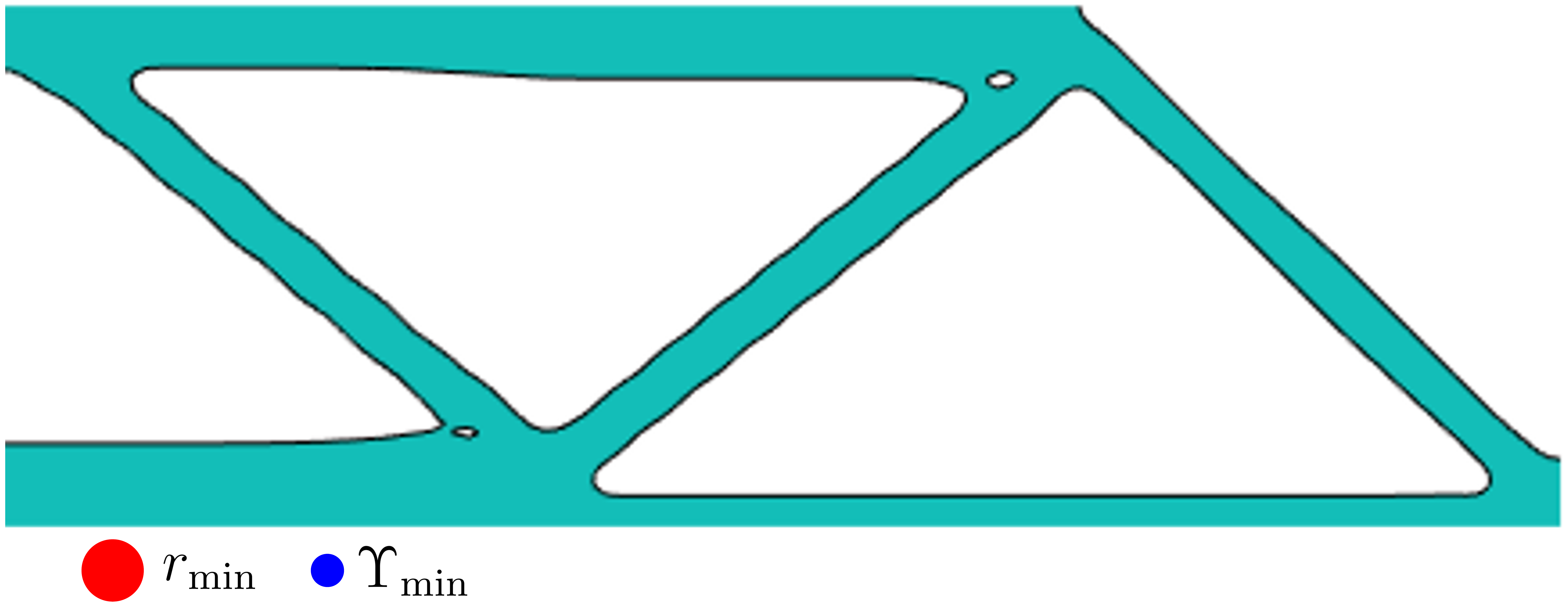}
		\caption{$r_{\min}=3$, $\Upsilon_{\min}=1.6$}
		\label{2DC316}
	\end{subfigure}
	\begin{subfigure}{0.32\textwidth}
		\centering
		\includegraphics [width=145 pt] {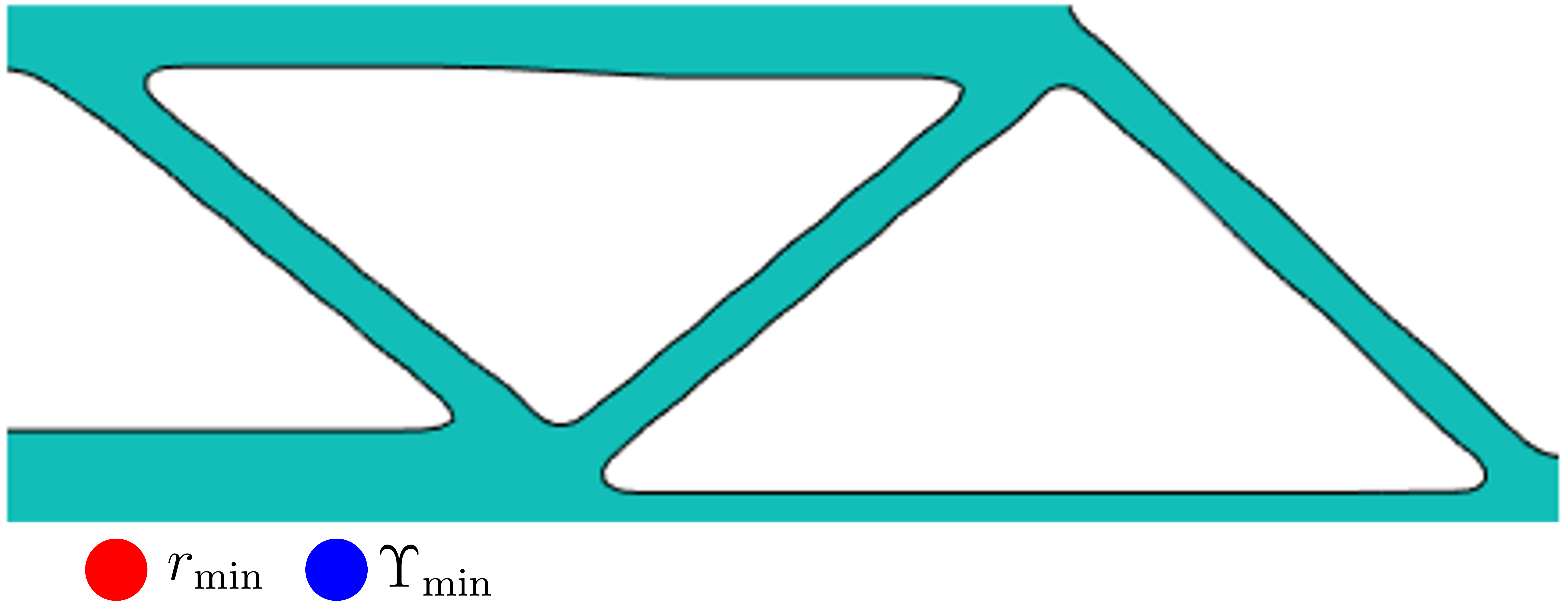}
		\caption{$r_{\min}=3$, $\Upsilon_{\min}=3$}
		\label{2DC33}
	\end{subfigure}
	\caption{Optimized topologies with different combinations of $r_{\min}$ and $\Upsilon_{\min}$ for MBB beam case}
	\label{Fig: TDCR}
\end{figure}

\begin{figure}
	\centering
	\includegraphics[scale=1]{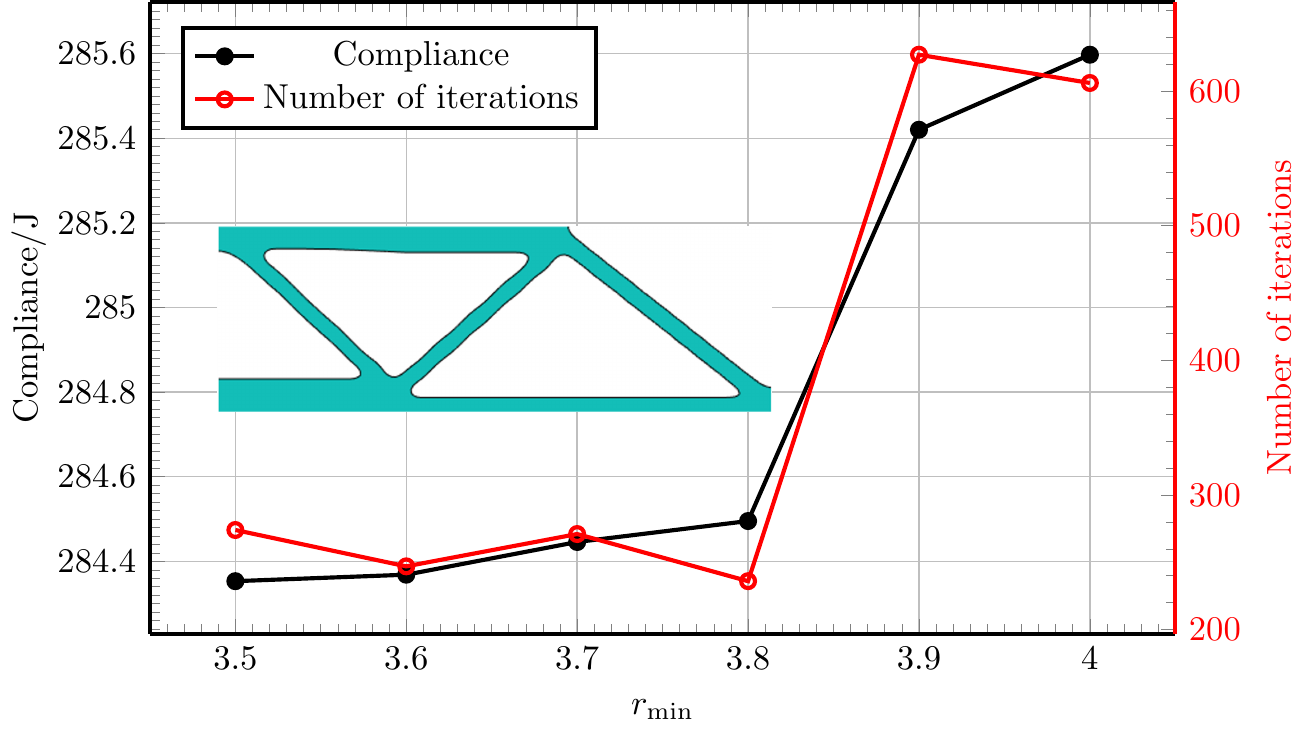}
	\caption{Compliance, convergence, and topological designs under large values of $r_{\min}$ and $\Upsilon_{\min}$=1}
	\label{Fig: RRCN}
\end{figure}

Topological designs with different combinations of $r_{\min}$ and $\Upsilon_{\min}$ are shown in Figure \ref{Fig: TDCR}. Length scales representing $r_{\min}$ and $\Upsilon_{\min}$ are added into Figure \ref{Fig: TDCR} as reference. Increasing either $r_{\min}$ or $\Upsilon_{\min}$ results in simpler topologies with less holes.  Small holes vanish when the combination of $r_{\min}=3$ and $\Upsilon_{\min}=3$ is used, which is beneficial to the manufacturability of optimized topologies. Figure \ref{Fig: RRCN} shows the compliance, convergence, and topological designs under large values of $r_{\min}$ and $\Upsilon_{\min}$=1. As topological designs from $r_{\min}=3.5$ have the similar structural layout with no small holes, only the topology at $r_{\min}=3.5$ and $\Upsilon_{\min}$=1 is given in Figure \ref{Fig: RRCN} for simplicity. Compliance and the number of iterations of the combination of $r_{\min}=3.5$ and $\Upsilon_{\min}=1$ are 284.3535 J and 274, respectively, which are close to those of the combination of $r_{\min}=3$ and $\Upsilon_{\min}=3$ (284.2814 J and 290, respectively). There is an overall tendency for compliance to increase with $r_{\min}$ (Figure \ref{Fig: RRCN}). When $r_{\min}$ reaches 3.9 and 4, large numbers of iterations (627 and 606, respectively) are required to reach convergence, as shown in Figure \ref{Fig: RRCN}.

Even though Equation \ref{Eq: Filter} is basically used for filtering elemental volume fractions in SEMDOT, and the main function of Equation \ref{Eq: FEN} is to assign elemental volume fractions to grid points, different combinations of $r_{\min}$ and $\Upsilon_{\min}$ can also be considered in SEMDOT to explore different topological designs with better performance, quick convergence, or both. Generally, the fixed value of $\Upsilon_{\min}$=1 is recommended for the convenience of implementation.

\subsection{Validation of sensitivity analysis method}
\begin{figure}[htbp!]
	\centering
	\includegraphics[width=350 pt]{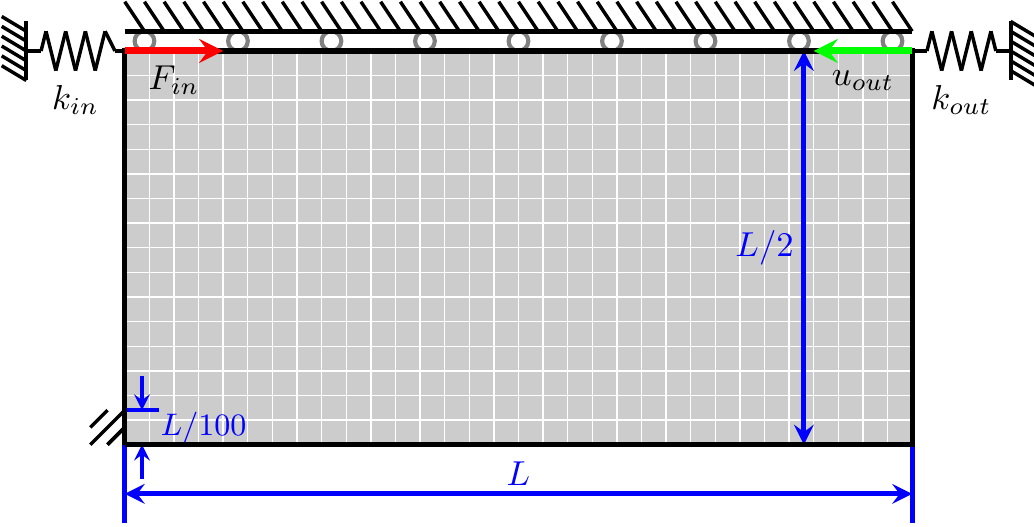}
	\caption{Force inverter design problem }
	\label{Fig: CMD}
\end{figure}

To further validate the effectiveness of the sensitivity analysis method (Equation \ref{Eq: boundaryES}), one typical compliant mechanism design problem (Figure \ref{Fig: CMD}) is considered. The input force $F_{in}$ is set to 1 N, and input ($k_{in}$) and output ($k_{out}$) spring stiffnesses are set to 1 and 0.001, respectively. The design domain is discretized by a  $80 \times 40$ finite element mesh, and the filter radius $r_{\min}$ is set to 2 time elements width ($r_{\min}$=2).

\begin{figure}[htbp!]
	\centering
	\begin{subfigure}{0.6\textwidth}
		\centering
		\includegraphics [width=285 pt] {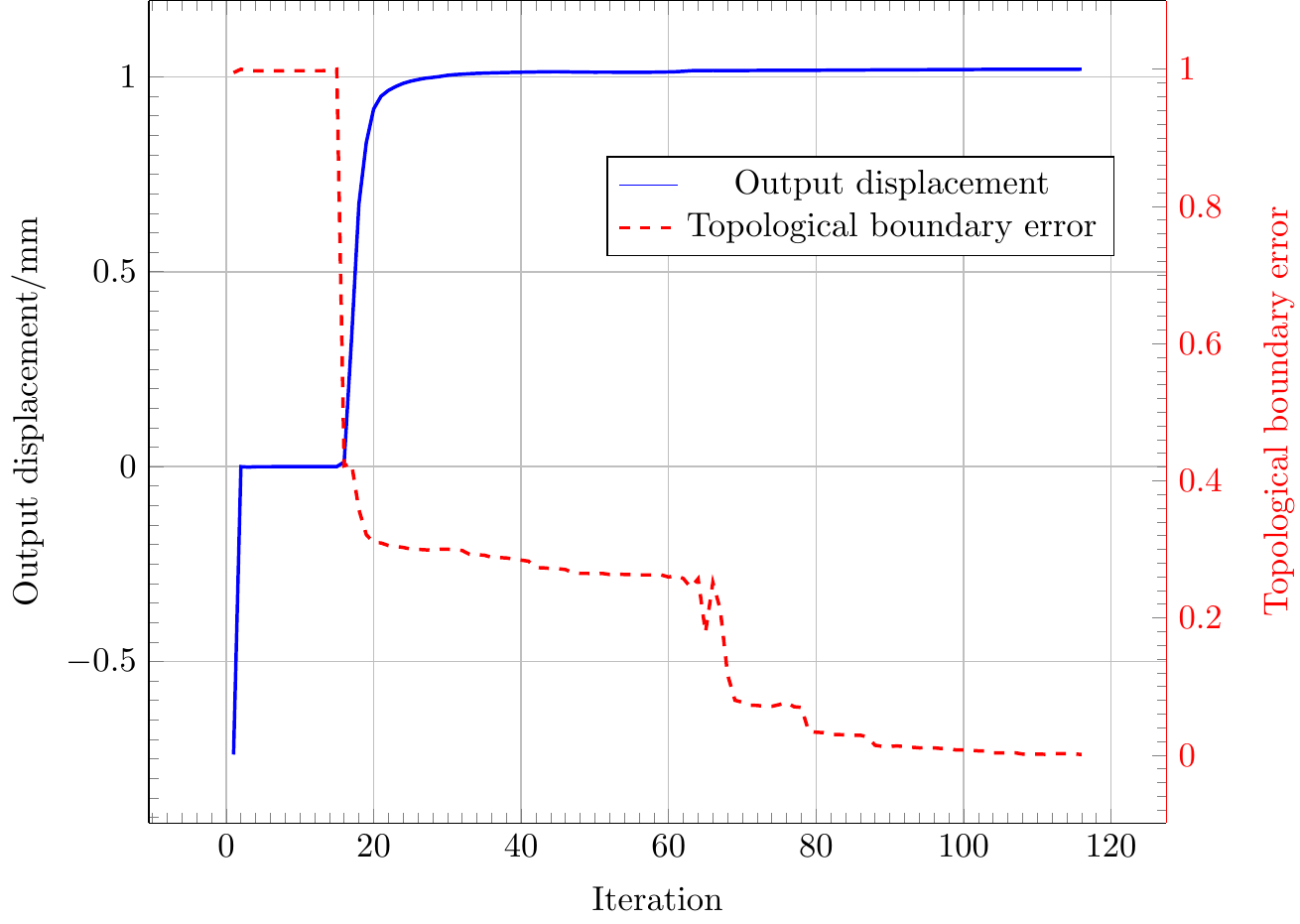}
		\caption{Convergence process}
		\label{sFig: OutputD}
	\end{subfigure}
	\begin{subfigure}{0.39\textwidth}
		\centering
		\includegraphics [width=200 pt] {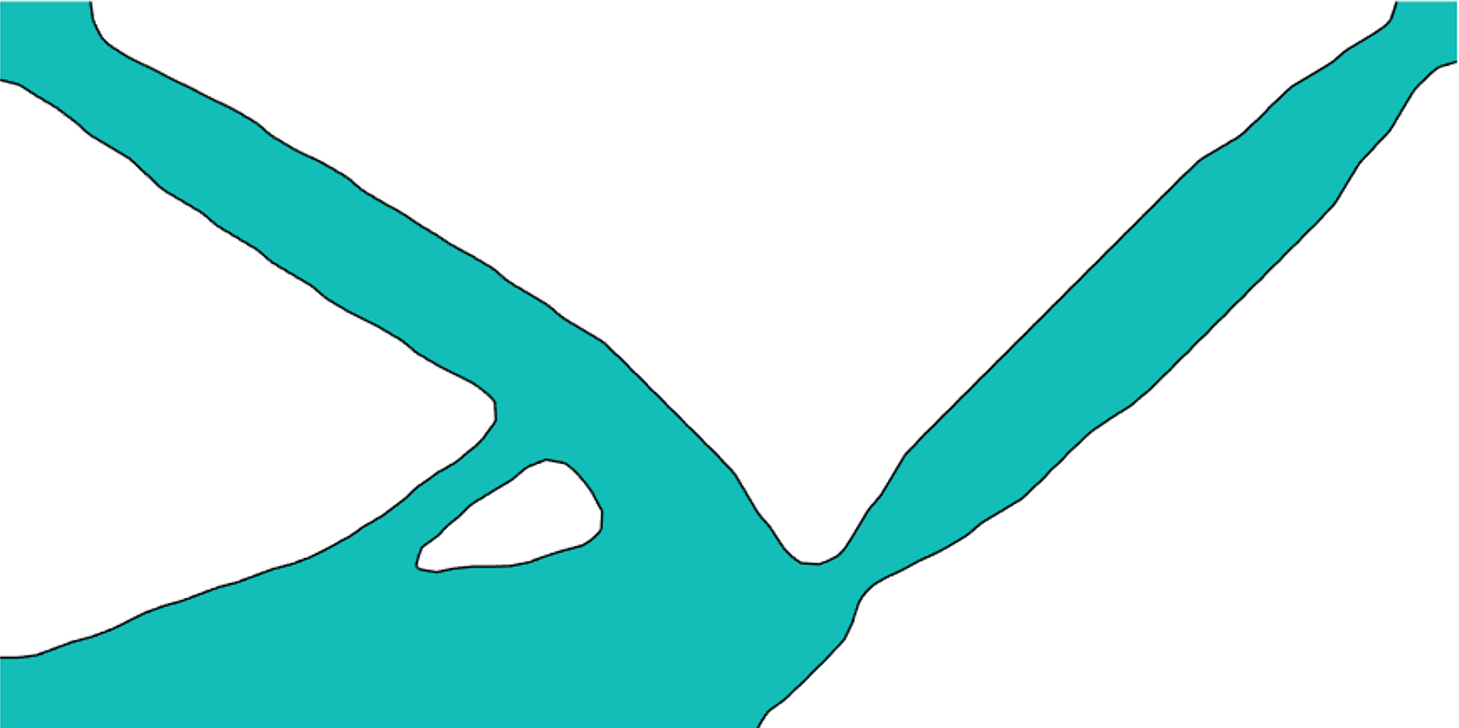}
		\caption{Optimized topology}
		\label{sFig: ComDT}
	\end{subfigure}
\caption{Convergence process and optimized topology for force inverter design case}
\label{Fig: CPOPFID}
\end{figure}

The optimization process converges at the output displacement of 1.0197 mm after 116 iterations (Figure \ref{sFig: OutputD}), and the correct topological design is obtained by SEMDOT (Figure \ref{sFig: ComDT}). Based on the results in Figure \ref{Fig: CPOPFID}, it can be concluded that the sensitivity analysis method adopted in SEMDOT is suitable for compliant mechanism design, which is more challenging than compliance minimization design. The effectiveness of the sensitivity analysis method in SEMDOT is therefore proved.

\subsection{Numerical comparisons} \label{Sec: NC}
Cantilever beam and L-bracket beam test cases are used to thoroughly compare SEMDOT with some well-established element-based algorithms in the ability of seeking the optimal solution and convergency.  Only methods with published source codes (i.e., SIMP \cite{andreassen2011efficient}, BESO \cite{huang2010further}, and ETO \cite{da2018evolutionary}) are selected in this comparison, together with SEMDOT with the single filtering step (Equation \ref{Eq: FEN}). In terms of SIMP, two typical filters: the density filter \cite{bourdin2001filters,bruns2001topology,sigmund2007morphology} and the Heaviside projection filter \cite{guest2004achieving} are considered. In the Heaviside projection filter, the parameter controlling the smoothness of the approximation is gradually increased from 1 to 128 by doubling its value every 25 iterations or when the change between two consecutive designs is less than 0.01. For simplicity, SIMP-D and SIMP-H are used to represent SIMP with density and Heaviside projection filters, respectively, in graph legends and captions. In addition, SEMDOT-S is used to represent SEMDOT with the sole heuristic filter in graph legends and captions. The penalty coefficient of 1.5 is used in SEMDOT, SEMDOT-S, and ETO, and the penalty coefficient of 3 is used in SIMP and BESO. For BESO and ETO, the evolution rate $er$ is set to 2$\%$. In addition, the maximum number of iterations is set to 300 for all methods.

\begin{figure}[htbp!]
	\centering
	\includegraphics[scale=0.8]{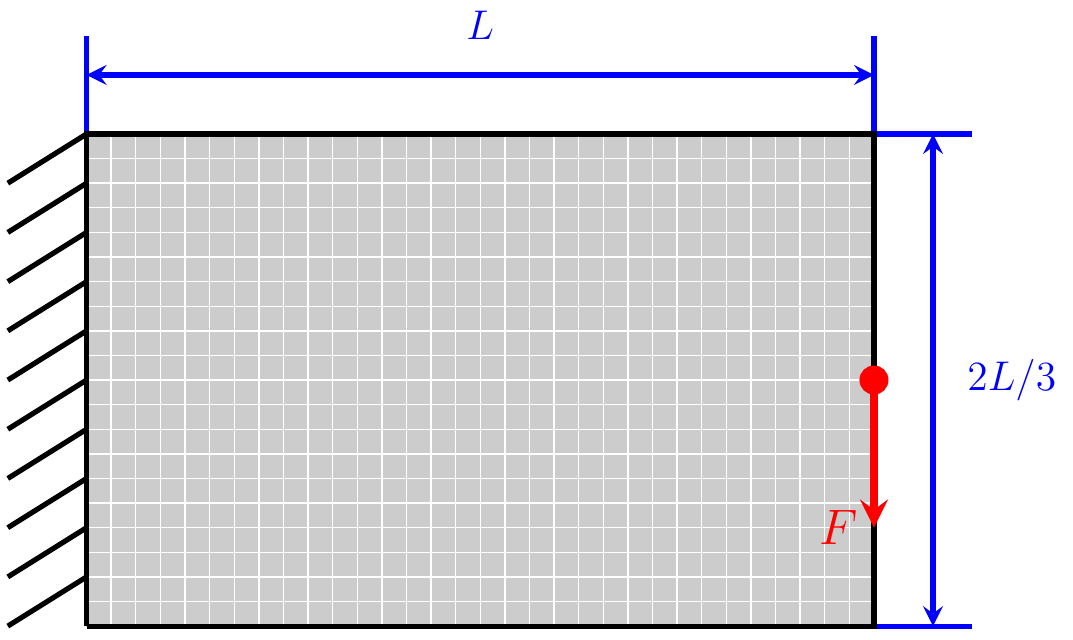}
	\caption{Design domain of a deep cantilever beam}
	\label{Fig: Cantilever}
\end{figure}

The design domain and boundary condition of a deep cantilever beam are shown in Figure \ref{Fig: Cantilever}. The left side is fixed and a unit vertical load ($F=-1\mathrm{N}$) is imposed at the center point of the right side. The design domain of this beam is discretized by a  $150 \times 100$ finite element mesh. The filter radius $r_{\min}$ is set to 2.5 time elements width ($r_{\min}$=2.5). 

\begin{figure}[htbp!]
	\centering
	\begin{subfigure}{\textwidth}
		\centering
		\includegraphics[scale=1]{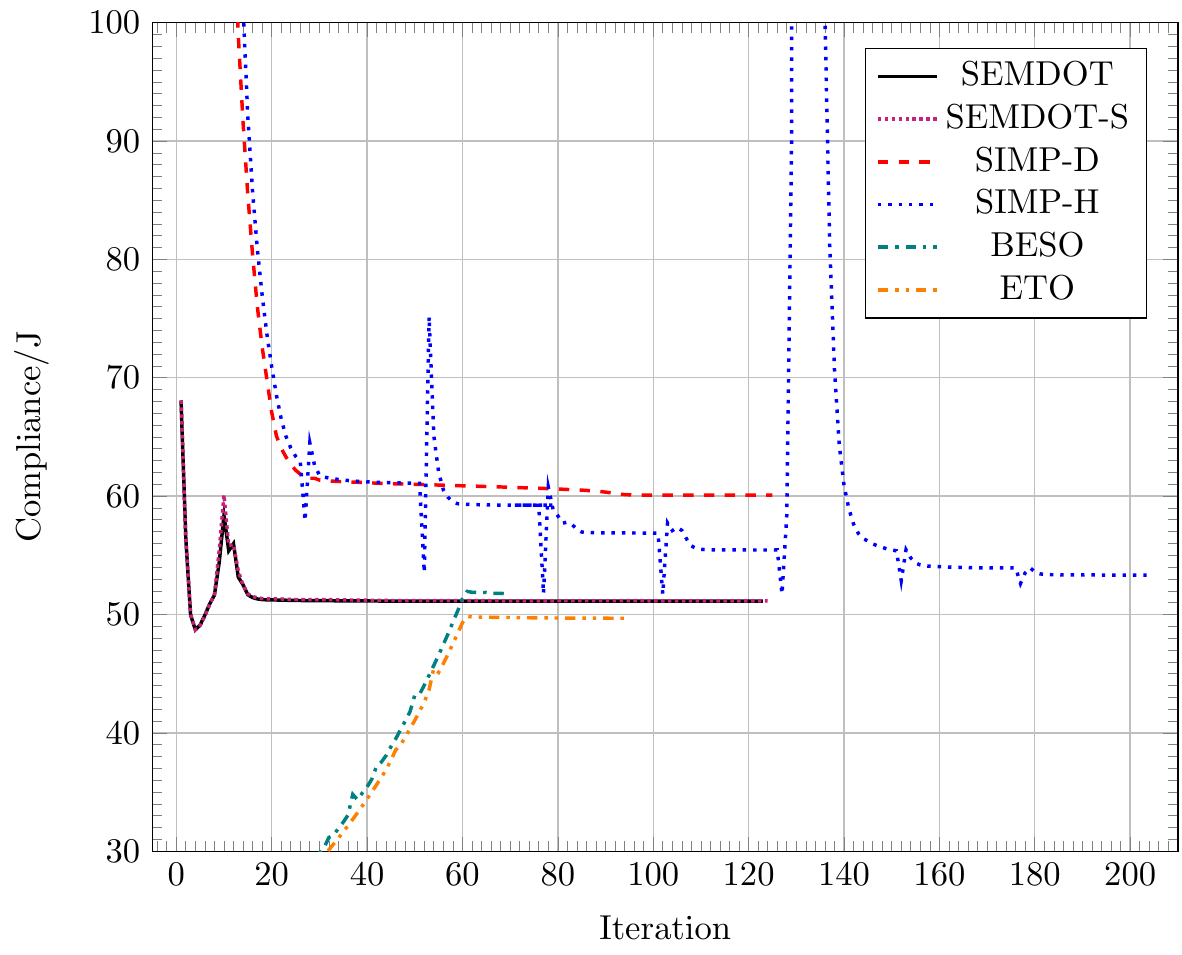}
		\caption{Convergence histories}
		\label{Fig: Ccompliance}
	\end{subfigure}
	\\
	\begin{subfigure}{0.32\textwidth}
		\centering
		\includegraphics [width=145 pt] {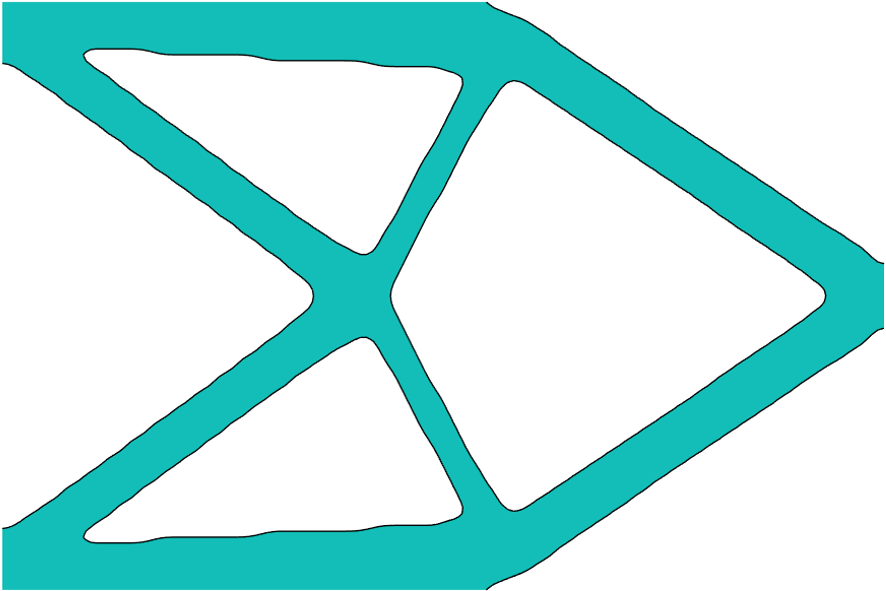}
		\caption{SEMDOT}
		\label{sFig: TSCTO}
	\end{subfigure}
	\begin{subfigure}{0.32\textwidth}
		\centering
		\includegraphics [width=145 pt] {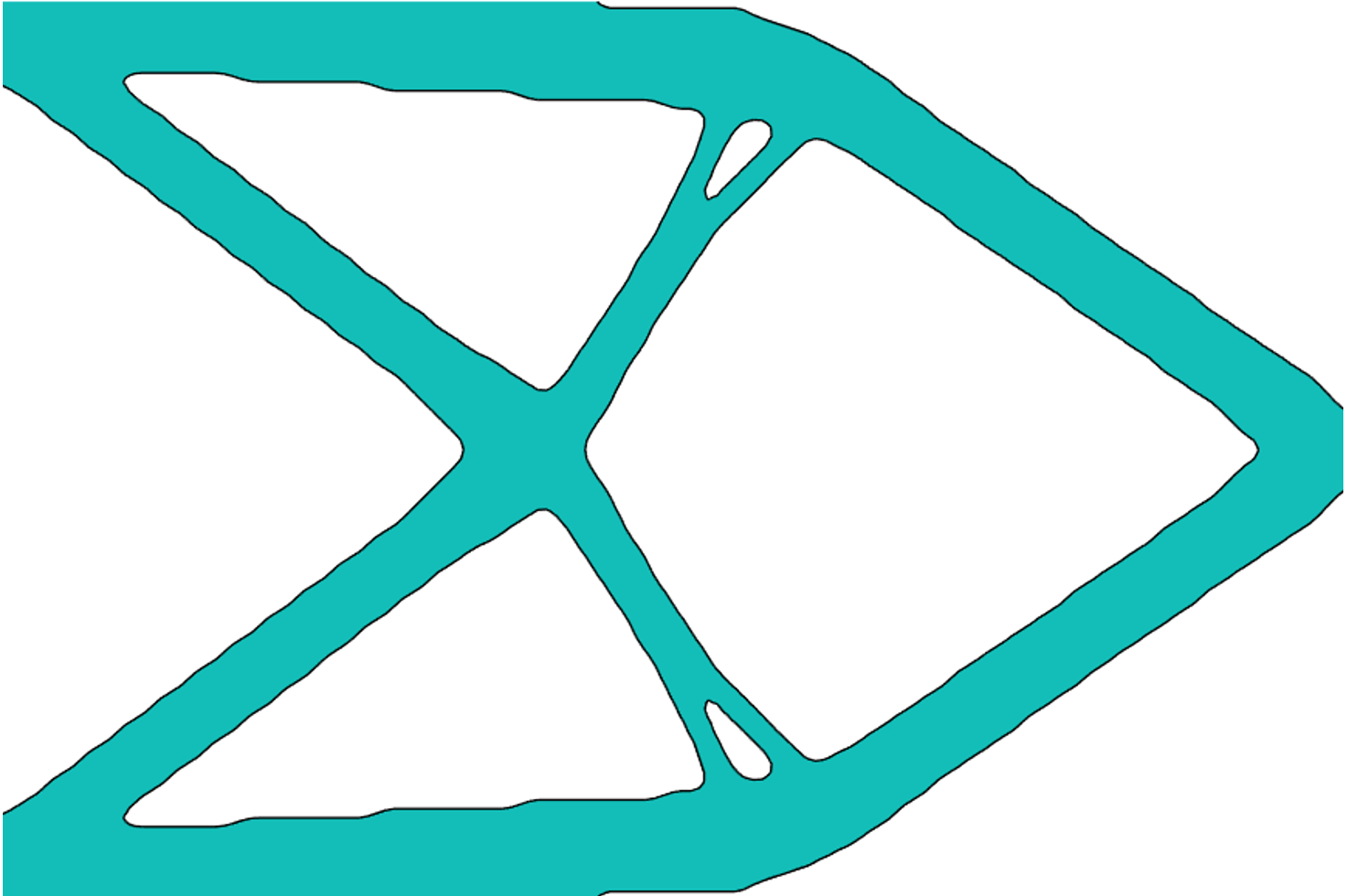}
		\caption{SEMDOT-S}
		\label{sFig: TSEMDOTS}
	\end{subfigure}
	\begin{subfigure}{0.32\textwidth}
		\centering
		\includegraphics [width=145 pt] {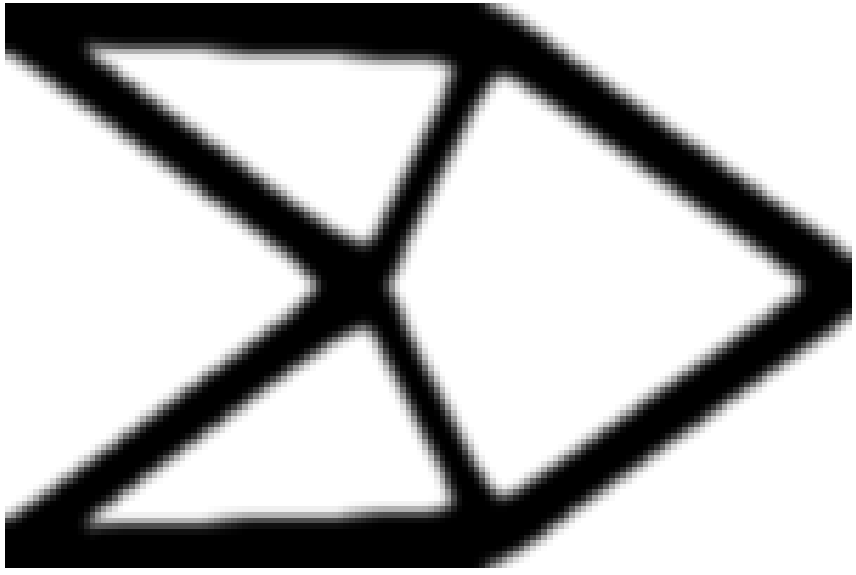}
		\caption{SIMP-D}
		\label{sFig: TDSIMP}
	\end{subfigure}
\\
	\begin{subfigure}{0.32\textwidth}
		\centering
		\includegraphics [width=145 pt] {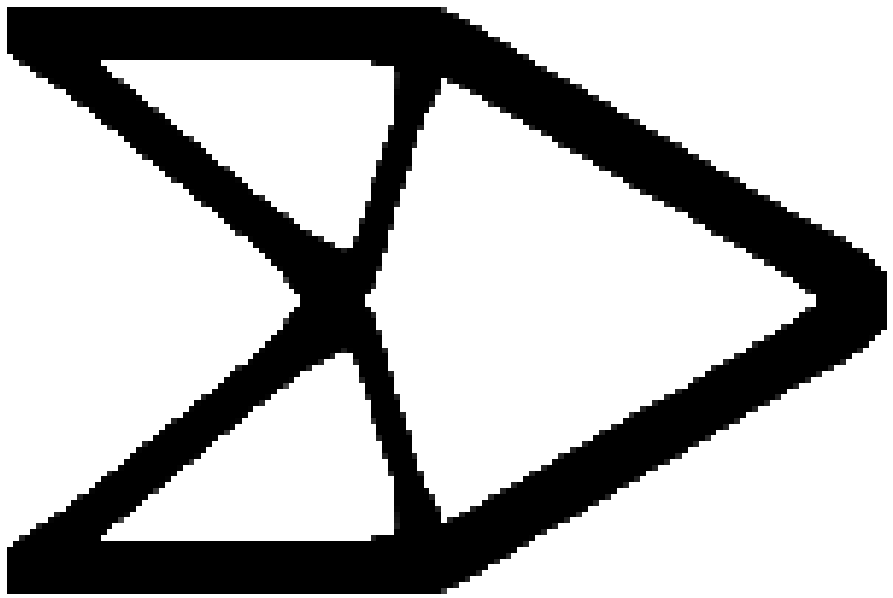}
		\caption{SIMP-H}
		\label{sFig: THSIMP}
	\end{subfigure}
	\begin{subfigure}{0.32\textwidth}
		\centering
		\includegraphics [width=145 pt] {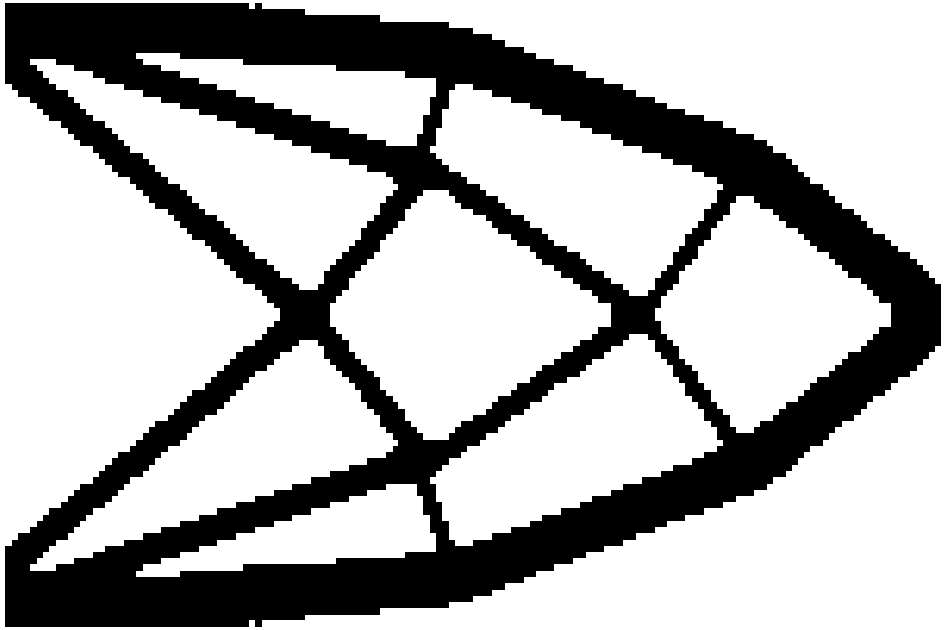}
		\caption{BESO}
		\label{sFig: TBESO}
	\end{subfigure}
	\begin{subfigure}{0.32\textwidth}
		\centering
		\includegraphics [width=145 pt] {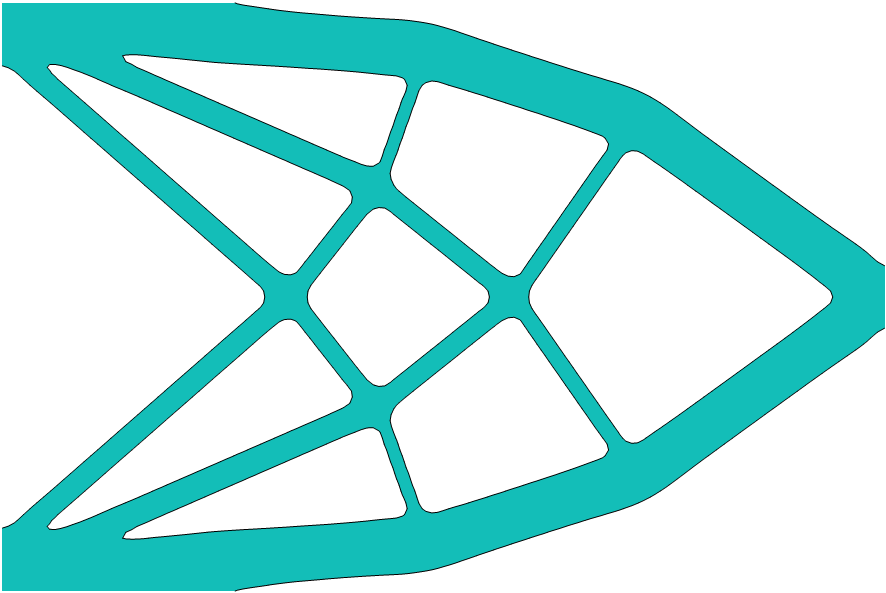}
		\caption{ETO}
		\label{sFig: TETO}
\end{subfigure}
	\caption{Comparisons of performance, convergency, and topological designs between different algorithms for deep cantilever beam case}
\end{figure}

Figure \ref{Fig: Ccompliance} shows that the best compliance (49.6856 J) is obtained by ETO, followed by 51.1240 J obtained by SEMDOT. The optimization process of ETO converges after 95 iterations, which is less than that of SEMDOT (123 iterations). SEMDOT-S converges at the compliance of 51.1558 J after 124 iterations, which is almost identical to SEMDOT, whereas two different topological designs are obtained (Figures \ref{sFig: TSCTO} and \ref{sFig: TSEMDOTS}). Even though both SEMDOT and ETO are based on elemental volume fractions, optimized topologies (Figures \ref{sFig: TSCTO} and \ref{sFig: TETO}) are different. The topological design obtained by SEMDOT is similar to those obtained by SIMP (Figures \ref{sFig: TDSIMP} and \ref{sFig: THSIMP}), and the topological design obtained by ETO is similar to that of BESO (Figure \ref{sFig: TBESO}). In this case, ETO is superior to SEMDOT in performance and convergence. The worst compliance (60.0841 J) is obtained by SIMP with the density filter. This is because intermediate elements are allowed to distribute across the whole design domain, and intermediate elements result in smaller improvement of stiffness per density due to the use of penalty factor. BESO converges at the compliance of 51.7538 J after 69 iterations, which is better than SIMP with the Heaviside projection filter with $C$=53.3355 J after 204 iterations.

Effects of different mesh sizes and domain aspect ratios are considered to provide statisticalcomparisons. As the element size is scaled with a certain ratio for test cases with different mesh sizes, the filter radius is scaled with the same ratio to ensure that its absolute value remains constant. Table \ref{tab:CaC} demonstrates that performance obtained by ETO is still the best for different mesh sizes. Performance obtained by SEMDOT-S is the second best for the mesh sizes ranging from 60$\times$40 to 120$\times$80, and performance obtained by SEMDOT is the second best for the mesh sizes ranging from 180$\times$120 to 270$\times$180. Table \ref{tab:CaI} demonstrates that BESO converges the fastest for different mesh sizes, followed by ETO. SEMDOT-S performs better than SEMDOT in terms of convergency. Both SIMP-D and SIMP-H face the convergence difficulty when the fine mesh is used.

\begin{table}[htbp!]
	\begin{center}
		\caption{Effects of mesh size on compliance obtained by different algorithms solving deep cantilever beam case}
		\label{tab:CaC}
		\begin{tabular}{cccccccc} % <-- Alignments: 1st column left, 2nd middle and 3rd right, with vertical lines in between
			\hline
			\hline		
			\multirow{2}{*}{Algorithms}	& \multicolumn{7}{c}{Mesh size}\\
			\cmidrule{2-8}
			 & 60$\times$40 & 90$\times$60 & 120$\times$80 & 180$\times$120 & 210$\times$140 & 240$\times$160 & 270$\times$180  \\
			\hline
			\rowcolor{black!10}
			SEMDOT & 51.0698  & 50.9852  & 50.9763  & 51.2065  & 51.2797  & 51.3474  & 51.3796 \\
			SEMDOT-S & 51.0698 & 50.9200 & 50.9716 & 51.3621 & 51.5000 & 51.5254 & 51.6761   \\
			\rowcolor{black!10}
			SIMP-D & 65.6168 & 59.8601 & 59.0882  & 59.7609  & 60.1124  & 60.1080 & 60.2780  \\
			SIMP-H & 61.9331 & 52.6167 & 205.9870 & 54.1442 & 53.4234  & 65.8000 & 64.4706 \\
			\rowcolor{black!10}
			BESO & 55.0152 & 52.7312 & 51.8462 & 51.2688 & 51.2836  & 51.3759 & 51.2489 \\
			ETO & 49.6404 & 49.2007 & 49.6575 & 49.8537 & 49.9187  & 50.0130 & 50.0750 \\
			\hline
			\hline
		\end{tabular}
	\end{center}
\end{table}

\begin{table}[htbp!]
	\begin{center}
		\caption{Effects of mesh size on number of iterations obtained by different algorithms solving deep cantilever beam case}
		\label{tab:CaI}
		\begin{tabular}{cccccccc} % <-- Alignments: 1st column left, 2nd middle and 3rd right, with vertical lines in between
			\hline
			\hline
			 \multirow{2}{*}{Algorithms}	& \multicolumn{7}{c}{Mesh size}\\
			\cmidrule{2-8}
			 & 60$\times$40 & 90$\times$60 & 120$\times$80 & 180$\times$120 & 210$\times$140 & 240$\times$160 & 270$\times$180  \\
			\hline
			\rowcolor{black!10}
			SEMDOT & 128 & 93 & 104 & 138 & 158 & 179 & 197\\
			SEMDOT-S & 128 & 79 & 165 & 132 & 152 & 173 & 196\\
			\rowcolor{black!10}
			SIMP-D & 36 & 98 & 102 & 215 & 300 & 300 & 300\\
			SIMP-H & 130 & 199 & 198 & 277 & 256 & 300  & 300\\
			\rowcolor{black!10}
			BESO & 70 & 70 & 69 & 81 & 70 & 70 & 70\\
			ETO & 78 & 99 & 97 & 86 & 93 & 94 & 99\\
			\hline
			\hline
		\end{tabular}
	\end{center}
\end{table}

In terms of different domain aspect ratios, the number of elements in the vertical direction is fixed at 60, and the filter radius $r_{\min}$ is set to 1.5 time elements width ($r_{\min}$=1.5). Table \ref{tab:CaCDR} demonstrates that ETO obtains better performance than both SEMDOT and SEMDOT-S. Performance obtained by SEMDOT-S is the second best for the domain aspect ratios of 0.5:1, 1:1, and 2.5:1, and performance obtained by SEMDOT is the second best for the domain aspect ratios of 2:1 and 3:1. Table \ref{tab:CaNDR} demonstrates that SEMDOT converges faster than ETO for the domain aspect ratios of 2.5:1 and 3:1, and SEMDOT-S converges faster than ETO for the domain aspect ratios of 0.5:1, 1:1, 2.5:1, and 3:1.

\begin{table}[htbp!]
	\begin{center}
		\caption{Effects of domain aspect ratio on compliance obtained by different algorithms solving deep cantilever beam case}
		\label{tab:CaCDR}
		\begin{tabular}{cccccccc} 
			\hline
			\hline		
			\multirow{2}{*}{Algorithms}	& \multicolumn{5}{c}{Domain aspect ratio}\\
			\cmidrule{2-6}
			& 0.5:1 & 1:1  & 2:1 & 2.5:1 & 3:1  \\
			\hline
			\rowcolor{black!10}
			SEMDOT &  7.8023 & 22.0224 & 94.7833 & 159.2461 & 248.6884 \\
			SEMDOT-S & 7.7981 & 21.9979 & 95.0301 & 159.1104 & 250.2365    \\
			\rowcolor{black!10}
			SIMP-D & 8.7245 & 25.7240 & 108.7384 & 187.2112 & 292.0282  \\
			SIMP-H & 7.9776 & 23.1224 & 98.0371 & 302.8705 & 260.9999  \\
			\rowcolor{black!10}
			BESO &  7.8691 & 22.9724 & 98.6805 & 168.8078 & 259.6505  \\
			ETO & 7.7621 & 21.7892 & 94.7019 & 159.2034 & 246.9935 \\
			\hline
			\hline
		\end{tabular}
	\end{center}
\end{table}

\begin{table}[htbp!]
	\begin{center}
		\caption{Effects of domain aspect ratio on number of iterations obtained by different algorithms solving deep cantilever beam case}
		\label{tab:CaNDR}
		\begin{tabular}{cccccccc} 
			\hline
			\hline		
			\multirow{2}{*}{Algorithms}	& \multicolumn{5}{c}{Domain aspect ratio}\\
			\cmidrule{2-6}
			& 0.5:1 & 1:1  & 2:1 & 2.5:1 & 3:1  \\
			\hline
			\rowcolor{black!10}
			SEMDOT &  91 & 158 & 125 & 126 & 110 \\
			SEMDOT-S & 66 & 88 & 112 & 84 & 91    \\
			\rowcolor{black!10}
			SIMP-D & 35 & 93 & 135 & 160 & 144  \\
			SIMP-H & 207  & 206 & 208 & 257 & 225  \\
			\rowcolor{black!10}
			BESO &  69 & 70 & 74 & 64  & 86  \\
			ETO & 75 & 95 & 98 & 171 & 180 \\
			\hline
			\hline
		\end{tabular}
	\end{center}
\end{table}

Even though SEMDOT-S performs better than SEMDOT for some test cases, it cannot be concluded that the use of multiple filtering step is not needed. This is because the rationality of obtained topologies is also an important assessment criteria. Table \ref{tab:CaTL} gives the comparisons of topological designs obtained by SEMDOT and SEMDOT-S for deep cantilever beam cases considering different domain aspect ratios. The topological boundary obtained by SEMDOT is smoother than SEMDOT-S for the test case with the domain aspect ratio of 1:1. SEMDOT-S forms tiny holes when the domain aspect ratio is 2:1 and discontinuous structures when the domain aspect ratio is 2.5:1. By contrast, SEMDOT does not generate such structural features.

\begin{table}[htbp!]
	\begin{center}
		\caption{Comparisons of topologies with different domain aspect ratios between SEMDOT and SEMDOT-S for deep cantilever beam case}
		\label{tab:CaTL}
		\begin{tabular}{ccc} 
			\hline
			\hline		
			\multirow{2}{*}{Domain aspect ratio}	& \multicolumn{2}{c}{Algorithms}\\
			\cmidrule{2-3}
			& SEMDOT & SEMDOT-S   \\
			\hline
			\rowcolor{black!10}
			0.5:1 & \raisebox{-.5\height}{\includegraphics [width=75 pt] {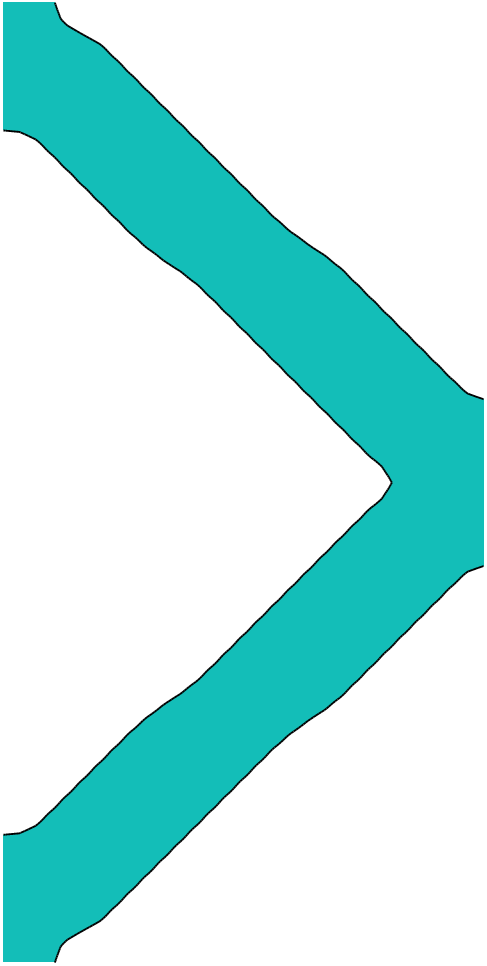}} & \raisebox{-.5\height}{\includegraphics [width=75 pt] {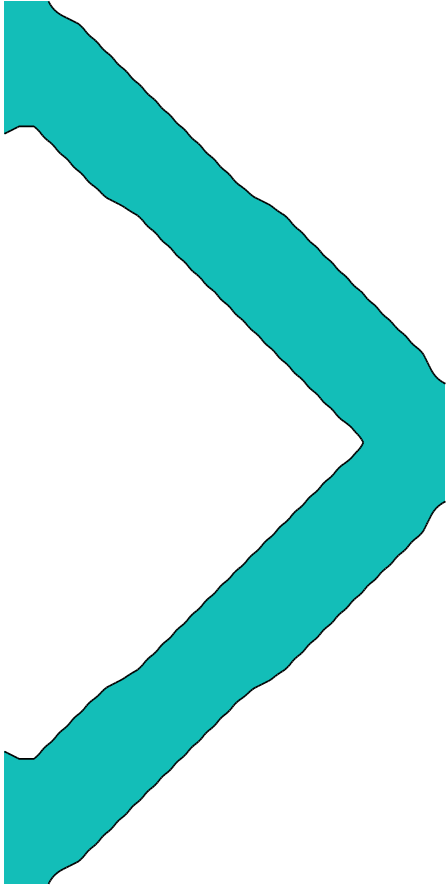}}\\
			1:1 & \raisebox{-.5\height}{\includegraphics [width=130 pt] {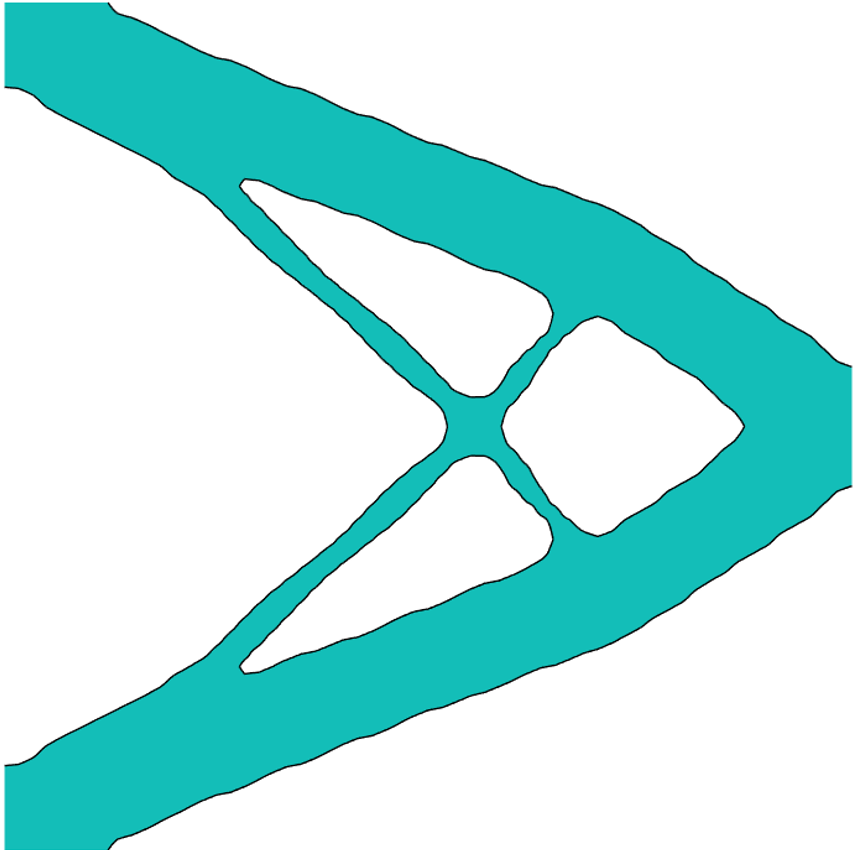}} & \raisebox{-.5\height}{\includegraphics [width=126 pt] {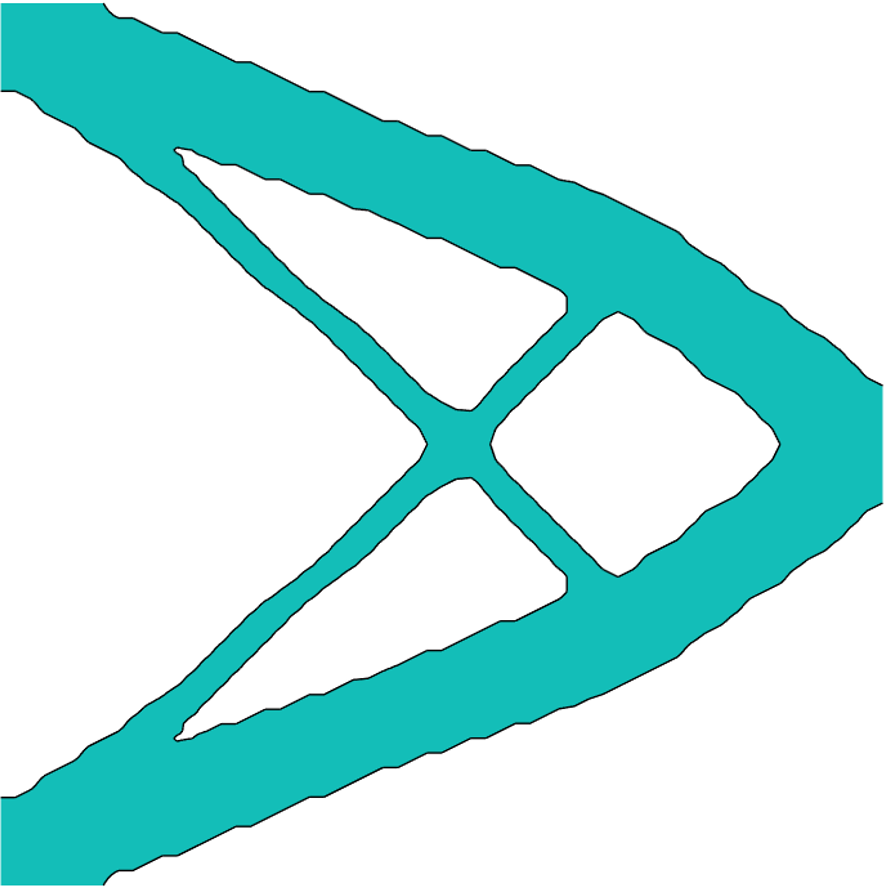}}\\
			\rowcolor{black!10}
			2:1 & \raisebox{-.5\height}{\includegraphics [width=160 pt] {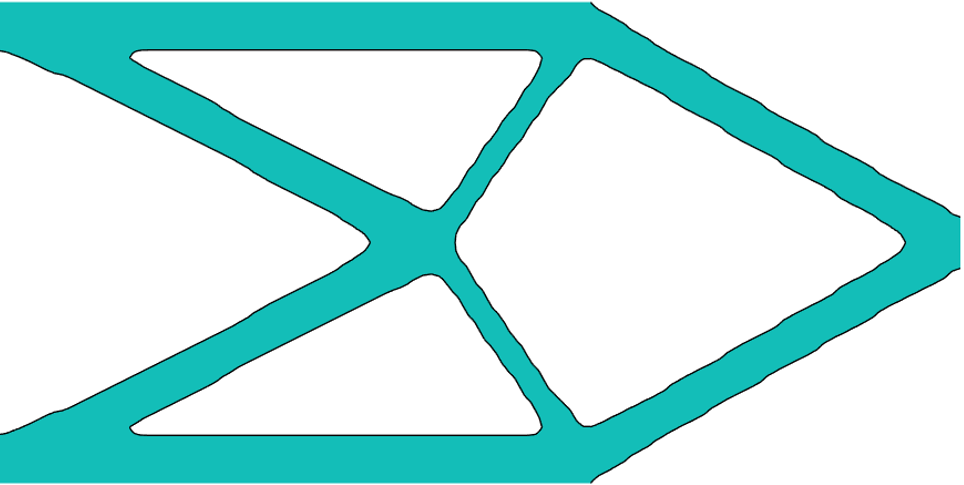}} & \raisebox{-.5\height}{\includegraphics [width=160 pt] {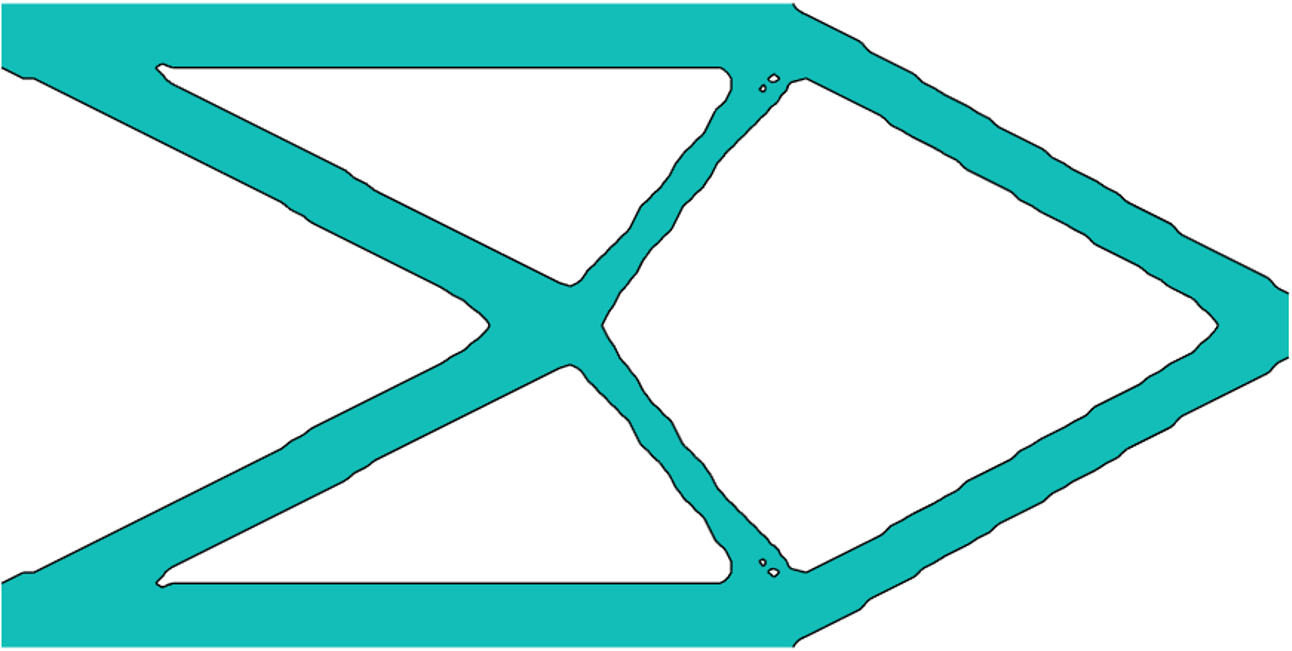}}\\
			2.5:1 & \raisebox{-.5\height}{\includegraphics [width=170 pt] {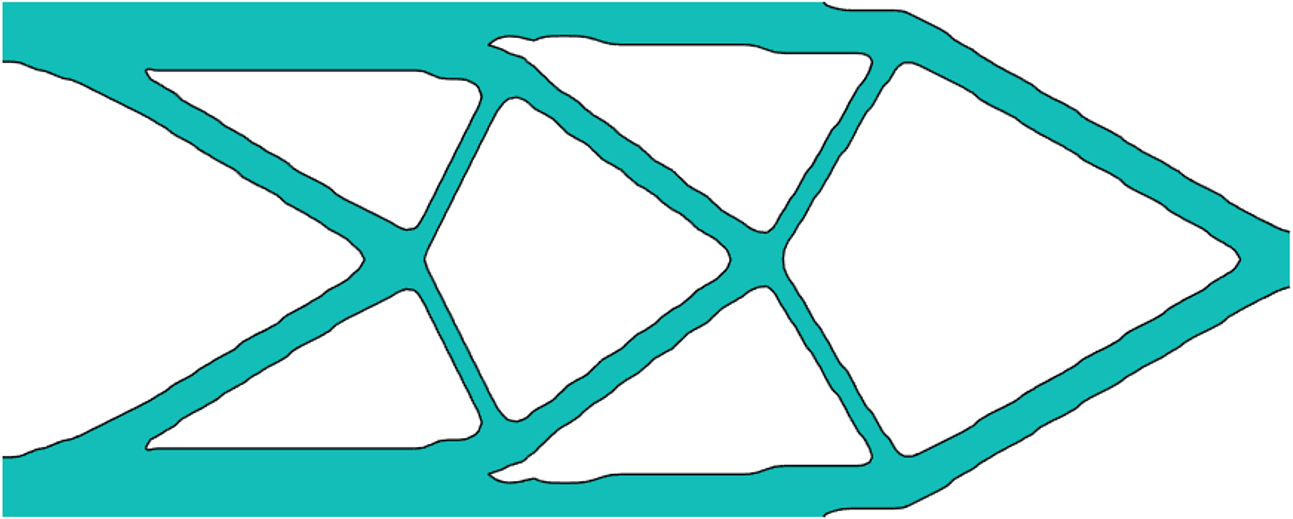}} & \raisebox{-.5\height}{\includegraphics [width=170 pt] {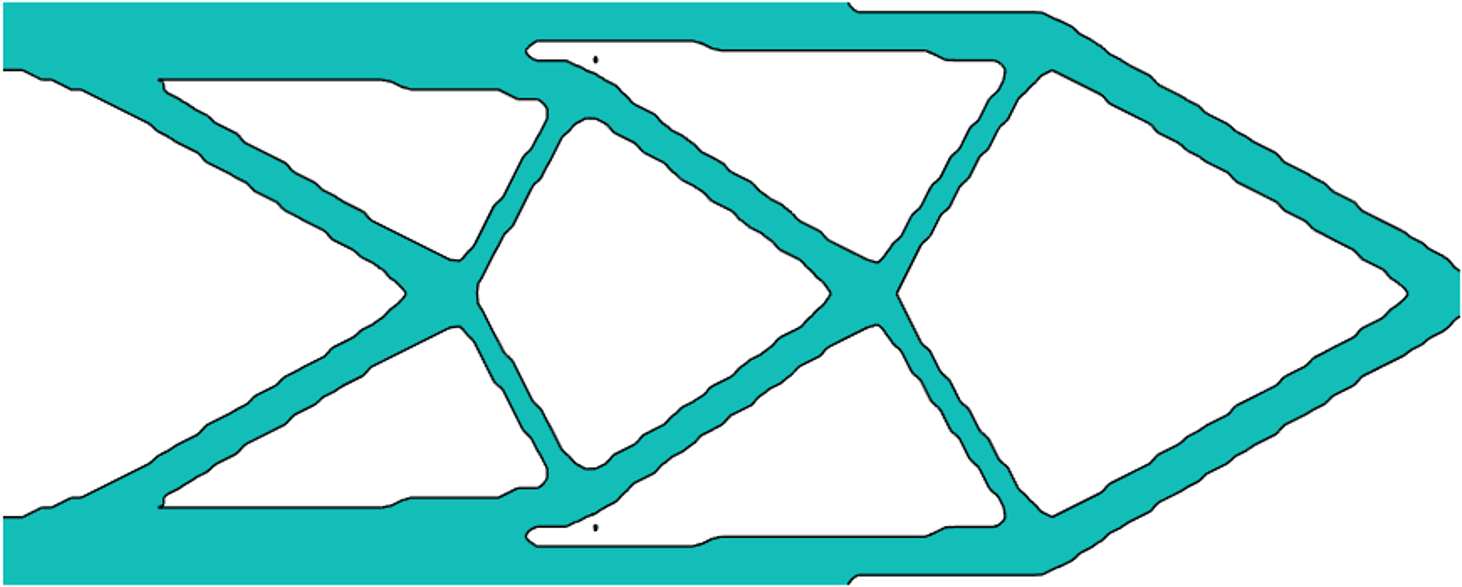}}\\
			\rowcolor{black!10}
			3:1 & \raisebox{-.5\height}{\includegraphics [width=185 pt] {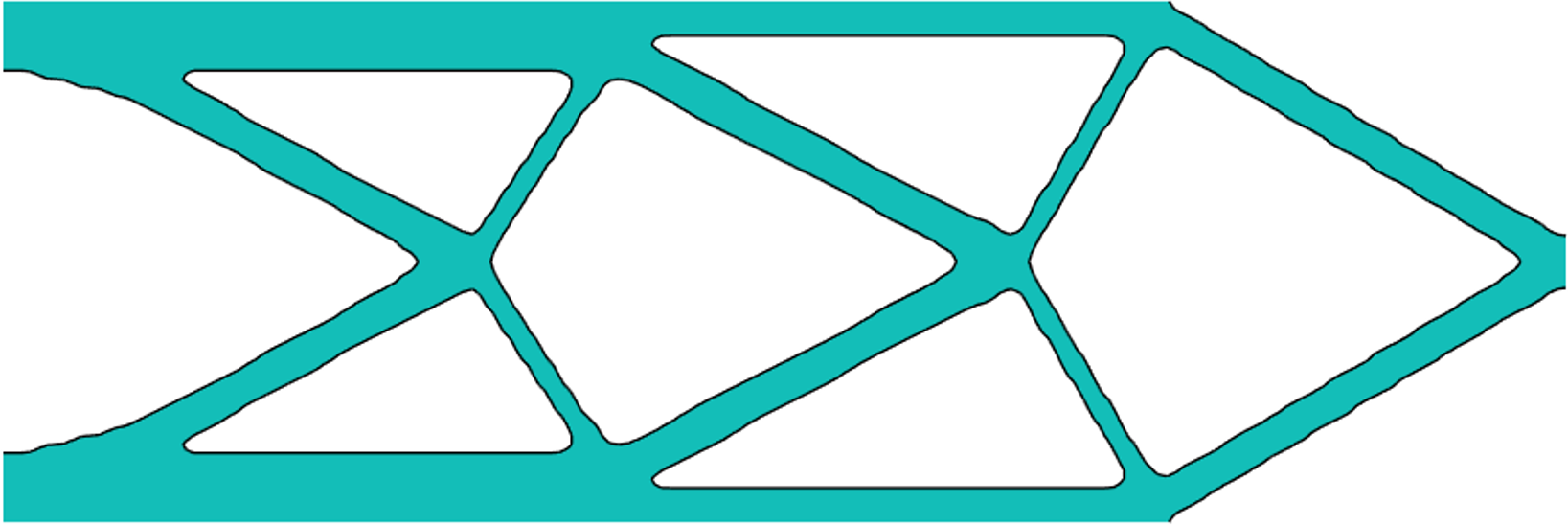}} & \raisebox{-.5\height}{\includegraphics [width=185   pt] {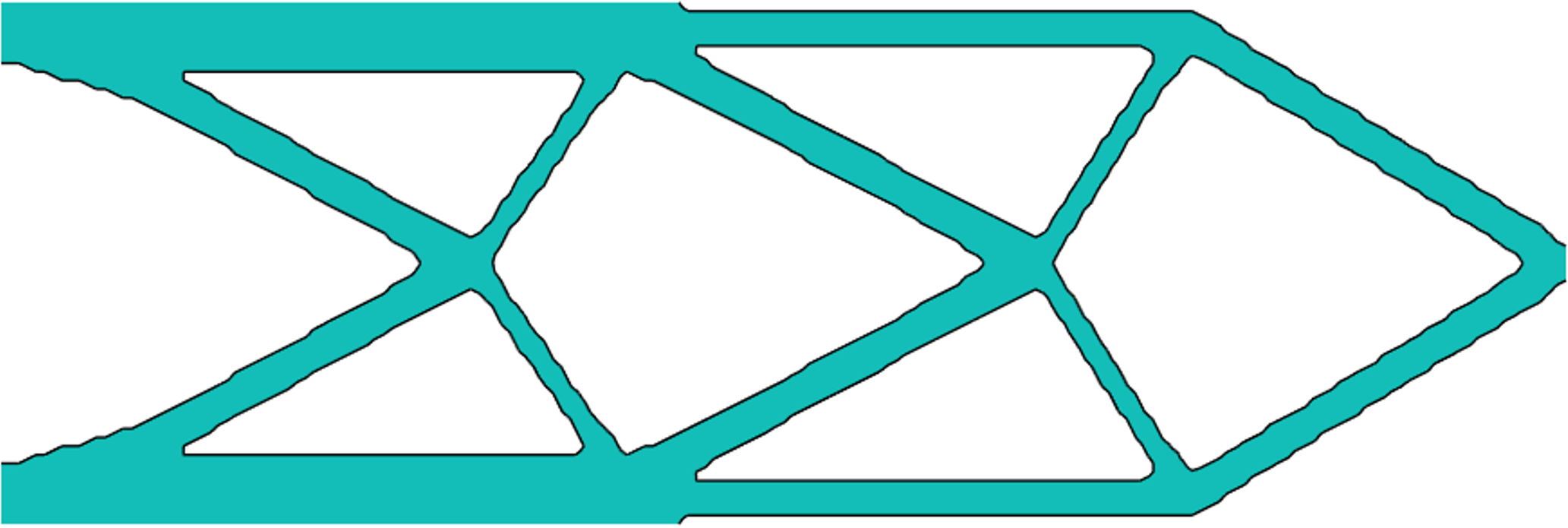}}\\
			\hline
			\hline
		\end{tabular}
	\end{center}
\end{table}

\begin{figure}[htbp!]
	\centering
	\includegraphics [scale=0.7] {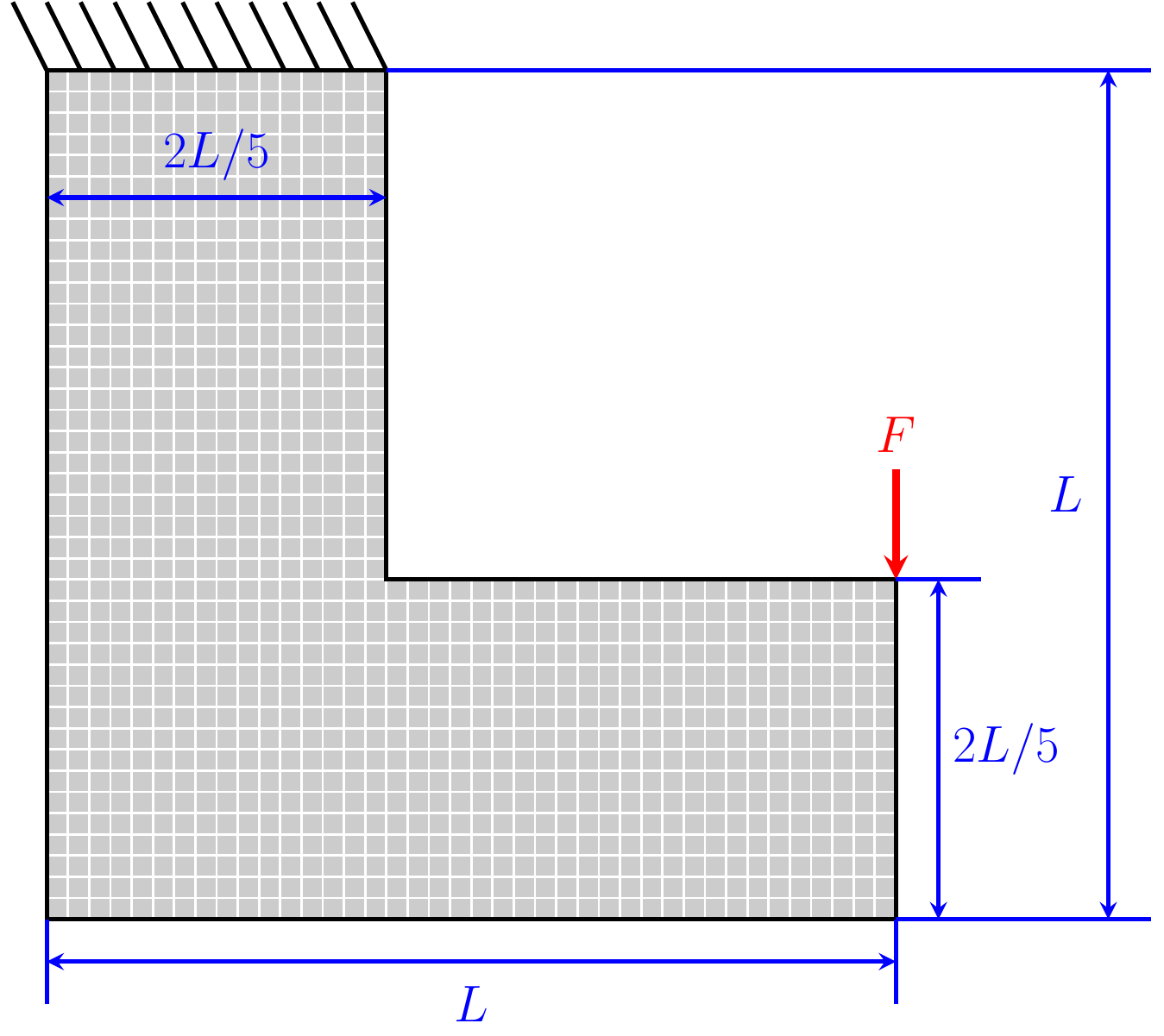}
	\caption{Design domain of an L-bracket beam}
	\label{sFig: Lbracket}
\end{figure}

The second test case is an L-bracket, for which the design domain and boundary condition are shown in Figure \ref{sFig: Lbracket}. Here $L$ is set to 400 element width. The top edge is fixed, and a unit vertical load ($F=-1\mathrm{N}$) is applied at the top corner of the right side. The filter radius $r_{\min}$ is set to 4 time elements width ($r_{\min}$=4). 

\begin{figure}[htbp!]
	\centering
	\begin{subfigure}{\textwidth}
		\centering
		\includegraphics [scale=1] {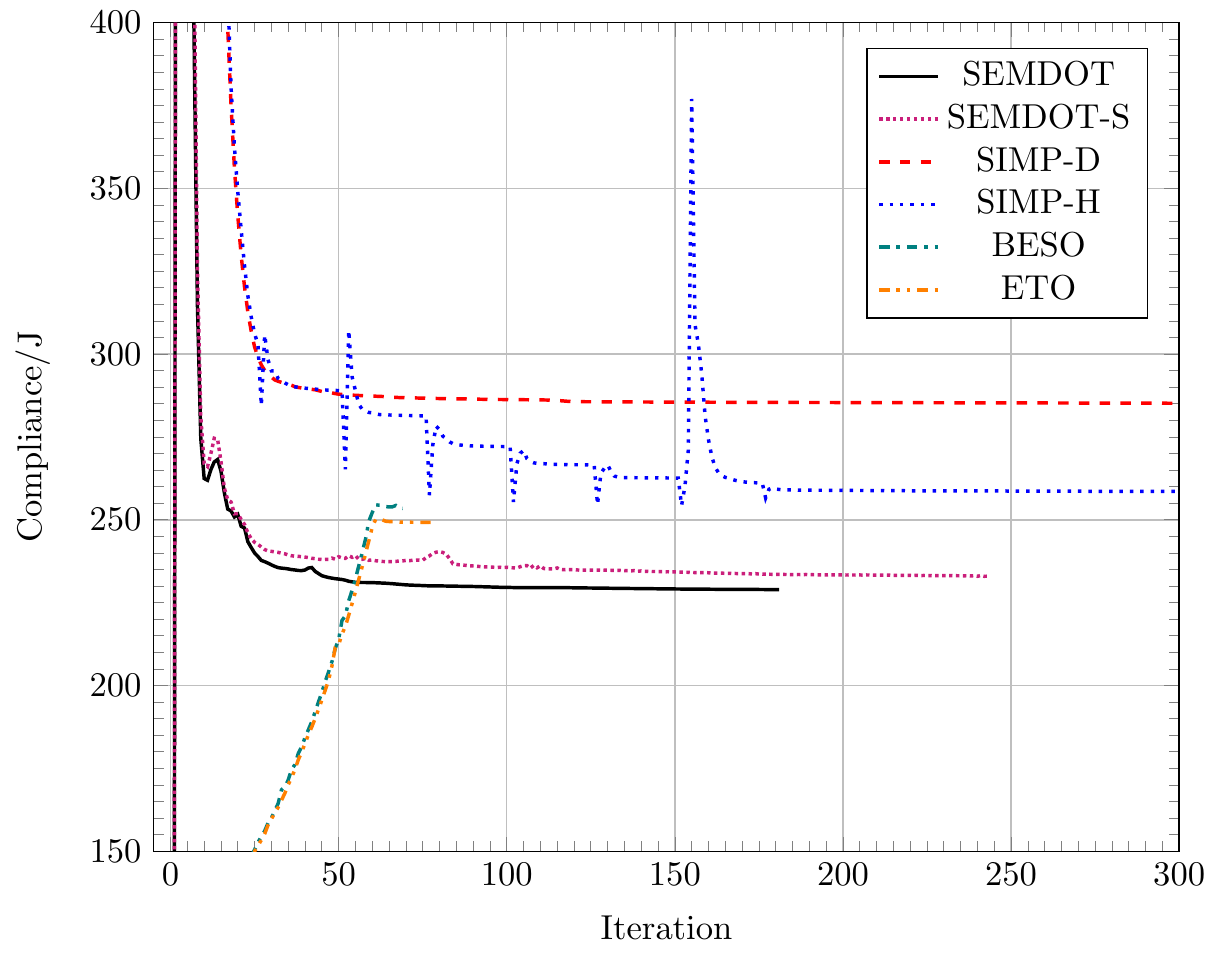}
		\caption{Convergence histories}
		\label{sFig: LCc}
	\end{subfigure}
	\\
	\begin{subfigure}{0.32\textwidth}
		\centering
		\includegraphics [width=145 pt] {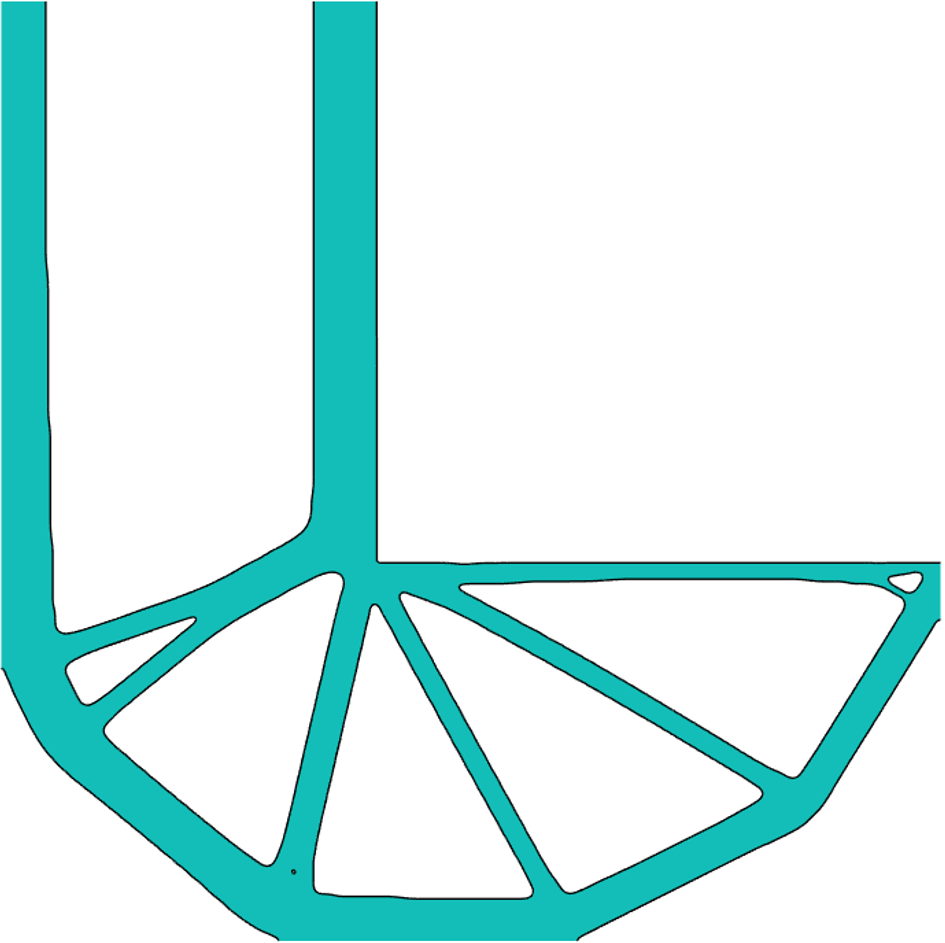}
		\caption{SEMDOT}
		\label{sFig: LSCTO}
	\end{subfigure}
	\begin{subfigure}{0.32\textwidth}
		\centering
		\includegraphics [width=145 pt] {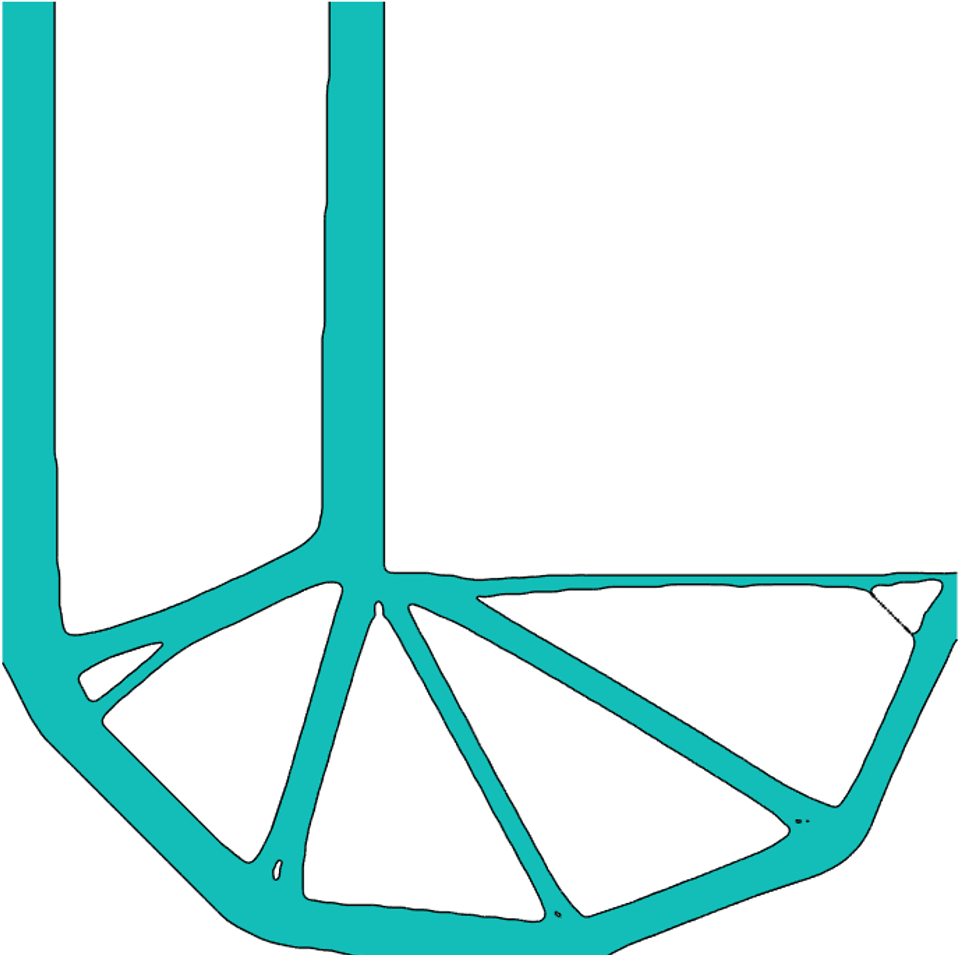}
		\caption{SEMDOT-S}
		\label{sFig: TLSEMDOTS}
	\end{subfigure}
	\begin{subfigure}{0.32\textwidth}
		\centering
		\includegraphics [width=145 pt] {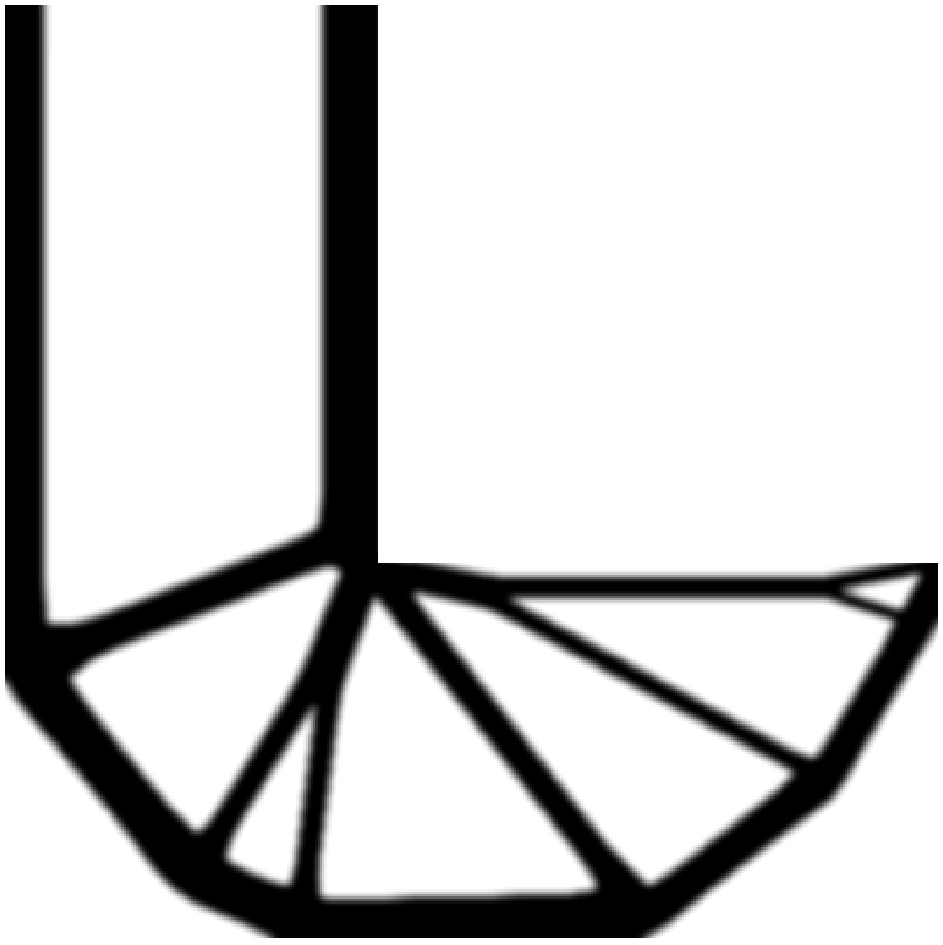}
		\caption{SIMP-D}
		\label{sFig: LDSIMP}
	\end{subfigure}
\\
	\begin{subfigure}{0.32\textwidth}
		\centering
		\includegraphics [width=145 pt] {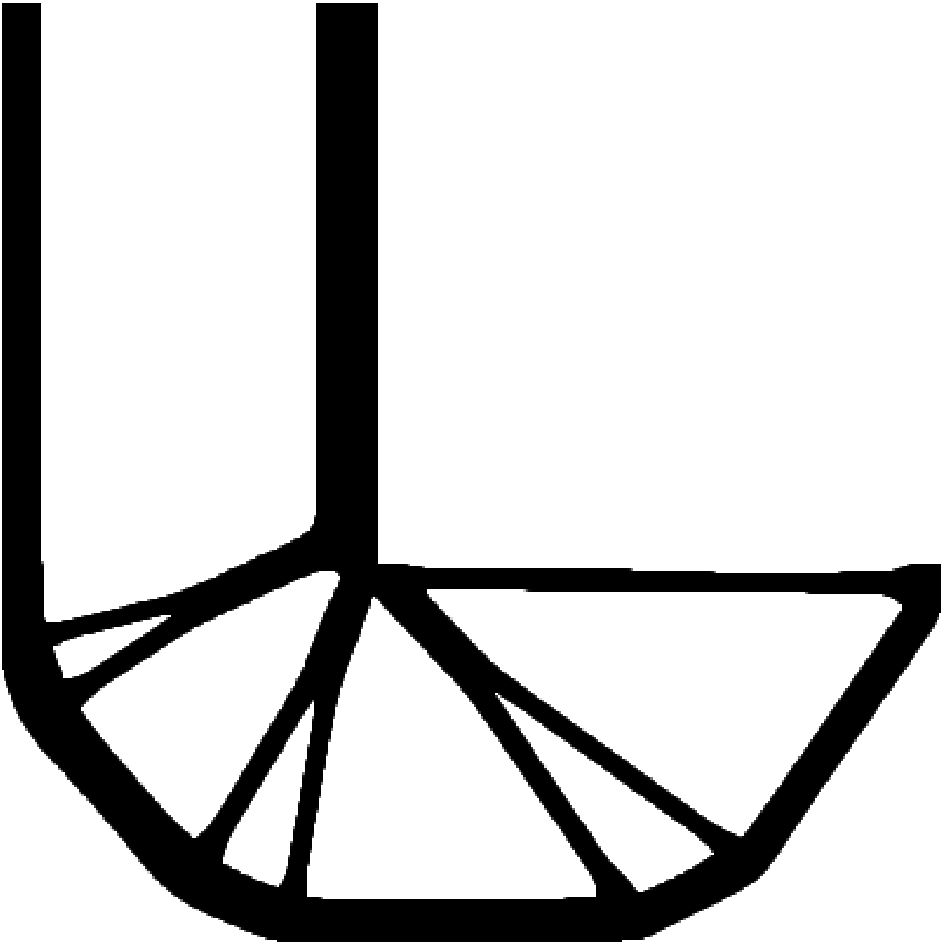}
		\caption{SIMP-H}
		\label{sFig: LHSIMP}
	\end{subfigure}
	\begin{subfigure}{0.32\textwidth}
		\centering
		\includegraphics [width=145 pt] {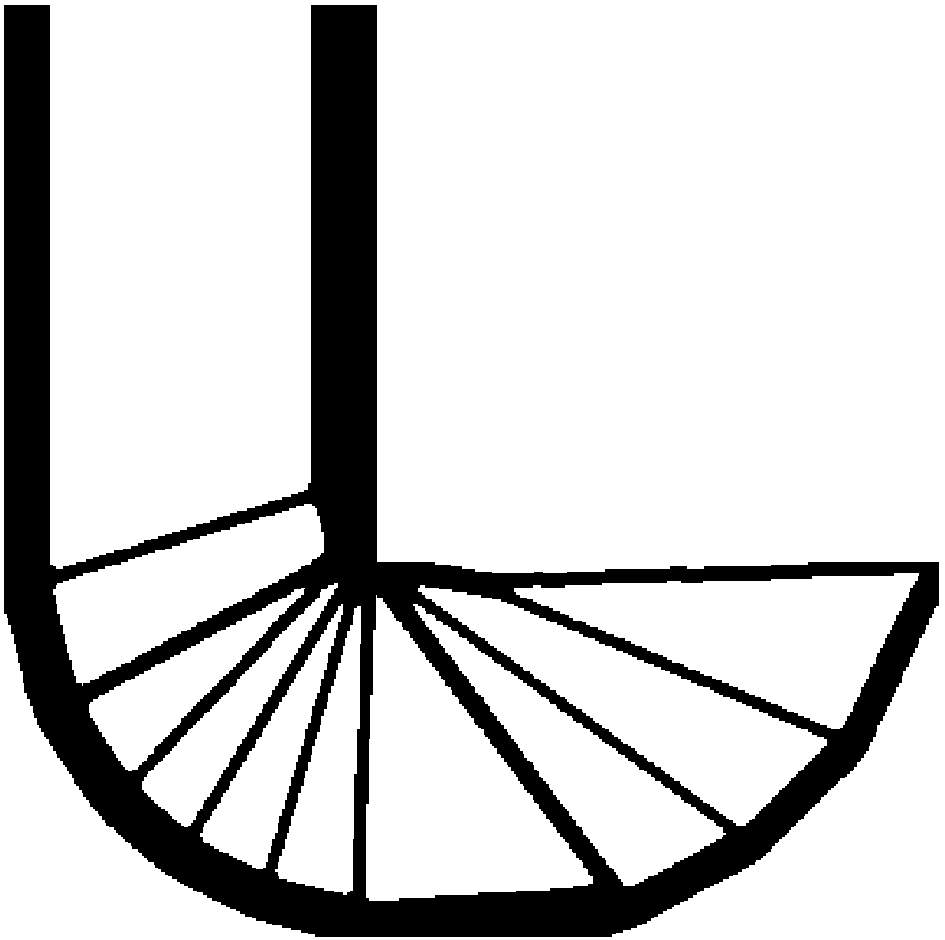}
		\caption{BESO}
		\label{sFig: LBESO}
	\end{subfigure}
	\begin{subfigure}{0.32\textwidth}
		\centering
		\includegraphics [width=145 pt] {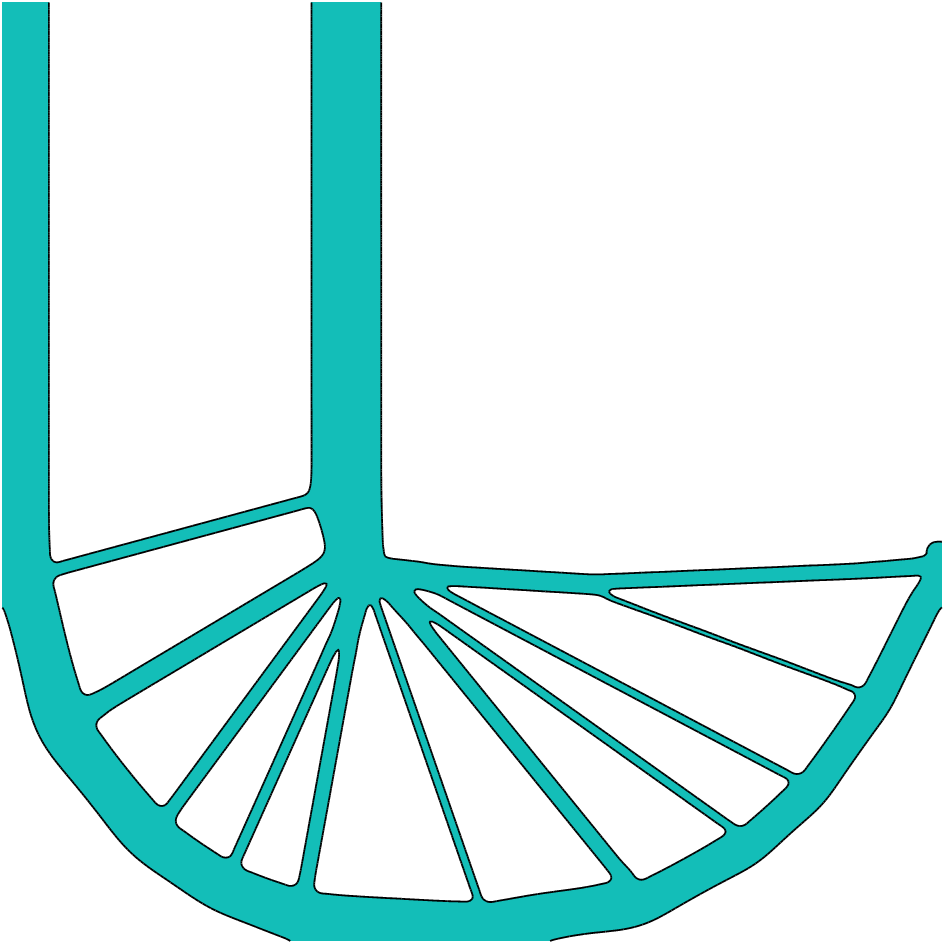}
		\caption{ETO}
		\label{sFig: LETO}
	\end{subfigure}
    \caption{Comparisons of performance, convergency, and topological designs between different algorithms for L-bracket beam}
\end{figure}

In this example, the best compliance (228.9422 J) is obtained by SEMDOT, followed by 232.9864 J obtained by SEMDOT-S (Figure \ref{sFig: LCc}). The optimization process of ETO converges at the compliance of 249.2654 J after 79 iterations where its number of iterations is still less than those of SEMDOT (181 iterations) and SEMDOT-S (243 iterations). Both SIMP algorithms face difficulties in convergence when the fine mesh is used, so optimization processes terminate after reaching the preset maximum number of iterations (300). By contrast, optimization processes of SEMDOT, SEMDOT-S, BESO, and ETO converge within 250 iterations. Interestingly, as is evident from Figure \ref{sFig: LCc}, numerical instabilities occur in the initial 10 iterations of the optimization processes of SEMDOT and SEMDOT-S despite using the Heaviside smooth function. This is because of the joint effects of the non-designable passive area and specific load and boundary conditions. Afterwards, the optimization process quickly settles to a steady pass. Unlike SEMDOT and SIMP (Figures \ref{sFig: LSCTO}, \ref{sFig: LDSIMP}, and \ref{sFig: LHSIMP}), SEMDOT-S, BESO, and ETO are prone to resulting in topological designs with thin features (Figures \ref{sFig: TLSEMDOTS}, \ref{sFig: LBESO}, and \ref{sFig: LETO}), which are not preferred from the manufacturing point of view despite using the same value for the filter radius.

Effects of different mesh sizes and domain aspect ratios are also investigated for the L-bracket beam case. Table \ref{tab:CaLB} demonstrates that performance obtained by SEMDOT is the best for the majority of test cases with different mesh sizes, followed by SEMDOT-S. ETO ranks the third position in terms of performance. Table \ref{tab:CaNI} demonstrates that BESO converges the fastest for different mesh sizes, followed by ETO. SEMDOT performs better than SEMDOT-S in the aspect of convergency for most of cases.
 
\begin{table}[htbp!]
	\begin{center}
		\caption{Effects of mesh size on compliance obtained by different algorithms solving L-bracket beam case}
		\label{tab:CaLB}
		\begin{tabular}{cccccc}
			\hline
			\hline			
			\multirow{2}{*}{Algorithms}	& \multicolumn{5}{c}{Mesh size}\\
			\cmidrule{2-6}
			& $L$=150 & $L$=200 & $L$=250 & $L$=300 & $L$=350  \\
			\hline
			\rowcolor{black!10}
			SEMDOT & 225.8848  & 225.4980  & 229.3275 & 228.3329 & 229.5550  \\
			SEMDOT-S & 225.9183 & 232.6018 & 229.7667 & 231.2625 & 229.9703 \\
			\rowcolor{black!10}
			SIMP-D & 289.0603 & 281.0499 & 286.5398 & 285.2692 & 286.9609 \\
			SIMP-H & 256.3416 & 255.9985 & 261.8570 & 267.4361 & 263.0003  \\
			\rowcolor{black!10}
			BESO & 257.7212 & 257.4816 & 256.4764 & 254.6842 & 253.9468  \\
			ETO & 245.8689 & 247.1456 & 248.7580 & 248.4174 & 249.1365 \\
			\hline
			\hline
		\end{tabular}
	\end{center}
\end{table}

\begin{table}[htbp!]
	\begin{center}
		\caption{Effects of mesh size on number of iterations obtained by different algorithms solving L-bracket beam case}
		\label{tab:CaNI}
		\begin{tabular}{cccccc} 
			\hline
			\hline		
			\multirow{2}{*}{Algorithms}	& \multicolumn{5}{c}{Mesh size}\\
			\cmidrule{2-6}
			& $L$=150 & $L$=200 & $L$=250 & $L$=300 & $L$=350  \\
			\hline
			\rowcolor{black!10}
			SEMDOT & 169 & 245  & 198 & 211 & 224  \\
			SEMDOT-S & 204 & 282  & 208 & 300 & 208  \\
			\rowcolor{black!10}
			SIMP-D & 300 & 300 & 300 & 300 & 300 \\
			SIMP-H & 248 & 205 & 212 & 300 & 300  \\
			\rowcolor{black!10}
			BESO & 68 & 69 & 70 & 74 & 72  \\
			ETO & 86 & 81 & 119 & 82 & 85 \\
			\hline
			\hline
		\end{tabular}
	\end{center}
\end{table}

To further demonstrate the difference between SEMDOT and SEMDOT-S, topological designs obtained by SEMDOT and SEMDOT-S for the test cases considering different mesh sizes are listed in Table \ref{tab:CaNIT}. Similar with the cantilever beam case, SEMDOT-S forms thin features when mesh sizes are $L$=250 and $L$=300, and discontinuous structures when mesh sizes are $L$=150 and $L$=350. The mentioned structural features are not observed in the topological designs obtained by SEMDOT.

\begin{table}[htbp!]
	\begin{center}
		\caption{Comparisons of topologies with different mesh sizes between SEMDOT and SEMDOT-S for L-bracket beam case}
		\label{tab:CaNIT}
		\begin{tabular}{ccc} 
			\hline
			\hline		
			\multirow{2}{*}{Mesh size}	& \multicolumn{2}{c}{Algorithms}\\
			\cmidrule{2-3}
			& SEMDOT & SEMDOT-S   \\
			\hline
			\rowcolor{black!10}
			$L$=150 & \raisebox{-.5\height}{\includegraphics [width=120 pt] {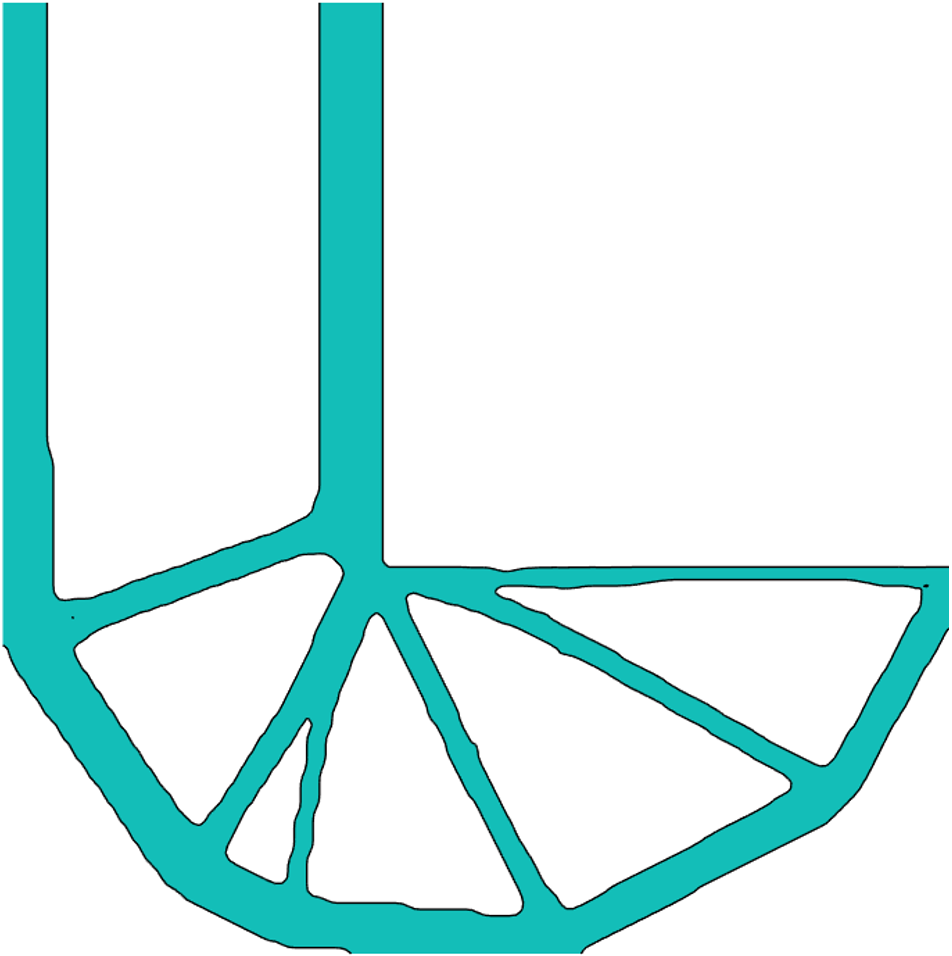}} & \raisebox{-.5\height}{\includegraphics [width=120 pt] {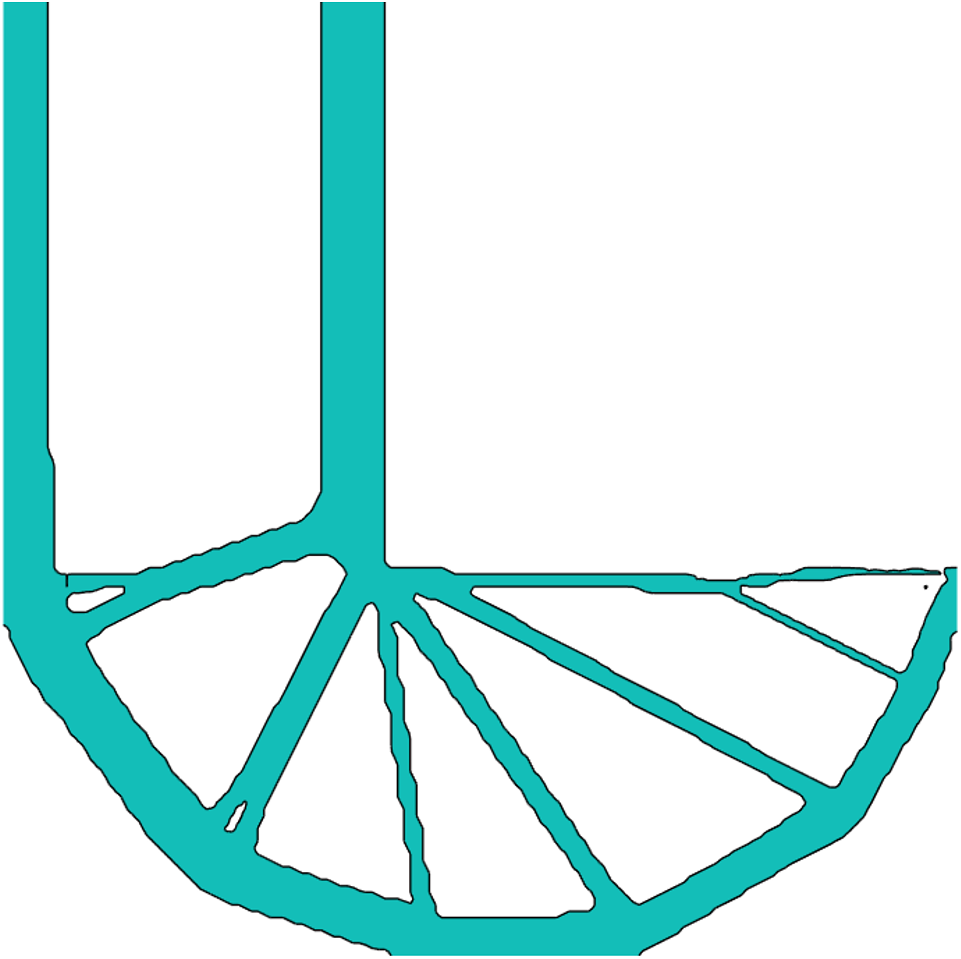}}\\
			$L$=200 & \raisebox{-.5\height}{\includegraphics [width=120 pt] {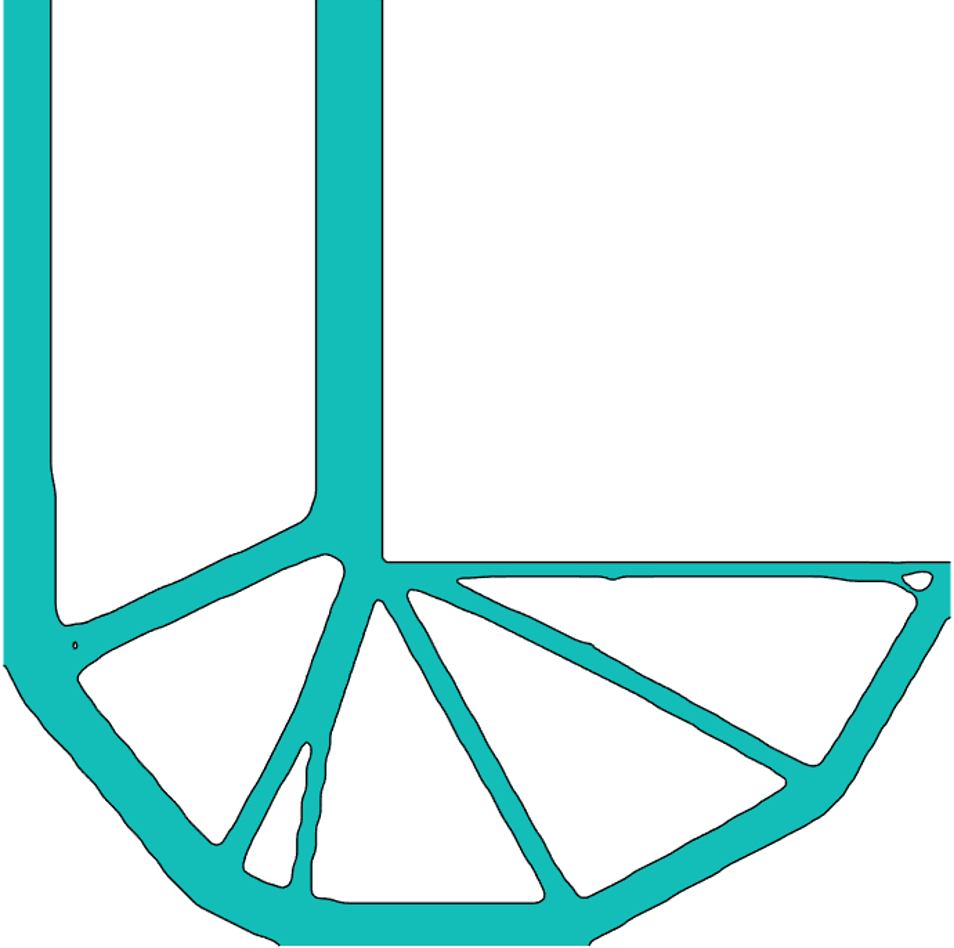}} & \raisebox{-.5\height}{\includegraphics [width=120 pt] {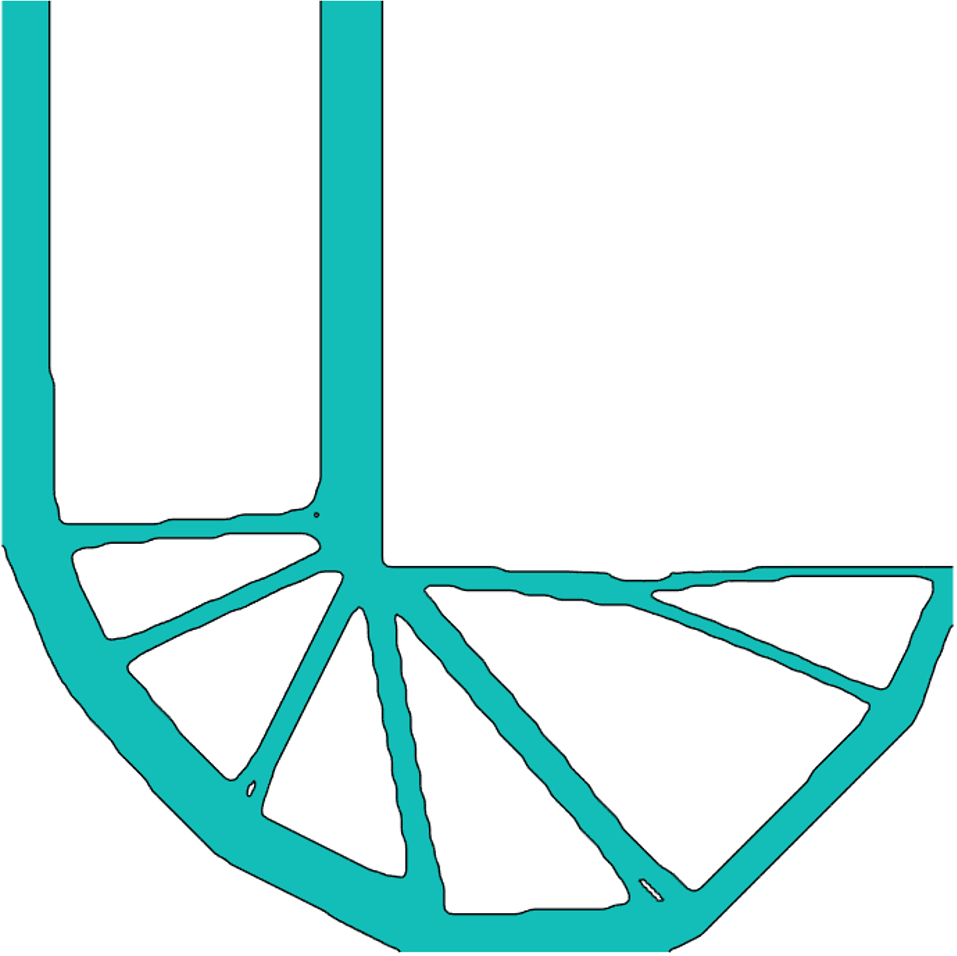}}\\
			\rowcolor{black!10}
			$L$=250 & \raisebox{-.5\height}{\includegraphics [width=120 pt] {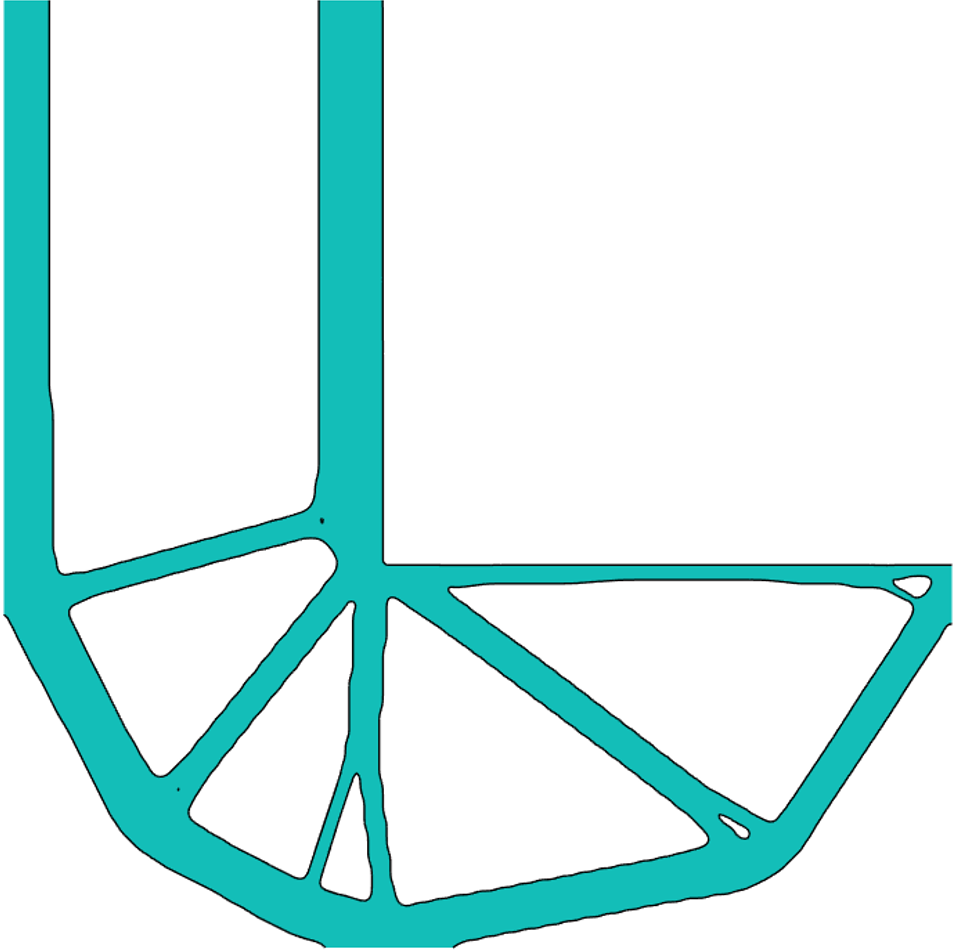}} & \raisebox{-.5\height}{\includegraphics [width=120 pt] {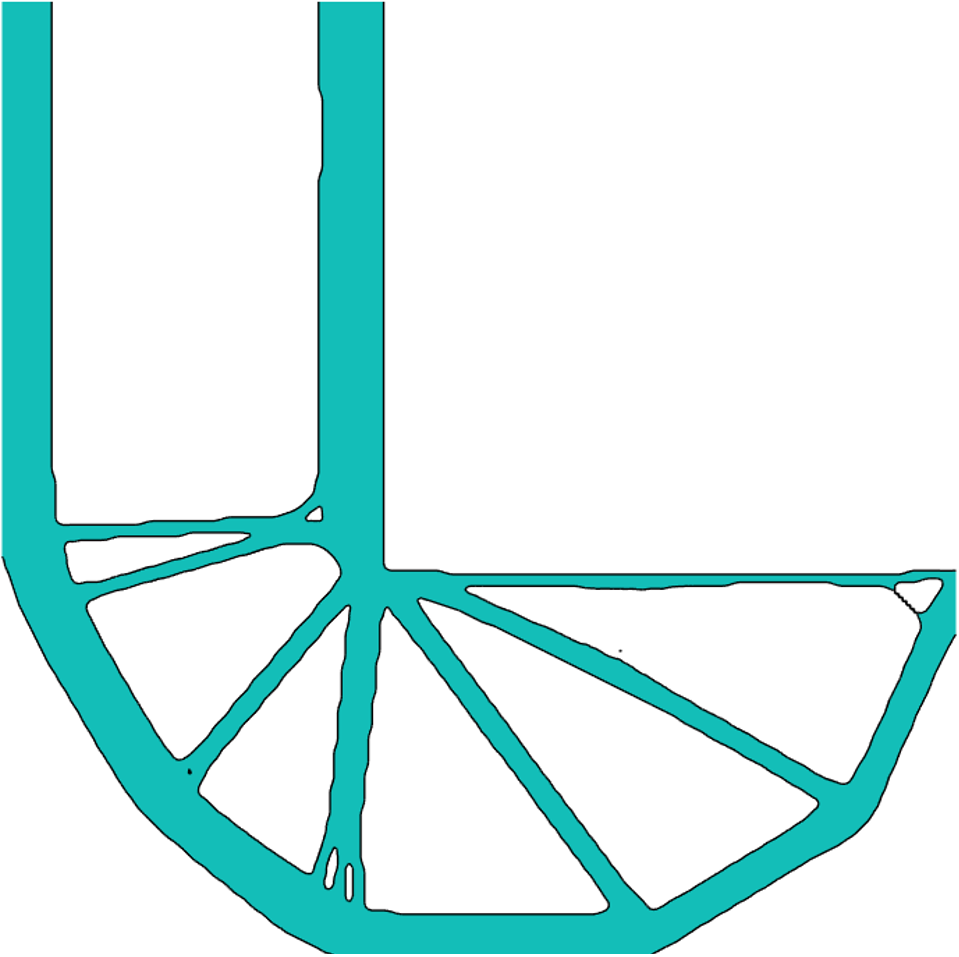}}\\
			$L$=300 & \raisebox{-.5\height}{\includegraphics [width=120 pt] {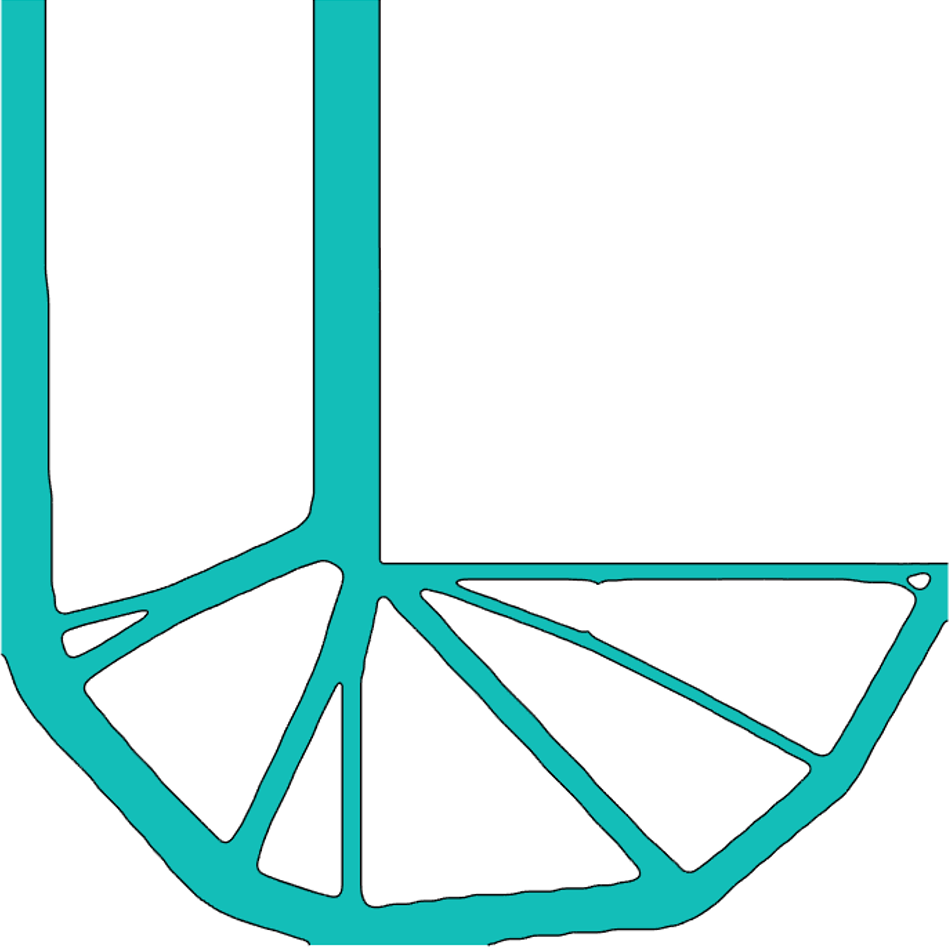}} & \raisebox{-.5\height}{\includegraphics [width=120 pt] {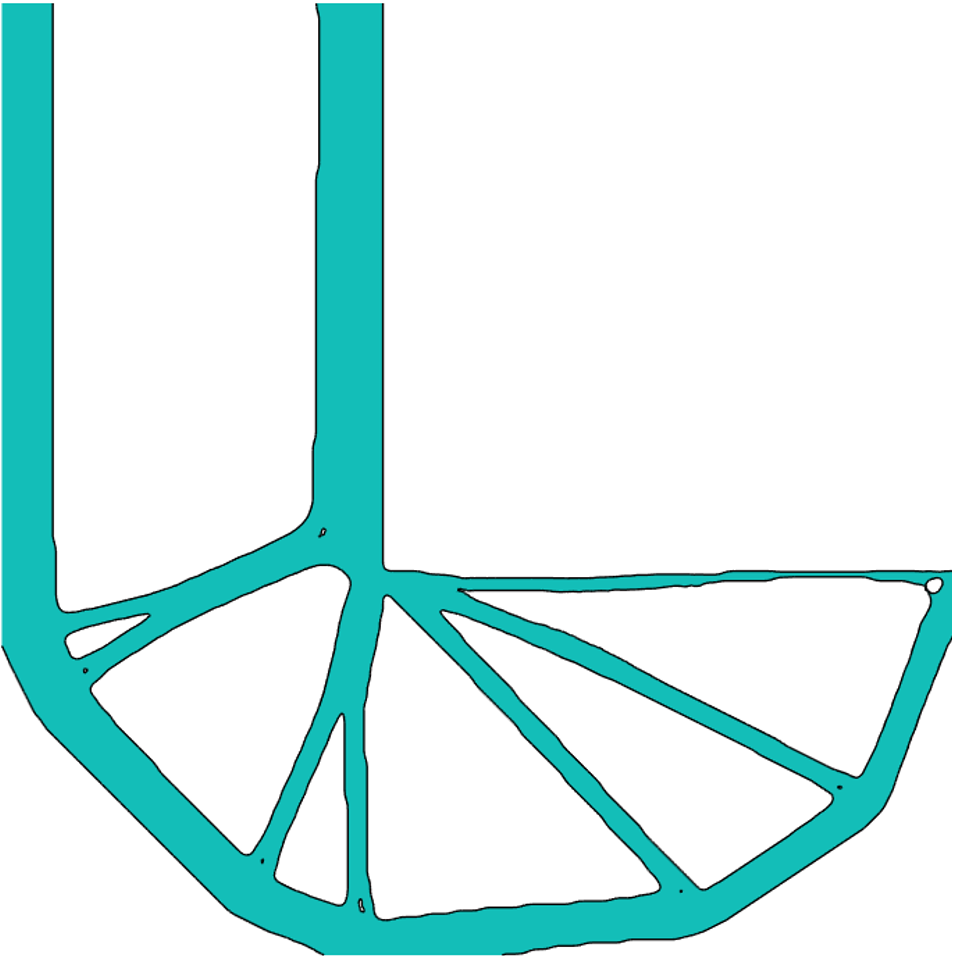}}\\
			\rowcolor{black!10}
			$L$=350 & \raisebox{-.5\height}{\includegraphics [width=120 pt] {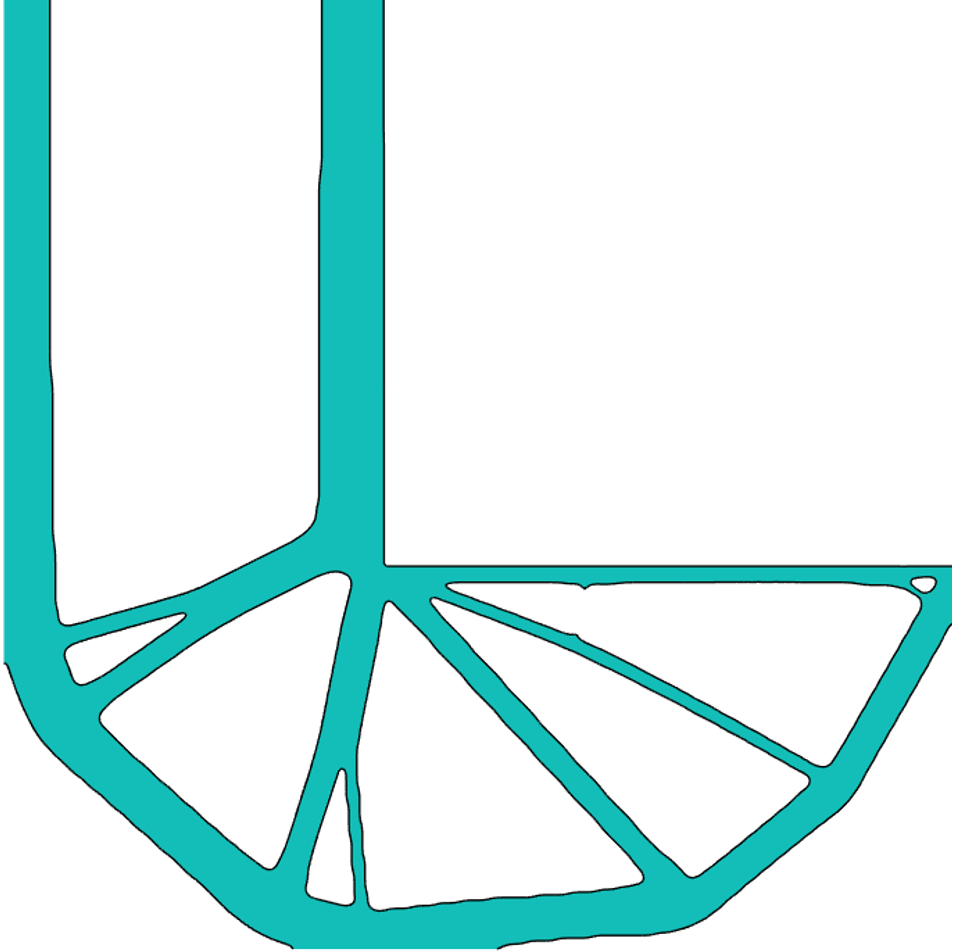}} & \raisebox{-.5\height}{\includegraphics [width=120 pt] {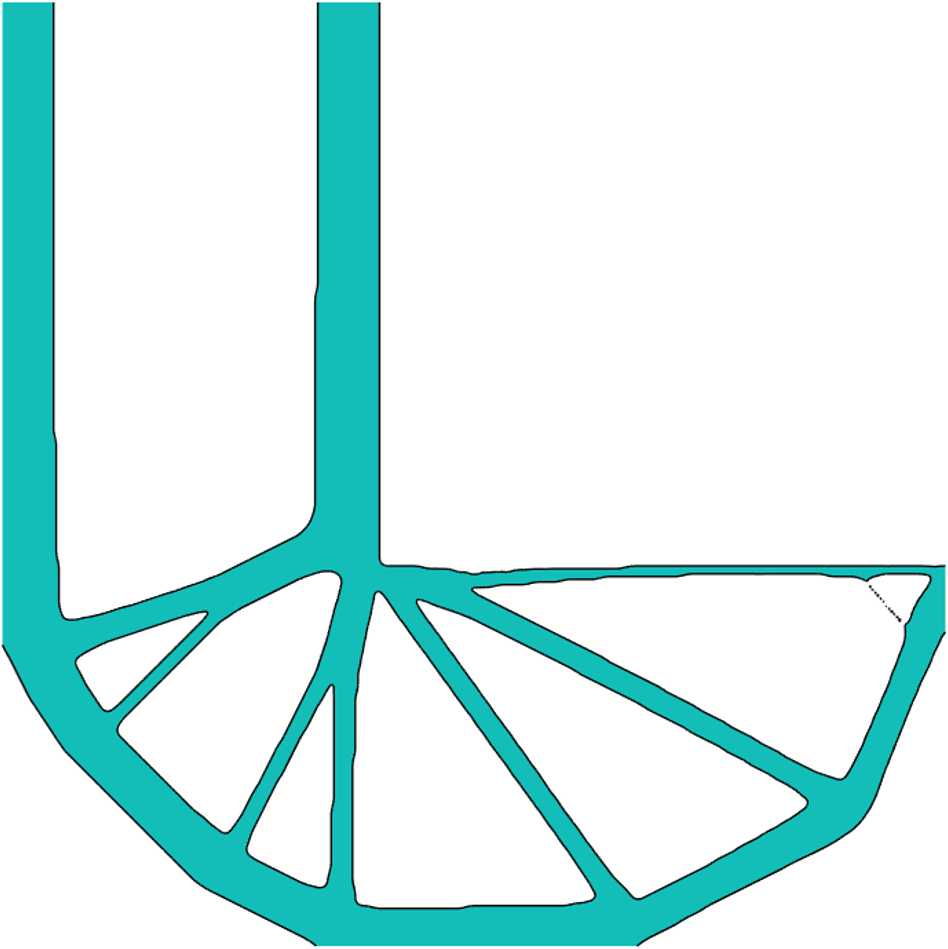}}\\
			\hline
			\hline
		\end{tabular}
	\end{center}
\end{table}

In terms of different domain aspect ratios, the number of elements in the vertical direction is fixed at 200, and the filter radius $r_{\min}$ is set to 2 time elements width ($r_{\min}$=2). Table \ref{tab:LCDR} demonstrates that performance obtained by SEMDOT and SEMDOT-S is better than ETO, and SEMDOT generally obtains better or comparable performance compared to SEMDOT-S. Table \ref{tab:NLCDR} demonstrates that SEMDOT performs better than SEMDOT-S in terms of convergency for most of test cases, and BESO and ETO generally converge faster than other algorithms.

\begin{table}[htbp!]
	\begin{center}
		\caption{Effects of domain aspect ratio on compliance obtained by different algorithms solving L-bracket beam case}
		\label{tab:LCDR}
		\begin{tabular}{cccccccc} 
			\hline
			\hline		
			\multirow{2}{*}{Algorithms}	& \multicolumn{6}{c}{Domain aspect ratio}\\
			\cmidrule{2-7}
			& 0.5:1 & 0.75:1  & 1.25:1 & 1.5:1 & 1.75:1 & 2:1 \\
			\hline
			\rowcolor{black!10}
			SEMDOT &  254.4356  & 229.8695 & 245.7824 & 285.4968 & 314.0988 & 358.8990  \\
			SEMDOT-S & 256.0683 & 232.1893 & 243.5014 & 285.5149 & 318.7357 & 360.8157  \\
			\rowcolor{black!10}
			SIMP-D & 297.6136 & 281.7546 & 317.8464 & 372.5226 & 449.1221  & 542.3667  \\
			SIMP-H & 282.6777 & 322.5282 & 288.8244 & 355.5204 & 454.8071 & 498.2979  \\
			\rowcolor{black!10}
			BESO &  278.3044 &  253.3296 & 284.5204 & 335.2234 & 401.4278 & 490.0256  \\
			ETO & 269.4174 & 244.8819 & 276.9958 & 321.1281 & 387.5802 & 476.9035  \\
			\hline
			\hline
		\end{tabular}
	\end{center}
\end{table}

\begin{table}[htbp!]
	\begin{center}
		\caption{Effects of domain aspect ratio on number of iterations obtained by different algorithms solving L-bracket beam case}
		\label{tab:NLCDR}
		\begin{tabular}{cccccccc} 
			\hline
			\hline		
			\multirow{2}{*}{Algorithms}	& \multicolumn{6}{c}{Domain aspect ratio}\\
			\cmidrule{2-7}
			& 0.5:1 & 0.75:1  & 1.25:1 & 1.5:1 & 1.75:1 & 2:1 \\
			\hline
			\rowcolor{black!10}
			SEMDOT &  104  & 202 & 293 & 116 & 138 & 113  \\
			SEMDOT-S & 128 & 300 & 192 & 148 & 130 & 122   \\
			\rowcolor{black!10}
			SIMP-D & 300 & 300 & 300 & 300 & 300 & 300  \\
			SIMP-H & 181 & 234 & 213 & 201 & 214 & 230  \\
			\rowcolor{black!10}
			BESO &  71 &  68 & 78 & 77 & 78 & 68  \\
			ETO & 97 & 90 & 218 & 89 & 98 & 90  \\
			\hline
			\hline
		\end{tabular}
	\end{center}
\end{table}

Although BESO shows the fastest convergency in these two test cases, this may be because of the premature termination mentioned in Section \ref{Sec: CONC}, and the same issue seems to occur in ETO because of using the same termination criterion. Otherwise, BESO and ETO could have the potential to obtain better performance. ETO converges slower than BESO as it uses elemental volume fractions in the FEA model instead of elemental densities. On the contrary, SEMDOT converges faster than SIMP because of distributing intermediate elements only along boundaries and introducing new termination criteria (Equations \ref{Eq: Terror} and \ref{Eq: Berror}). Based on the above discussion, SEMDOT-S easily obtains thin features and discontinuous structures, and performance obtained by SEMDOT-S is generally worse than SEMDOT for test cases with the fine mesh (refer to Tables \ref{tab:CaC}, \ref{tab:CaCDR}, \ref{tab:CaLB}, and \ref{tab:LCDR}). Therefore, it can be concluded that the use of multiple filtering steps is more suitable than single filtering step in SEMDOT.

%%%%%%%%%%%%%%%%%%%%%%%%%%%%%%%%%%%%%%%%%%
\section{Conclusions} \label{Sec: Conclusions}
This paper explains the algorithm mechanism of SEMDOT using several numerical examples. Performance of SEMDOT is demonstrated through numerical comparisons with a range of well-established element-based algorithms. Concluding remarks can be stated as follows:

\begin{itemize}
	\item The Heaviside smooth function is more suitable than the Heaviside step function for SEMDOT to obtain a more robust algorithm. 
	
	\item The use of multiple filtering steps enhances the flexibility of SEMDOT in exploring better performance and different topological designs, and the use of multiple filtering step is more suitable for SEMDOT than the single filtering step. 
	
	\item The sensitivity analysis strategy adopted in SEMDOT is effective.
	
	\item Even though SEMDOT is developed based on the SIMP framework, its convergency is stronger than standard SIMP because of its improved termination criteria.
	
	\item SEMDOT is capable of obtaining topological designs comparable or better than standard element-based algorithms such as SIMP and BESO or the newly developed ETO.
	
\end{itemize}

Even though this paper shows some benefits of SEMDOT, it should be acknowledged that when the same number of elements is used, the computational cost of SEMDOT would be higher than that of SIMP because of having to deal with extra grid points. It also should be acknowledged that SEMDOT is not a pioneering algorithm like SIMP, BESO, level-set method, MMC-based method, and using the floating projection. Instead, SEMDOT is an easy-to-use, flexible, and efficient optimization platform, which can be easily integrated with some existing approaches and solutions, particularly the ones developed for SIMP. 

%\section*{\color{red} CRediT Authorship Contribution Statement} 
{\color{red} }
%%%%%%%%%%%%%%%%%%%%%%%%%%%%%%%%%%%%%%%%%%
\section*{Funding}
This research received no external funding.
%%%%%%%%%%%%%%%%%%%%%%%%%%%%%%%%%%%%%%%%%%
\section*{Acknowledgements}
The authors would like to thank Prof. Krister Svanberg for providing the MMA optimizer code. Mr. Yun-Fei Fu would like to thank Dr. Joe Alexandersen, University of Southern Denmark, for his suggestions on the mathematically statement of SEMDOT and Dr. Hongxin Wang, Hunan University, for his assistance on BESO test cases.
%%%%%%%%%%%%%%%%%%%%%%%%%%%%%%%%%%%%%%%%%%
\medskip
\bibliographystyle{elsarticle-num-names}
\bibliography{2DTO}
%%%%%%%%%%%%%%%%%%%%%%%%%%%%%%%%%%%%%%%%%%
\appendix
\section*{Appendix: Matlab Code of SEMDOT} \label{Sec: Appendix}
The 2D code of SEMDOT was written based on the Matlab codes presented by \citet{andreassen2011efficient} and \citet{huang2010further}. To use the MMA optimizer, two additional files: mmasub.m and subsolv.m, which can be obtained by contacting Prof. Krister Svanberg from KTH in Stockholm Sweden, are needed. The implementation of SEMDOT is not limited to the code provided here. The code should be regarded as a reference, and therefore the readers are encouraged to improve the code based on the idea of SEMDOT. In addition, the 2D Matlab code of SEMDOT can be easily extended to a 3D version based on the 169 line code presented by \citet{liu2014efficient}.

\begin{lstlisting}
%%%% A Matlab Code for SEMDOT Algorithm %%%%
function SEMDOT(nelx,nely,vol,rmin,nG)
% Author One: Yun-Fei Fu <fuyunf@deakin.edu.au>
% Topology Optimization Group
% School of Engineering, Deakin University
% Author Two: Prof. Xiaodong Huang <xhuang@swin.edu.au>
% School of Engineering, Swinburne University of Technology
% Advisors: Dr. Kazem Ghabraie; Prof. Bernard Rolfe; Dr. Yanan Wang;
% Dr. Louis N.S. Chiu
% Date Created: 12/08/18
% Date last modified: 03/01/20
% Description:
% This function SEMDOT implements Smooth-Edged Material Distribution for
% Optimizing Topology (SEMDOT) algorithm

%%% Disclaimer:
%%%% This code is provided for educational or academic purposes and is not
%%%% guaranteed to be free of errors. The authors are not liable for any
%%%% problems caused by the use of this code.

% INPUT:
% nelx: The number of elements in the horizontal direction.
% nely: The number of elements in the vertical direction.
% vol: The prescribed value of the allowable volume.
% rmin: The predefined filter radius.
% nG: The number of grid points assigned to each element.

% OUTPUT:
% It.: The iteration number.
% Obj.: The value of compliance.
% Vol.: The volume fraction.
% ch.: The computational error.
% Topo.: The topological error.
% contourf(fnx, flipud(fny), xg-ls, [0 0]) shows optimized topologies.

%% INITIALIZATION
vx = repmat(vol,nely,nelx); vxPhys = vx; change = 1; loop = 0;
Emin = 0.001; maxloop = 1000; ngrid = nG-1; tolx = 0.001;
E0 = 1; nu = 0.3; penal = 1.5; rnmin = 1;
%% INITIALIZE MMA OPTIMIZER
nele = nely*nelx; m = 1; n = nely*nelx; nelm = nely*nelx;
vxmin = 1e-3*ones(nelm,1); vxmax = ones(nelm,1);
vxold1 = reshape(vx,nelm,1); vxold2 = vxold1;
a0 = 1; ai = 0; ci = 1000; di = 0; low = ones(nelm,1); upp = 1;
%% INITIALIZE HEAVISIDE SMOOTH FUNCTION
beta = 0.5; ER = 0.5;
%% PREPARE FINITE ELEMENT ANALYSIS
A11 = [12  3 -6 -3;  3 12  3  0; -6  3 12 -3; -3  0 -3 12];
A12 = [-6 -3  0  3; -3 -6 -3 -6;  0 -3 -6  3;  3 -6  3 -6];
B11 = [-4  3 -2  9;  3 -4 -9  4; -2 -9 -4 -3;  9  4 -3 -4];
B12 = [ 2 -3  4 -9; -3  2  9 -2;  4  9  2  3; -9 -2  3  2];
KE = 1/(1-nu^2)/24*([A11 A12;A12' A11]+nu*[B11 B12;B12' B11]);
nodenrs = reshape(1:(1+nelx)*(1+nely),1+nely,1+nelx);
edofVec = reshape(2*nodenrs(1:end-1,1:end-1)+1,nelx*nely,1);
edofMat = repmat(edofVec,1,8)+repmat([0 1 2*nely+[2 3 0 1] -2 -1],nelx*nely,1);
iK = reshape(kron(edofMat,ones(8,1))',64*nelx*nely,1);
jK = reshape(kron(edofMat,ones(1,8))',64*nelx*nely,1);
%% ELEMENTAL NODES AND COORDINATES
[nodex,nodey] = meshgrid(0:nelx,0:nely);
[fnx,fny] = meshgrid(0:1/ngrid:nelx,0:1/ngrid:nely);
%% DEFINE LOADS AND SUPPORTS (CANTILEVER BEAM)
F = sparse(2*(nely+1)*(nelx+1)-nely,1,-1,2*(nely+1)*(nelx+1),1);
fixeddofs = (1:2*(nely+1));
U = zeros(2*(nely+1)*(nelx+1),1);
alldofs = [1:2*(nely+1)*(nelx+1)];
freedofs = setdiff(alldofs,fixeddofs);
%% PREPARE FILTER FOR ELEMENT
iH = ones(nelx* nely*(2*(ceil(rmin)-1)+1)^2,1);
jH = ones(size(iH));
sH = zeros(size(iH));
k = 0;
for i1 = 1:nelx
    for j1 = 1:nely
        e1 = (i1-1)*nely+j1;
        for i2 = max(i1-(ceil(rmin)-1),1):min(i1+(ceil(rmin)-1),nelx)
            for j2 = max(j1-(ceil(rmin)-1),1):min(j1+(ceil(rmin)-1),nely)
                e2 = (i2-1)*nely+j2;
                k = k+1;
                iH(k) = e1;
                jH(k) = e2;
                sH(k) = max(0,rmin-sqrt((i1-i2)^2+(j1-j2)^2));
            end
        end
    end
end
H = sparse(iH,jH,sH); Hs = sum(H,2);
%% PREPARE FILTER FOR NODALS
inH = ones((nelx+1)*(nely+1)*(2*(ceil(rnmin)+1))^2,1);
jnH = ones(size(inH)); snH = zeros(size(inH)); k =0;
[elex,eley] = meshgrid(1.5:nelx+0.5,1.5:nely+0.5);
for in1 = 1:nelx+1
    for jn1 = 1:nely+1
        en1 = (in1-1)*(nely+1)+jn1;
        for in2 = max(in1-ceil(rnmin),1):min(in1+ceil(rnmin)-1,nelx)
            for jn2 = max(jn1-ceil(rnmin),1):min(jn1+ceil(rnmin)-1,nely)
                en2 = (in2-1)*nely+jn2; k = k+1; inH(k) = en1;
                jnH(k) = en2;
                snH(k) = max(0,rnmin-sqrt((in1-elex(jn2,in2))^2+(jn1-eley(jn2,in2))^2));
            end
        end
    end
end
Hn = sparse(inH,jnH,snH); Hns = sum(Hn,2);
%% START ITERATION
while (change > tolx || tol>0.001) && loop < maxloop
    loop = loop+1;
    %% FE-ANALYSIS
    sK = reshape(KE(:)*(vxPhys(:)'*E0+(1-vxPhys(:))'*(Emin^penal*E0)),64*nelx*nely,1);
    K = sparse(iK,jK,sK); K = (K+K')/2;
    U(freedofs) = K(freedofs,freedofs)\F(freedofs);
    %% OBJECTIVE FUNCTION AND SENSITIVITY ANALYSIS
    ce = reshape(sum((U(edofMat)*KE).*U(edofMat),2),nely,nelx);
    c(loop) = sum(sum((vxPhys.*E0+(1-vxPhys).*(Emin^penal*E0)).*ce));
    dc = -penal*((1-vxPhys)*Emin.^(penal-1)+vxPhys).*E0.*ce;
    dv = ones(nely,nelx);
    %% FILTERING/MODIFICATION OF SENSITIVITIES
    dc(:) = H*(dc(:)./Hs);
    dv(:) = H*(dv(:)./Hs);
    %%  UPDATE DESIGN VARIABLES AND FILTERED ELEMENTAL VOLUME FRACTIONS
    vxval = reshape(vx,nelm,1);
    fval = sum(vxPhys(:))/(vol*nely*nelx)-1;
    dfdx = dv(:)'/(vol*nely*nelx);
    f0val = c;
    df0dx = dc(:);
    [vxmma,~,~,~,~,~,~,~,~,low,upp] = ...
        mmasub(m,n,loop,vxval,vxmin,vxmax,vxold1,vxold2,f0val,df0dx,fval,dfdx,low,upp,a0,ai,ci,di);
    vxnew = reshape(vxmma,nely,nelx);
    vxPhys(:) = (H*vxnew(:))./Hs;
    vxold2 = vxold1(:);
    vxold1 = vx(:);
    %% ASSIGN FILTERED ELEMENTAL VOLUME FRACTIONS TO NODAL DENSITIES
    xn = reshape((Hn*vxPhys(:)./Hns),nely+1,nelx+1);
    %% UPDATE POINT DESNIGY BY A HEAVISIDE SMOOTH/STEP FUNCTION
    xg = interp2(nodex,nodey,xn,fnx,fny,'linear');
    l1 =0; l2 = 1;
    while (l2-l1) > 1.0e-5
        ls = (l1+l2)/2.0;
        xgnew = max(0.001,(tanh(beta*ls)+tanh(beta*(xg-ls)))/(tanh(beta*ls)+tanh(beta*(1-ls))));
        if sum(sum(xgnew))/((ngrid*nelx+1)*(ngrid*nely+1)) - sum(vxPhys(:))/(nelx*nely) > 0
            l1 = ls;
        else
            l2 = ls;
        end
    end
    %% ASSEMBLE GRID POINTS TO ELEMENTS
    vxPhys(:) = 0;
    Terr = 0;
    Tm=[];
    for i = 1:nelx
        for j = 1:nely
            e = (i-1)*nely + j;
            for i1 = ngrid*(i-1)+1:ngrid*i+1
                for j1 = ngrid*(j-1)+1:ngrid*j+1
                    Tm = [Tm;xgnew(j1,i1)];
                    vxPhys(e) = vxPhys(e)+xgnew(j1,i1);
                end
            end
            if min(Tm)>0.001 && max(Tm)<1
                Terr = Terr+1;
            end
            Tm = [];
        end
    end
    vxPhys = vxPhys/(ngrid+1)^2;
    %% CHECK CONVERGENCE
    change = sum(abs(vxnew(:)-vx(:)))/(vol*nely*nelx);
    tol = Terr/nele;
    vx = vxnew;
    %% PLOT RESULTS
    fprintf('It.:%3i Obj.:%8.4f Vol.:%4.3f ch.:%4.5f Topo.:%7.5f\n\n',loop,c(loop),mean(vxPhys(:)),change,tol);
    contourf(fnx, flipud(fny), xg-ls, [0 0]);
    axis equal;  axis tight; axis off; pause(1e-6);
    %% UPDATE HEAVISIDE SMOOTH FUNCTION
    beta = beta+ER; fprintf('Parameter beta increased to %g.\n',beta);
end

\end{lstlisting}

\end{document}